\documentclass[superscriptaddress,preprintnumbers,amssymb]{revtex4}
\usepackage{graphicx}
\usepackage{epsfig}
\usepackage{dcolumn}
\usepackage{bm}
\usepackage{amsmath}
\usepackage{amssymb}
\usepackage[svgnames]{xcolor}

\def\be{\begin{equation}}
\def\ee{\end{equation}}
\def\bea{\begin{eqnarray}}
\def\eea{\end{eqnarray}}
\def\nnb{\nonumber}

\newcommand{\f}{\frac}

\def\lsim{\:\raisebox{-0.5ex}{$\stackrel{\textstyle<}{\sim}$}\:}
\def\gsim{\:\raisebox{-0.5ex}{$\stackrel{\textstyle>}{\sim}$}\:}

\def\gsim{\lower0.5ex\hbox{$\:\buildrel >\over\sim\:$}}
\def\lsim{\lower0.5ex\hbox{$\:\buildrel <\over\sim\:$}}

\begin{document}

\preprint{SI-HEP-2012-13}

\title{Two Higgs doublets, a 4th generation and a 125 GeV Higgs: a review}

\author{Shaouly Bar-Shalom}
\email{shaouly@physics.technion.ac.il}
\affiliation{Physics Department, Technion-Institute of Technology, Haifa 32000, Israel}
\author{Michael Geller}
\email{mic.geller@gmail.com}
\affiliation{Physics Department, Technion-Institute of Technology, Haifa 32000, Israel}
\author{Soumitra Nandi}
\email{soumitra.nandi@gmail.com}
\affiliation{Theoretische Elementarteilchenphysik,
Naturwissenschaftlich Technische Fakult\"at, \\
Universit\"at Siegen, 57068 Siegen, Germany}
\author{Amarjit Soni}
\email{soni@bnl.gov}
\affiliation{Theory Group, Brookhaven National Laboratory, Upton, NY 11973, USA}

\date{\today}

\begin{abstract}
We review the possible role that multi-Higgs models may play in our
understanding of the dynamics of a heavy 4th sequential generation of fermions.
We describe the underlying ingredients of such models, focusing on two Higgs doublets,
 and discuss how they may effectively accommodate the low energy phenomenology of such
new heavy fermionic degrees of freedom.
We also discuss the constraints on these models from precision electroweak data
as well as from flavor physics and the implications for collider searches of the Higgs particles
and of the 4th generation fermions, bearing in mind the recent observation of a light Higgs with a mass of $\sim$ 125 GeV.
\end{abstract}


\maketitle

\bigskip
\bigskip

\begin{center}
Invited Review:\\ to appear in a special issue of Advances in High Energy Physics (AHEP) on Very Heavy Quarks at the LHC
\end{center}
\newpage

\section{Introduction - the ``need" of a multi-Higgs setup for the 4th generation}

The minimal and perhaps the simplest framework for incorporating 4th generation
fermions can be constructed by adding to the Standard Model (SM)
a 4th sequential generation of fermion (quarks and leptons) doublets
(for reviews see \cite{sher0,hou2009-rev,SM4proc}). This framework, which is widely known as the SM4,
can already address some of the leading theoretical challenges in particle physics:
\begin{itemize}
\item The hierarchy problem \cite{DEWSB0,DEWSB,holdom-new,hung1}.
\item The origin of matter - anti matter asymmetry in the universe \cite{baryo-ref,fok}.
\item Flavor physics and CKM anomalies \cite{SAGMN08,SAGMN10,ajb10B,buras_charm,4Gflavor}.
\end{itemize}

Unfortunately, the current bounds on the masses of the 4th generation quarks within the
SM4 are rather high - reaching up to $\sim 600$ GeV \cite{CMS_limit,CMS_limit_dilepton,Atlas_bprime,CMS_bprime}, i.e., around the
unitarity bounds on quark masses \cite{unitarity}. The implications of such a ``super-heavy" 4th generation spectrum are far reaching. In fact, the SM4 as such is also strongly disfavored from searches at the LHC \cite{CMSHiggs,ATLASHiggs} and Tevatron
 \cite{TevatronHiggs} of the single Higgs particle of this model,
 essentially excluding the SM4 Higgs with masses up to
600 GeV \cite{SM4-Higgs-bound} and, thus, making it incompatible with the recent
observation/evidence of a light Higgs
with a mass of $ \sim 125$ GeV \cite{Lenz2,Nir}
(for a recent comprehensive analysis of the SM4
status in light of the latest Higgs results and electroweak precision data (EWPD), we refer
the reader to \cite{12091101}).
These rather stringent limits on the SM4
raise several questions at the fundamental level:

\begin{enumerate}
\item Are super-heavy fermionic degrees of freedom a surprise or
is that expected once new physics (NP), beyond the SM4 (BSM4), is assumed to enter at the TeV-scale?
\item Are such heavy fermions linked to strong dynamics and/or to compositeness at the near by TeV-scale?
\item What sub-TeV degrees of freedom should we expect if indeed such heavy fermions are found? and what is the
proper framework/effective theory required to describe the corresponding low energy dynamics?
\item How do such heavy fermions effect Higgs physics?
\item Can one construct a natural framework for 4th generation heavy fermions with a mass in the range $400 - 600$
GeV that is consistent with EWPD and that is not excluded by the
recent direct measurements from present high energy colliders?
\item What type of indirect hints for BSM4 dynamics can we expect in low energy flavor physics?
\end{enumerate}

In this article we will try to address these questions by considering
a class of BSM4 low energy effective theories which are based on multi-Higgs models.

Let us start by studying the hints for BSM4 and strong dynamics from the evolution
of the 4th generation Yukawa coupling $y_4$, under some simplifying assumptions.
In particular, one can write the RGE of $y_4$ assuming SM4 dynamics and neglecting the
gauge  and the top-Yukawa couplings and taking all 4th generation Yukawa couplings equal \cite{hashimoto}:
\begin{eqnarray}
\left( 16 \pi^2 \right) \mu \frac{\partial}{\partial \mu} y_4 \simeq
(2 y_4)^3
\end{eqnarray}
This yields a Landau Pole (defined by $1/y_4^2(\mu = \Lambda_y) \to 0$)
at
$\Lambda_y \simeq m_4 e^{\frac{\pi^2 v^2}{2 m_4^2}}$, giving $\Lambda_y \sim 8, 3, 2$ TeV for
$m_4 \sim 300, ~400, ~500$ GeV. Therefore, within the SM4, the 4th generation Yukawa couplings are expected to
``run into" a Landau Pole at the near by TeV-scale.

In fact, there are additional strong indications from the Higgs sector that
a heavy 4th generation of fermions is tied with new strong dynamics at the near by TeV-scale and that
the SM4 is not the adequate framework to describe the new TeV-scale physics:

\begin{enumerate}
\item {\bf The Higgs mass correction due to such heavy fermions is pushed to the cutoff scale:} \\
    To see that, one can calculate the self-energy 1-loop correction to the Higgs mass with the
exchange of a heavy fermion $q^\prime$ and set the cutoff to $\Lambda > m_{q^\prime}$, obtaining:
\begin{eqnarray}
\delta m_H^2 \sim \left( \frac{m_{q^\prime}}{400~{\rm GeV}} \right)^2 \cdot \Lambda^2 ~,
\end{eqnarray}
indicating that a heavy 4th family fermion with a mass around 400 GeV cannot co-exist with
the recently observed single light Higgs, since in the absence of fine tuning, the Higgs mass
should be pushed up to the
cutoff scale where the NP enters (in which case the definition of the Higgs particle
becomes meaningless).

\item {\bf The SM4 Higgs quartic coupling ($\lambda$) and a heavy Higgs:} \\
    One can again study the RGE for $\lambda$, assuming SM4 dynamics and neglecting
the gauge and the top-Yukawa couplings and taking all 4th generation Yukawa couplings equal.
One then obtaines \cite{hashimoto}:
\begin{eqnarray}
\left( 16 \pi^2 \right) \mu \frac{\partial}{\partial \mu} \lambda \simeq
24 \lambda^2 + 16 y_4^2 \left(2 \lambda - y_4^2 \right)
\theta(\mu - m_4)~,
\end{eqnarray}
giving a Landau Pole (i.e., $\lambda(\mu = \Lambda_\lambda) \to \infty$)
at
$\Lambda_\lambda \sim 4. 3, 2.5, 2.1$ TeV for
$m_H \sim 500, ~600, ~700$ GeV and, thus, indicating that
a light Higgs is not consistent with the SM4 if the NP scale
is at the few TeV range.
Indeed, solving the full RGE for the SM4 one finds that $m_H \gsim m_{q^\prime}$ when the cutoff
of the theory is set to the TeV-scale, i.e., to the proper cutoff of the SM4
when $m_{q^\prime} \sim {\cal O}(500)$ GeV \cite{hashimoto}.
The implications of a heavy Higgs in this mass range was considered e.g., in \cite{heavyH1,heavyH2,heavyH3,heavyH4},
claiming that the heavy SM4 Higgs case can relax the currently reported exclusion on the SM4.
However, the heavy SM4 Higgs scenario is now in contradiction with the recent measurements of
the two experiments at the LHC, which observe
a light Higgs boson with a mass of $ \sim 125 ~{\rm GeV}$ \cite{CMSHiggs,ATLASHiggs}.
On the other hand, as will be shown in this paper
(and was also demonstrated before in \cite{hashimoto} for the case of the popular 2HDM of type II
with a 4th generation of doublets),
a multi-Higgs setup for the 4th generation theory can relax the constraint $m_H \gsim m_{q^\prime}$.

\end{enumerate}

Thus, under the assumption that heavy 4th generation quarks exist, if one
assumes a light Higgs with a mass around 125 GeV and
seriously takes into account the fact that
low energy 4th generation theories
posses a new threshold/cutoff (or a fixed point, see e.g., \cite{hung3,hung2}) at the TeV-scale,
then one is forced to consider extensions of the naive SM4 with more than one Higgs doublet which, in turn,
leads to the possibility that the Higgs particles (or some of the Higgs particles) may be
composites primarily of the 4th generation fermions (see e.g., \cite{wise1,alfonso1,gustavo1,gustavo2,hashimoto1}),
with condensates $<Q_{L}^\prime t_R^\prime> \neq 0$, $<Q_{L}^\prime b_R^\prime> \neq 0$ (and possibly
also $<L_L^\prime \nu_R^\prime> \neq 0$, $<L_L^\prime \tau_R^\prime> \neq 0$). These condensates
then induce EWSB and
generate a dynamical mass for the condensing fermions.
This viewpoint in fact dates back to an ``old" idea suggested more than two decades ago \cite{DEWSB0}; that
a heavy top-quark may be used to form a $t \bar t$ condensate which could
trigger dynamical EWSB. Although, this top-condensate mechanism led to the prediction of a too large $m_t$,
this idea ignited further thoughts and studies towards the possibility that 4th generation fermions may play an
important role in dynamical EWSB \cite{DEWSB0,DEWSB}. In particular, due to the presence of
such heavy fermionic degrees of freedom, some form of strong dynamics
and/or compositeness may occur at the near by TeV-scale.

In this article, we will review the above viewpoint which was also adopted in Ref.~\cite{4G2HDM}: that theories which contain such heavy fermionic states are inevitably cutoff at the near by TeV-scale, and are, therefore, more naturally embedded at
low energies in multi-Higgs models, which are the proper low-energy effective frameworks
for describing the sub-TeV dynamics of 4th generation fermions. As
mentioned above, in this picture,
the Higgs particles are viewed as the
composite scalars that emerge as manifestations of the different possible bound states of the
fundamental heavy fermions. This approach was considered already 20 years ago by Luty
\cite{luty} and more recently in \cite{hashimoto1}, where an attempt to put 4th degeneration heavy fermions
into an effective multi (composite) Higgs doublets model was made,
using a Nambu-Jona-Lasinio (NJL) type approach.

The phenomenology of multi-Higgs models with a 4th family of fermions
was studied to some extent recently in \cite{sher1,sher2HDM,Bern,Gunion,valencia,He2,soninew,hashimoto} and
within a SUSY framework in \cite{fok,shersusy,dawson,rizzo}.
In this article, we will further study the phenomenology of 2HDM frameworks with a
4th family of fermions, focusing on a new class of 2HDM's ``for the 4th generation"
(named hereafter 4G2HDM) that can effectively address the low-energy phenomenology
of a TeV-scale dynamical EWSB scenario, which is possibly triggered by the
condensates of the 4th generation fermions.

We will first describe a few viable manifestations of a 2HDM framework with a 4th generation
of fermions, focusing on the 4G2HDM framework of Ref.~\cite{4G2HDM}.
 We will then discuss the constraints on such 4th generation
2HDM models from PEWD as well as from flavor physics. We will end
by studying the expected
implication of such 2HDM frameworks on direct searches for the 4th generation fermions and for the Higgs particle(s), assuming the existence
of a light Higgs with a mass of 125 GeV.

\section{2HDM's and 4th generation fermions \label{sec2}}

Assuming a common generic 2HDM potential, the phenomenology of 2HDM's is generically encoded in the texture of the Yukawa interaction Lagrangian.
The simplest variant of a 2HDM with 4th generations of fermions, can be constructed based on the so called type II 2HDM
(which we denote hereafter by 2HDMII), in which one of the Higgs doublets couples only to up-type fermions and the
other to down-type ones. This setup ensures the absence of tree-level flavor changing neutral currents (FCNC)
and is, therefore, widely favored
when confronted with low energy flavor data. The Yukawa terms of the 2HDMII, extended to include the extra
4th generation quark doublet is (and similarly in the leptonic sector):
\begin{eqnarray}
\mathcal{L}_{Y}= -\bar{Q}_{L} \Phi_{d} F_d d_{R}
-\bar{Q}_{L} \tilde\Phi_{u} F_u u_{R} + h.c.\mbox{ ,}
\label{eq:LYII}
\end{eqnarray}
where $f_{L(R)}$ are left(right)-handed fermion fields, $Q_{L}$ is the left-handed
$SU(2)$ quark doublet and $F_d,F_u$ are general $4\times4$
Yukawa matrices in flavor space. Also, $\Phi_{d,u}$ are the Higgs doublets:
\begin{eqnarray*}
\Phi_i & =\left(\begin{array}{c}
\phi^{+}_i\\
\frac{v_i+\phi^{0}_i}{\sqrt{2}}\end{array}\right),\quad\tilde{\Phi_i}=\left(\begin{array}{c}
\frac{v_i^{*}+\phi^{0*}_i}{\sqrt{2}}\\
-\phi^{-}_i\end{array}\right) ~,
\end{eqnarray*}

Motivated by the idea that the low energy scalar degrees of freedom may be the composites of the heavy
4th generation fermions, it is possible to construct a new class of 2HDM's that effectively parameterize
4th generation condensation by giving a special status to the 4th family fermions.
This was done in \cite{4G2HDM}, where (in the spirit of
the Das and Kao 2HDM that was based on the SM's three families of fermions \cite{Das})
one of the Higgs fields ($\phi_h$ - call it the ``heavier" field) was assumed to
couple only to heavy fermionic states, while the second Higgs field ($\phi_\ell$ - the ``lighter" field)
is responsible for the mass generation of all other (lighter) fermions.
The possible viable variants of this approach can be parameterized as \cite{4G2HDM} (and similarly in the leptonic sector):
\begin{widetext}
\begin{eqnarray}
\mathcal{L}_{Y}= -\bar{Q}_{L}
\left( \Phi_{\ell}F_d \cdot \left( I-{\cal I}_d^{\alpha_d \beta_d}
\right) +
\Phi_{h}F_d \cdot {\cal I}_d^{\alpha_d \beta_d} \right) d_{R}
-\bar{Q}_{L}
\left( \tilde\Phi_{\ell} F_u \cdot \left( I - {\cal I}_u^{\alpha_u \beta_u} \right) +
\Phi_{h} F_u \cdot {\cal I}_u^{\alpha_u \beta_u} \right)
u_{R} + h.c.\mbox{ ,}
\label{eq:LY4G}
\end{eqnarray}
\end{widetext}
where $\Phi_{\ell,h}$ are the two Higgs doublets,
$I$ is the identity matrix and ${\cal I}_q^{\alpha_q \beta_q}$ ($q=d,u$)
are diagonal $4\times4$ matrices defined by
${\cal I}_q^{\alpha_q \beta_q} \equiv {\rm diag}\left(0,0,\alpha_q,\beta_q\right)$.

The Yukawa interaction Lagrangian of (\ref{eq:LY4G}) can lead to several interesting textures
that can be realized in terms of a $Z_2$-symmetry under which the fields transform as follows:
\begin{eqnarray}
 && \Phi_{\ell}\to-\Phi_{\ell},~ \Phi_{h}\to+\Phi_{h},~ Q_{L}\to+Q_{L}, \nonumber \\
 && d_{R}\to-d_{R}\;(d=d,s),~ u_{R}\to-u_{R}\;(u=u,c),\nonumber \\
&& b_{R}\to (-1)^{1+\alpha_d} b_{R},~ b^\prime_{R}\to (-1)^{1+\beta_d} b^\prime_{R},\nonumber \\
&& t_{R}\to (-1)^{1+\alpha_u} t_{R},~ t^\prime_{R}\to (-1)^{1+\beta_u} t^\prime_{R}~,
\label{eq:z2}
 \end{eqnarray}

which allows us to construct several models that have a non-trivial Yukawa structure and that are
potentially associated with the compositeness scenario:

\begin{itemize}
\item {\bf type I 4G2HDM}: denoted hereafter by {\bf 4G2HDMI} and defined by $\left(\alpha_d,\beta_d,\alpha_u,\beta_u\right)=\left(0,1,0,1\right)$,
in which case $\Phi_h$ gives masses only to $t^\prime$ and $b^\prime$, while $\Phi_\ell$
generates masses for all other quarks (including the top-quark).
For this model, which seems to be the natural choice for the leptonic sector, we expect:
\begin{eqnarray}
\tan\beta \equiv \frac{v_h}{v_\ell} \approx \frac{m_{q^\prime}}{m_t} \sim {\cal O}(1)~.
\end{eqnarray}

\item {\bf type II 4G2HDM}: denoted hereafter by {\bf 4G2HDMII} and defined by $\left(\alpha_d,\beta_d,\alpha_u,\beta_u\right)=\left(1,1,1,1\right)$,
in which case the heavy condensate $\Phi_h$ couples to the heavy
quarks states of both the 3rd and 4th generations $t$ and $b$-quarks, whereas $\Phi_\ell$
couples to the light quarks of the 1st and 2nd generations. For this model one expects $\tan\beta \gg 1$.
\item {\bf type III 4G2HDM}: denoted hereafter by {\bf 4G2HDMIII} and defined by $\left(\alpha_d,\beta_d,\alpha_u,\beta_u\right)=\left(0,1,1,1) \right)$,
in which case $m_t,~m_{b^\prime}$ and $m_{t^\prime} \propto v_h$, so that only quarks with
masses at the EW-scale are coupled to the heavy doublet $\Phi_h$.
Here also one expects $\tan\beta \gg 1$.
\end{itemize}

The Yukawa interactions for these models are given by \cite{4G2HDM}:
\begin{widetext}
\begin{eqnarray}
{\cal L}(h q_i q_j) &=& \frac{g}{2 m_W} \bar q_i \left\{ m_{q_i} \frac{s_\alpha}{c_\beta} \delta_{ij}
- \left( \frac{c_\alpha}{s_\beta} + \frac{s_\alpha}{c_\beta} \right) \cdot
\left[ m_{q_i} \Sigma_{ij}^q R + m_{q_j} \Sigma_{ji}^{q \star} L \right] \right\} q_j h \label{Sff1}~, \\
{\cal L}(H q_i q_j) &=& \frac{g}{2 m_W} \bar q_i \left\{ -m_{q_i} \frac{c_\alpha}{c_\beta} \delta_{ij}
+ \left( \frac{c_\alpha}{c_\beta} - \frac{s_\alpha}{s_\beta} \right) \cdot
\left[ m_{q_i} \Sigma_{ij}^q R + m_{q_j} \Sigma_{ji}^{q \star} L \right] \right\} q_j H ~, \\
{\cal L}(A q_i q_j) &=& - i I_q \frac{g}{m_W} \bar q_i \left\{ m_{q_i} \tan\beta \gamma_5 \delta_{ij}
- \left( \tan\beta + \cot\beta \right) \cdot
\left[ m_{q_i} \Sigma_{ij}^q R - m_{q_j} \Sigma_{ji}^{q \star} L \right] \right\} q_j A ~, \\
{\cal L}(H^+ u_i d_j) &=& \frac{g}{\sqrt{2} m_W} \bar u_i \left\{
\left[ m_{d_j} \tan\beta \cdot V_{u_id_j} - m_{d_k} \left( \tan\beta + \cot\beta \right) \cdot
V_{ik} \Sigma^{d}_{kj} \right] R \right. \nonumber \\
&& \left. + \left[ -m_{u_i} \tan\beta \cdot V_{u_id_j} + m_{u_k} \left( \tan\beta + \cot\beta \right) \cdot
\Sigma^{u \star}_{ki} V_{kj} \right] L
 \right\} d_j H^+ \label{Sff2}~,
\end{eqnarray}
\end{widetext}
where $V$ is the $4 \times 4$ CKM matrix, $q=d$ or $u$ for down or up-quarks with weak Isospin $I_d=-\frac{1}{2}$ and $I_u=+\frac{1}{2}$, respectively,
and $R(L)=\frac{1}{2}\left(1+(-)\gamma_5\right)$.
Also, the 4G2HDM type, i.e., the 4G2HDMI, 4G2HDMII and 4G2HDMIII, as well as FCNC effects are all encoded in $\Sigma^d$ and $\Sigma^u$,
which are new mixing matrices in the down(up)-quark sectors, obtained after diagonalizing the quarks mass matrices:
\begin{widetext}
\begin{eqnarray}
\Sigma_{ij}^d &=& \Sigma_{ij}^d(\alpha_d,\beta_d,D_R) = \alpha_d D_{R,3i}^\star D_{R,3j} + \beta_d D_{R,4i}^\star D_{R,4j}~, \nonumber \\
\Sigma_{ij}^u &=& \Sigma_{ij}^u(\alpha_u,\beta_u,U_R) = \alpha_u U_{R,3i}^\star U_{R,3j} + \beta_u U_{R,4i}^\star U_{R,4j} ~, \label{sigma}
\end{eqnarray}
\end{widetext}
depending on $D_R,U_R$ which are the rotation (unitary) matrices of the right-handed down and up-quarks, respectively, and on
whether $\alpha_q$ and/or $\beta_q$ are ``turned on". This is in contrast to
``standard" frameworks such as the SM4 and the 2HDM's of types I and II, where the right-handed mixing matrices
$U_R$ and $D_R$ are non-physical being ``rotated away" in the diagonalization procedure of the quark masses.
Indeed, in the 4G2HDM's described above some elements of $D_R$ and $U_R$
can, in principle, be measured in Higgs-fermion systems, as we will later show.

In particular, inspired by the working assumption of the 4G2HDM's
and by the observed flavor pattern in the up and down-quark sectors,
it was shown in \cite{4G2HDM} that the new mixing matrices $\Sigma^d$ and $\Sigma^u$ are expected to have the following form:
\begin{widetext}
\begin{eqnarray}
\Sigma^u &=& \left(\begin{array}{cccc}
0 & 0 & 0 & 0 \\
0 & \alpha_u |\epsilon_c|^2 & \alpha_u \epsilon_c^\star \left( 1- \frac{|\epsilon_t|^2}{2} \right) & - \alpha_u \epsilon_c^\star \epsilon_t^\star \\
0 & \alpha_u \epsilon_c \left( 1- \frac{|\epsilon_t|^2}{2} \right) & \alpha_u \left( 1- \frac{|\epsilon_t|^2}{2} \right) + \beta_u |\epsilon_t|^2 & (\beta_t -\alpha_t) \epsilon_t^\star \left( 1- \frac{|\epsilon_t|^2}{2} \right) \\
0 & - \alpha_u \epsilon_c \epsilon_t &  (\beta_u -\alpha_u) \epsilon_t \left( 1- \frac{|\epsilon_t|^2}{2} \right) & \alpha_u |\epsilon_t|^2 + \beta_u \left( 1- \frac{|\epsilon_t|^2}{2} \right)
\end{array}\right), \label{sigsimple}
\end{eqnarray}
\end{widetext}
and similarly for $\Sigma^d$ by replacing $\alpha_u,\beta_u \to \alpha_d,\beta_d$ and $\epsilon_c,\epsilon_t \to \epsilon_s,\epsilon_b$.
 The new parameters $\epsilon_c,~\epsilon_t$ are free parameters that effectively control the mixing between the 4th generation $t^\prime$ and the 2nd and 3rd generation quarks $c$ and $t$, respectively. Thus,
A natural choice which will be adopted here in some instances is
$|\epsilon_t| = \sim m_t/m_{t^\prime}$,
$|\epsilon_b| = \sim m_b/m_{b^\prime}$ and
$\epsilon_s,~\epsilon_c \to 0$.

\section{Constraints on 2HDM's with a 4th generation of fermions \label{sec3}}

\subsection{Constraints from electroweak precision data: oblique parameters}

The sensitivity of EWPD to 4th generation fermions within the minimal SM4 framework was extensively analyzed
in the past decade \cite{PDG,polonsky,novikov,Kribs_EWPT,langacker,chenowitz,chenowitz1}.
Here we are interested instead on the constraints that EWPD impose on 2HDM's
with a 4th generation family.
As usual, the effects of the NP can be divided into the effects
of the heavy NP which does and which does not couple directly
to the ordinary SM fermions. For the former, the leading effect comes from the decay
$Z \to b \bar b$, which is mainly sensitive to the
$H^+ t^\prime b$ and $W^+ t^\prime b$ couplings through one-loop exchanges
of $H^+$ and $W^+$ shown in Fig.~\ref{diagrams}, and which was
analyzed in detail in \cite{4G2HDM}.

On the other hand, the effects which do not involve direct couplings to the ordinary fermions,
can be analyzed by the quantum oblique corrections to the gauge-bosons 2-point functions,
which can be parameterized in terms of the oblique parameters S,T and U
\cite{peskin}.
For the oblique parameters the effects of a 2HDM with a 4th generation are common
to any variant of a 2HDM framework (including the 2HDMII and the 4G2HDMI, 4G2HDMII and 4G2HDMIII described in the previous section),
since the $Hff$ Yukawa interactions of any 2HDM
do not contribute at 1-loop to the gauge-bosons self energies.

In particular, apart from the pure 1-loop Higgs exchanges,
one also has to include the new contributions from
$t^\prime$ and $b^\prime$ exchanges which shift the T parameter
($\Delta T_f$) and which involve the new SM4-like diagonal coupling
$W t^\prime b^\prime$ as well as the $W t^\prime b$ and $W t b^\prime$ off-diagonal
vertices (see e.g., \cite{chenowitz}):
\begin{widetext}
\begin{eqnarray}
\Delta T_f = \frac{3}{8 \pi s_W^2 c_W^2}
\left( |V_{t^\prime b^\prime}|^2 F_{t^\prime b^\prime} +
|V_{t^\prime b}|^2 F_{t^\prime b} +
|V_{t b^\prime}|^2 F_{t b^\prime}  -
|V_{t b}|^2 F_{t b} +
\frac{1}{3} F_{\ell_4 \nu_4} \right) ~,
\end{eqnarray}
\end{widetext}
with
\begin{eqnarray}
F_{ij} = \frac{x_i+x_j}{2} - \frac{x_i x_j}{x_i-x_j} {\rm log}\frac{x_i}{x_j}~,
\end{eqnarray}
and $x_k \equiv (m_k/m_Z)^2$.

\begin{figure}[htb]
\begin{center}
\epsfig{file=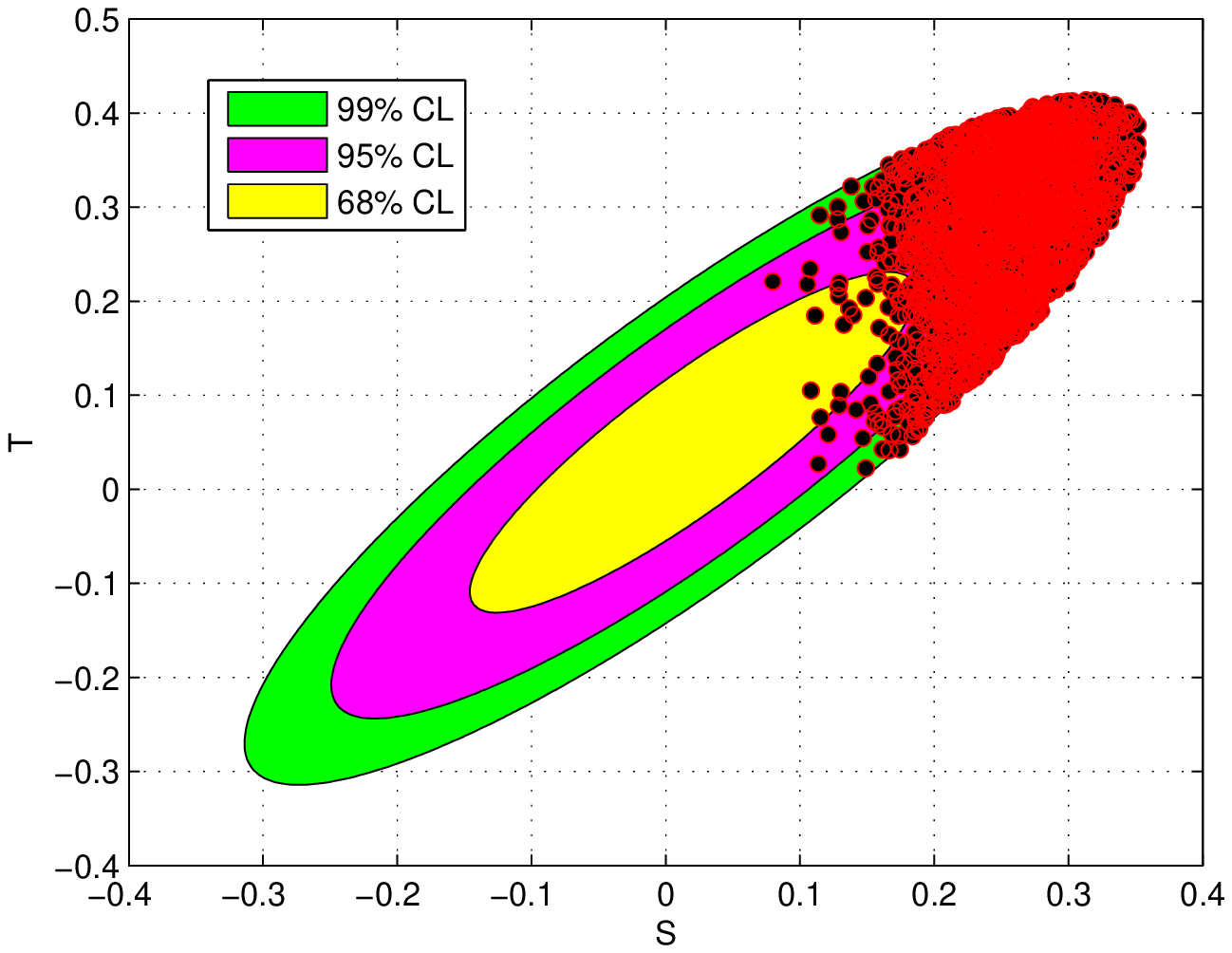,height=5cm,width=5cm,angle=0}
\epsfig{file=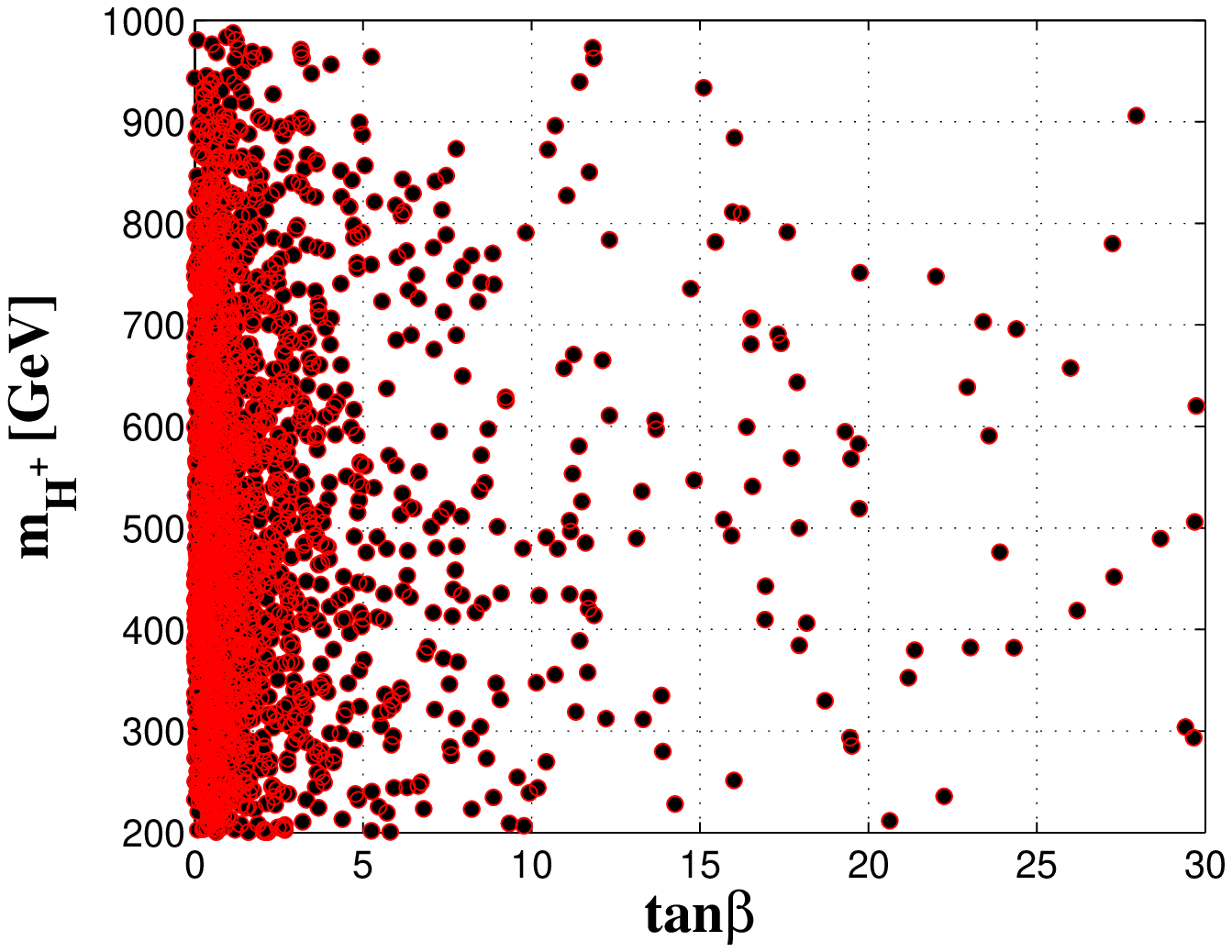,height=5cm,width=5cm,angle=0}
\epsfig{file=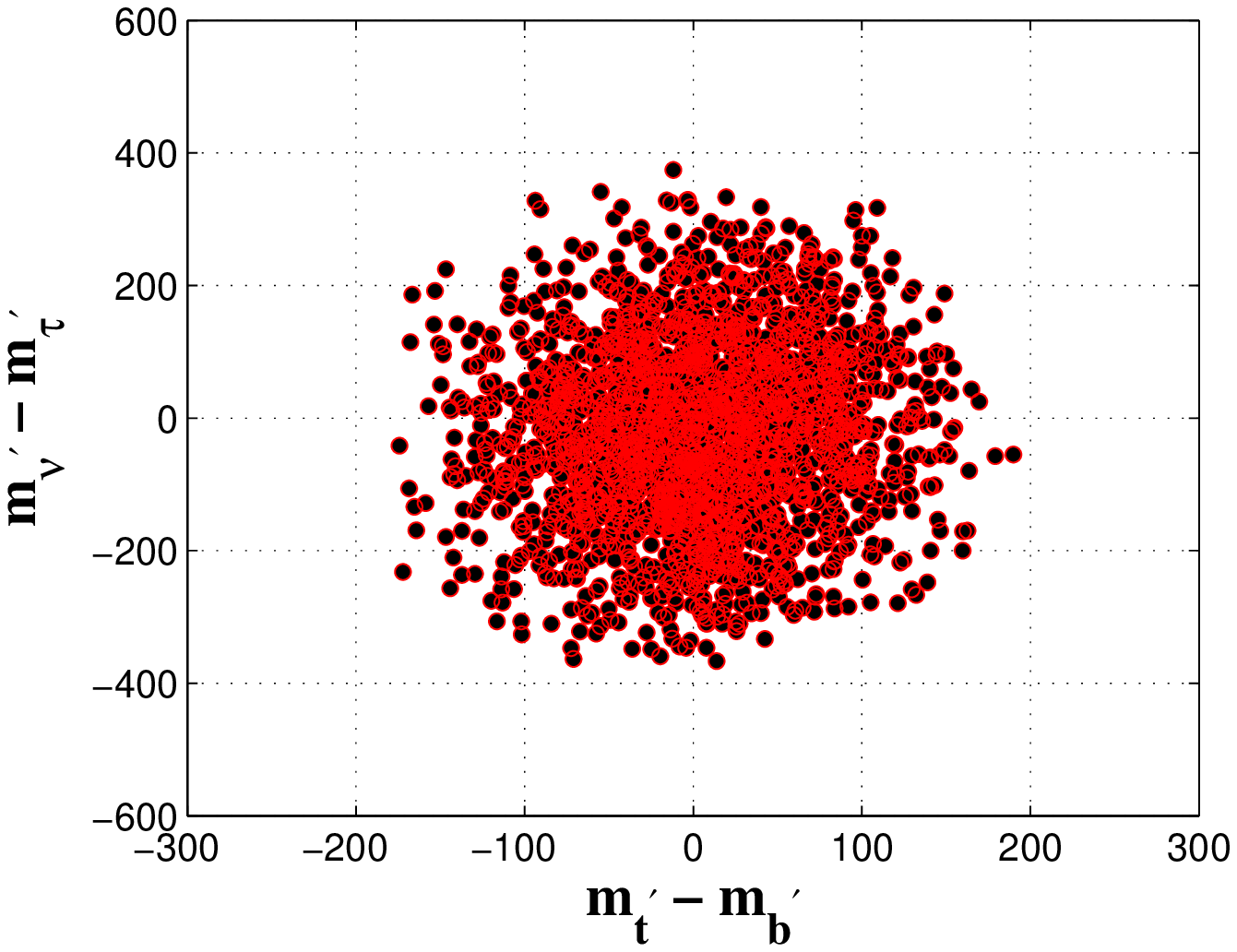,height=5cm,width=5cm,angle=0}
\caption{\emph{Left plot: the
allowed points in parameter space projected onto the
68\%, 95\% and 99\% allowed contours in the S-T plane.
Middle plot: 95\% CL allowed range in the $m_{H^+} - \tan\beta$ plane.
Right plot: allowed region in the
$\Delta m_{q^\prime} - \Delta m_{\ell^\prime}$ plane
within the $95\%$CL contour in the S-T plane.
All plots are for any 2HDM setup (such as the 2HDMII and the three types of the 4G2HDM, see text)
and with 100000 data points setting the light Higgs mass to $m_h=125$ GeV and varying the rest
of the parameters  in the ranges: $\tan\beta \leq 30$, $\theta_{34} \leq 0.3$,
$150~{\rm GeV} \leq m_{H} \leq 1~{\rm TeV}$,
$150~{\rm GeV} \leq m_{A} \leq 1~{\rm TeV}$,
$200~{\rm GeV} \leq m_{H^+} \leq 1~{\rm TeV}$,
$400~{\rm GeV} \leq m_{t^\prime},m_{b^\prime} \leq 600~{\rm GeV}$,
$100~{\rm GeV} \leq m_{\nu^\prime},m_{\tau^\prime} \leq 1.2~{\rm TeV}$
and the CP-even neutral Higgs mixing angle in the range
$0 \lsim \alpha \lsim 2 \pi$.}}
\label{blindscan}
\end{center}
\end{figure}

The complete set of corrections to the S and T parameters within a 2HDM with a 4th generation
of fermions was considered in \cite{polonsky,4G2HDM,new2HDM4G}.
Following the recent analysis in \cite{4G2HDM}, we show in
Fig.~\ref{blindscan} the results of ``blindly" (randomly) scanning the
relevant parameter space with 100000 models, where we set
the light Higgs mass to be $m_h=125$ GeV and vary the rest of
the relevant parameters within the ranges:
$\tan\beta \leq 30$, $\theta_{34} \leq 0.3$,
$150~{\rm GeV} \leq m_{H} \leq 1~{\rm TeV}$,
$150~{\rm GeV} \leq m_{A} \leq 1~{\rm TeV}$,
$200~{\rm GeV} \leq m_{H^+} \leq 1~{\rm TeV}$,
$400~{\rm GeV} \leq m_{t^\prime},m_{b^\prime} \leq 600~{\rm GeV}$,
$100~{\rm GeV} \leq m_{\nu^\prime},m_{\tau^\prime} \leq 1.2~{\rm TeV}$,
and the CP-even neutral Higgs mixing angle in the range
$0 \leq \alpha \leq 2 \pi$.
In particular, we plot in Fig.~\ref{blindscan} the
allowed points in parameter space projected onto the
68\%, 95\% and 99\% allowed contours in the S-T plane,
the 95\% CL allowed range in the $m_{H^+} - \tan\beta$
and the
$\Delta m_{q^\prime} - \Delta m_{\ell^\prime}$ planes,
corresponding to the $95\%$CL contour in the S-T plane

We find that compatibility with PEWD mostly requires
$\tan\beta \sim {\cal O}(1)$ with a small number of points in parameter space
having $\tan\beta \gsim 5$.
We also find that the 2HDM frameworks
allow 4th generation quarks and leptons mass splittings extended
to: $-200~{\rm GeV} \lsim \Delta m_{q^\prime} \lsim 200~{\rm GeV}$
and $-400~{\rm GeV} \lsim \Delta m_{\ell^\prime} \lsim 400~{\rm GeV}$,
and ``solutions" where both the
quarks and the leptons of the 4th generation doublets are degenerate.
For the cases of a small (or no) 4th generation fermion mass
splitting, the amount of isospin breaking required to compensate for the effect of the
extra fermions and Higgs particles on S and T is provided by a mass splitting among the
Higgs particles, see \cite{4G2HDM}.

\subsection{Constraints from electroweak precision data: $Z \to b \bar b$}

\begin{figure}[htb]
\begin{center}
\epsfig{file=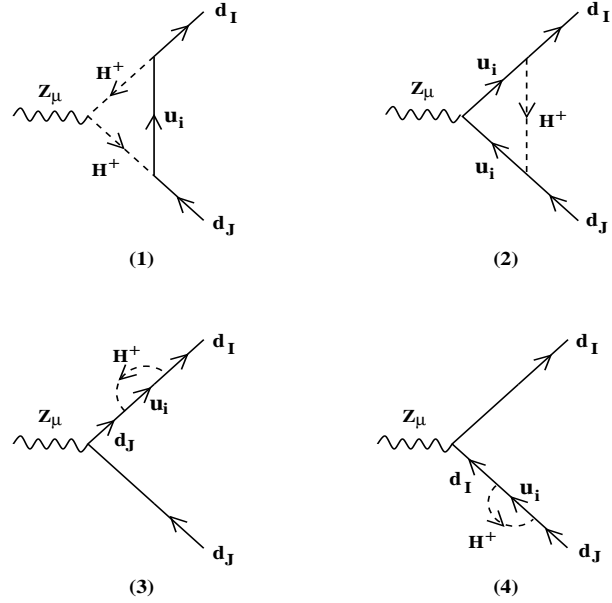,height=8cm,width=8cm,angle=0}
\caption{\emph{One-loop diagrams for corrections to $Z \to d_I \bar d_J$ from charged Higgs loops.
Similar diagrams with $W-t^\prime$ loops contribute as well.}}
\label{diagrams}
\end{center}
\end{figure}
The effects of the NP in $Z \to b \bar b$,
is best studied via the well measured quantity $R_b$:
\begin{eqnarray}
R_b \equiv \frac{\Gamma(Z \to b \bar b)}{\Gamma(Z \to {\rm hadrons})}~,
\end{eqnarray}
which is a rather clean test of the SM, since
being a ratio between two hadronic rates,
most of the electroweak, oblique and QCD corrections cancel between numerator and
denumerator.

Following \cite{4G2HDM}, the effects of NP in $R_b$ can be parameterized
in terms of the corrections $\delta_b$ and $\delta_c$ to
the decays $Z \to b \bar b$ and $Z \to c \bar c$, respectively:
\begin{eqnarray}
R_b = R_b^{SM} \frac{1+\delta_b}{1+ R_b^{SM} \delta_b + R_c^{SM} \delta_c} \label{Rb}~,
\end{eqnarray}
where $R_b^{SM}$ and $R_c^{SM}$
are the corresponding 1-loop quantities calculated
in the SM, and $\delta_q$ are the NP corrections defined in terms of the
$Z q \bar q$ couplings as:
\begin{eqnarray}
\delta_q = 2 \frac{ g_{qL}^{SM} g_{qL}^{new} + g_{qR}^{SM} g_{qR}^{new} }
{ \left( g_{qL}^{SM} \right)^2 + \left( g_{qR}^{SM} \right)^2 }~,
\end{eqnarray}
where
\begin{eqnarray}
V_{qqZ} \equiv -i \frac{g}{c_W} \bar q \gamma_\mu \left( \bar g_{qL} L + \bar g_{qR} R \right) q Z^\mu ~,
\end{eqnarray}
with $s_W(c_W)=\sin\theta_W(\cos\theta_W)$, $L(R) = \left( 1-(+) \gamma_5 \right)/2$ and
$\bar g_{qL,R} = g_{qL,R}^{SM} + g_{qL,R}^{new}$,
so that $g_{qL,R}^{SM}$ are the SM (1-loop) quantities and $g_{qL,R}^{new}$ are
the NP 1-loop corrections.

The corrections to $R_b$ from the 4th generation quarks in the 4G2HDMI, 4G2HDMII and 4G2HDMIII
are of three types (see \cite{4G2HDM}), where in all cases one finds that $\delta_c \ll \delta_b$,
so that one can safely neglect the effects from $Z \to c \bar c$:

\begin{figure}[htb]
\begin{center}
\epsfig{file=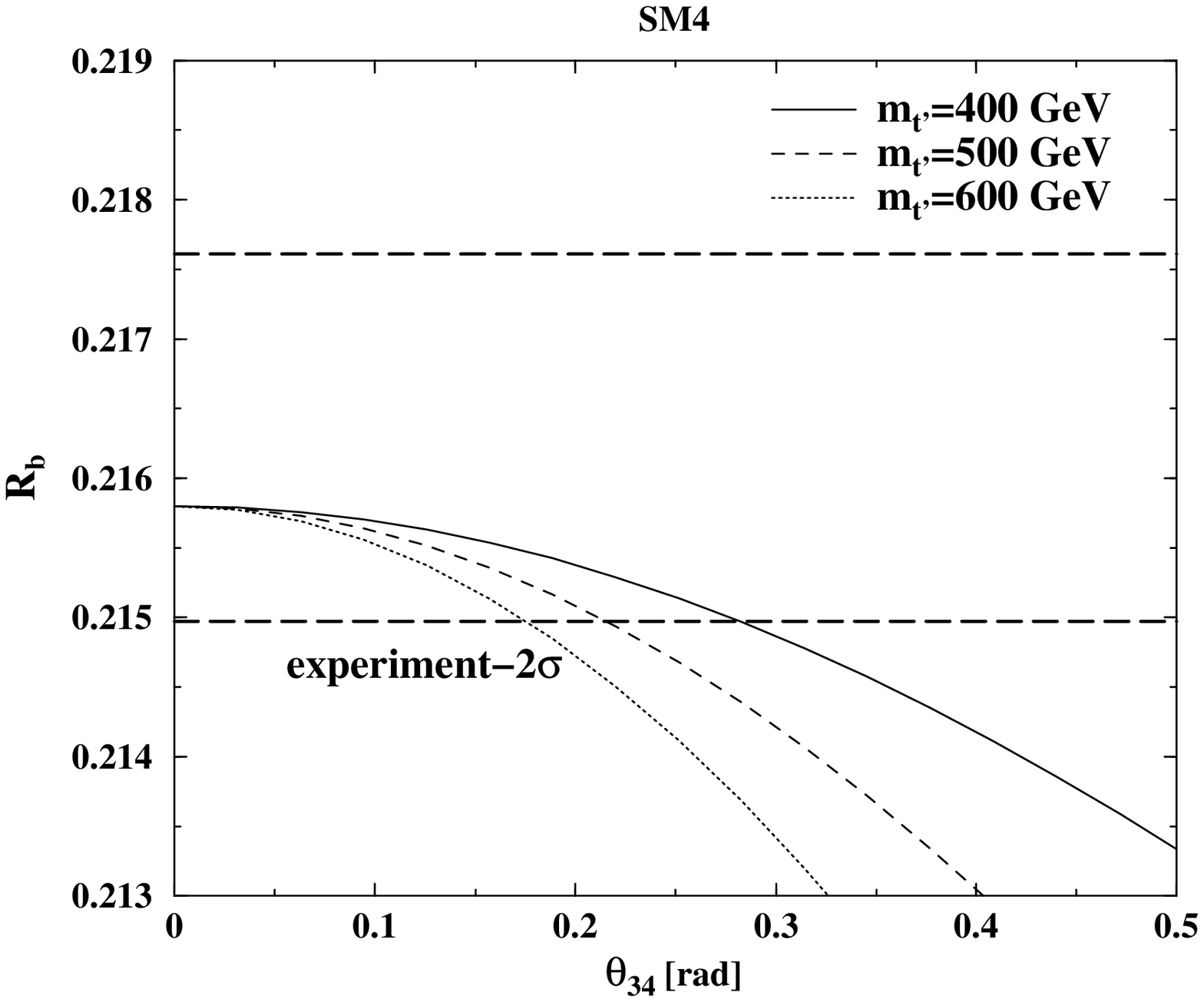,height=8cm,width=8cm,angle=0}
\epsfig{file=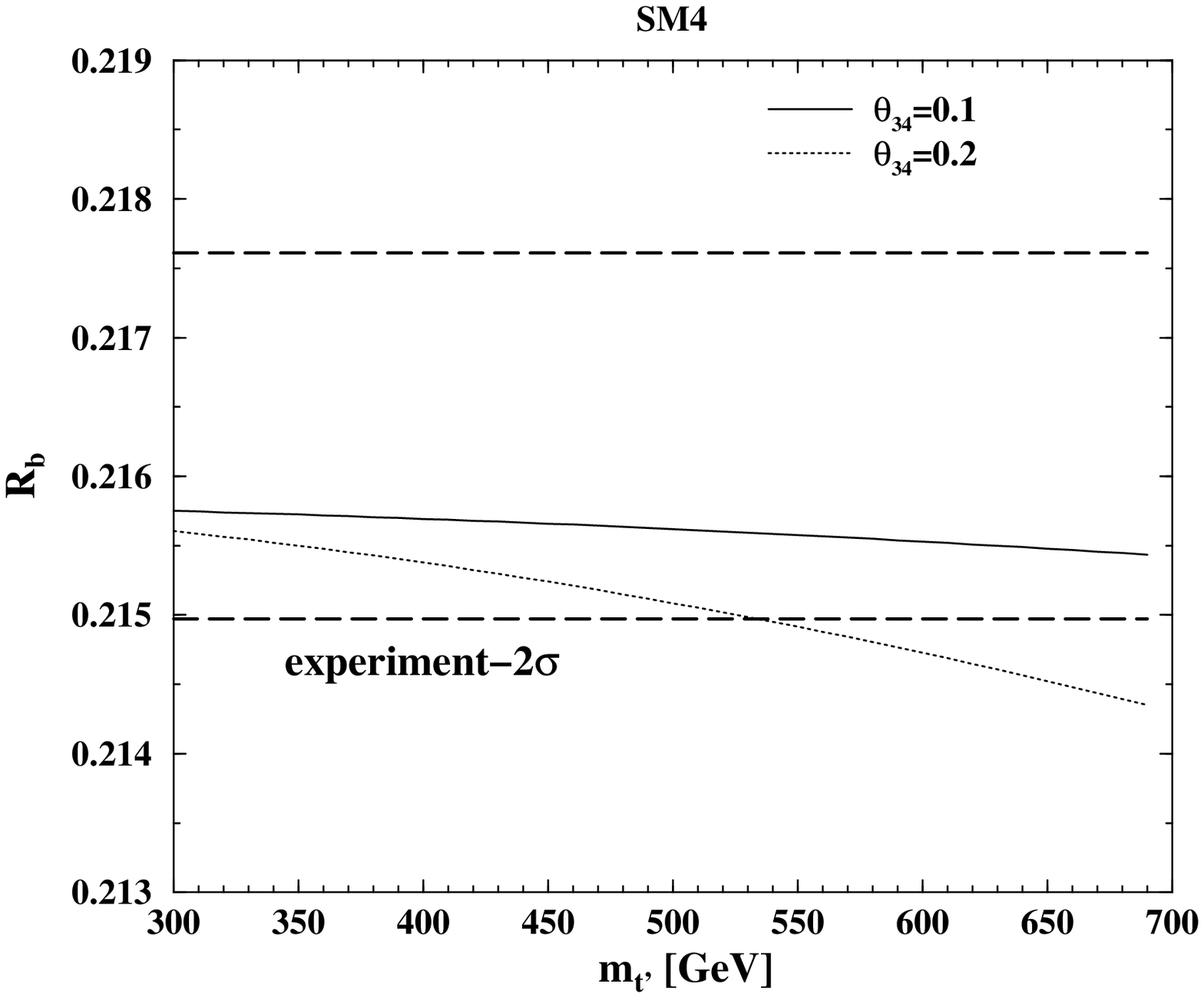,height=8cm,width=8cm,angle=0}\\
\caption{\emph{$R_b$ in the SM4,
as a function of $\theta_{34}$ for several values
of the $t^\prime$ mass (left) and as a function
of $m_{t^\prime}$ for $\theta_{34}=0.1$ and $0.2$ (right). Figure taken from \cite{4G2HDM}.}}
\label{figRbSM4}
\end{center}
\end{figure}
\begin{figure}[htb]
\begin{center}
\epsfig{file=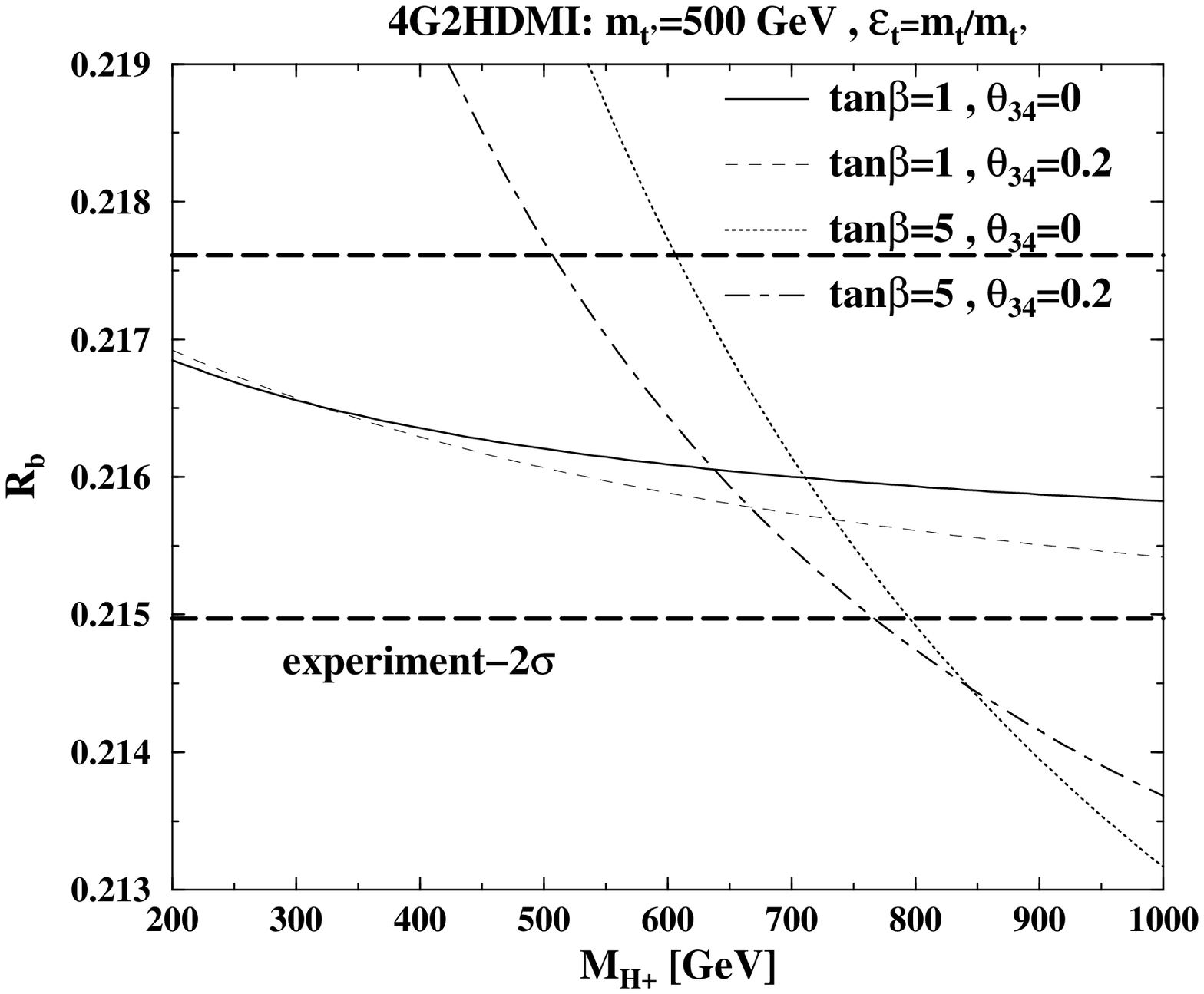,height=8cm,width=8cm,angle=0}
\epsfig{file=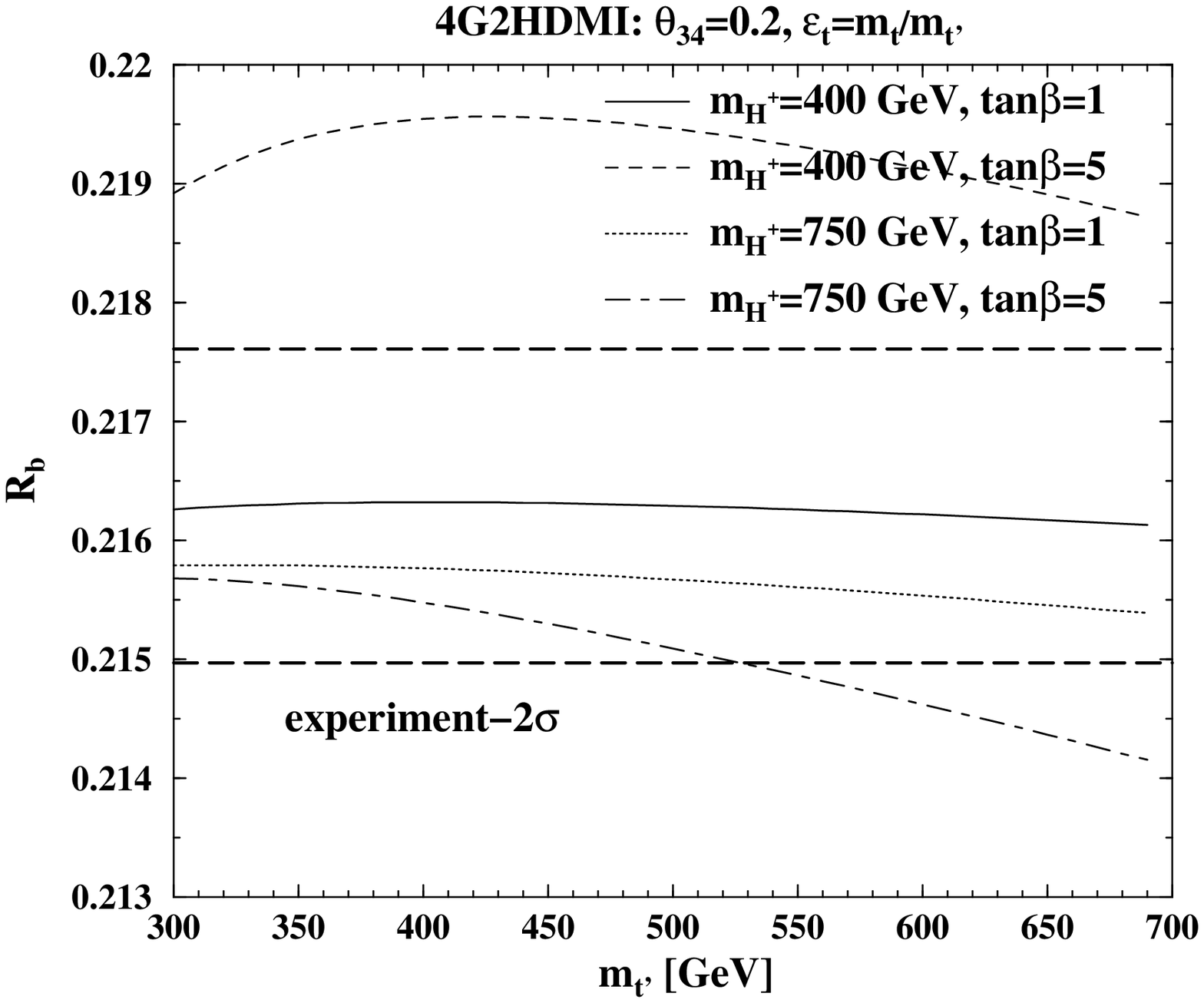,height=8cm,width=8cm,angle=0}\\
\epsfig{file=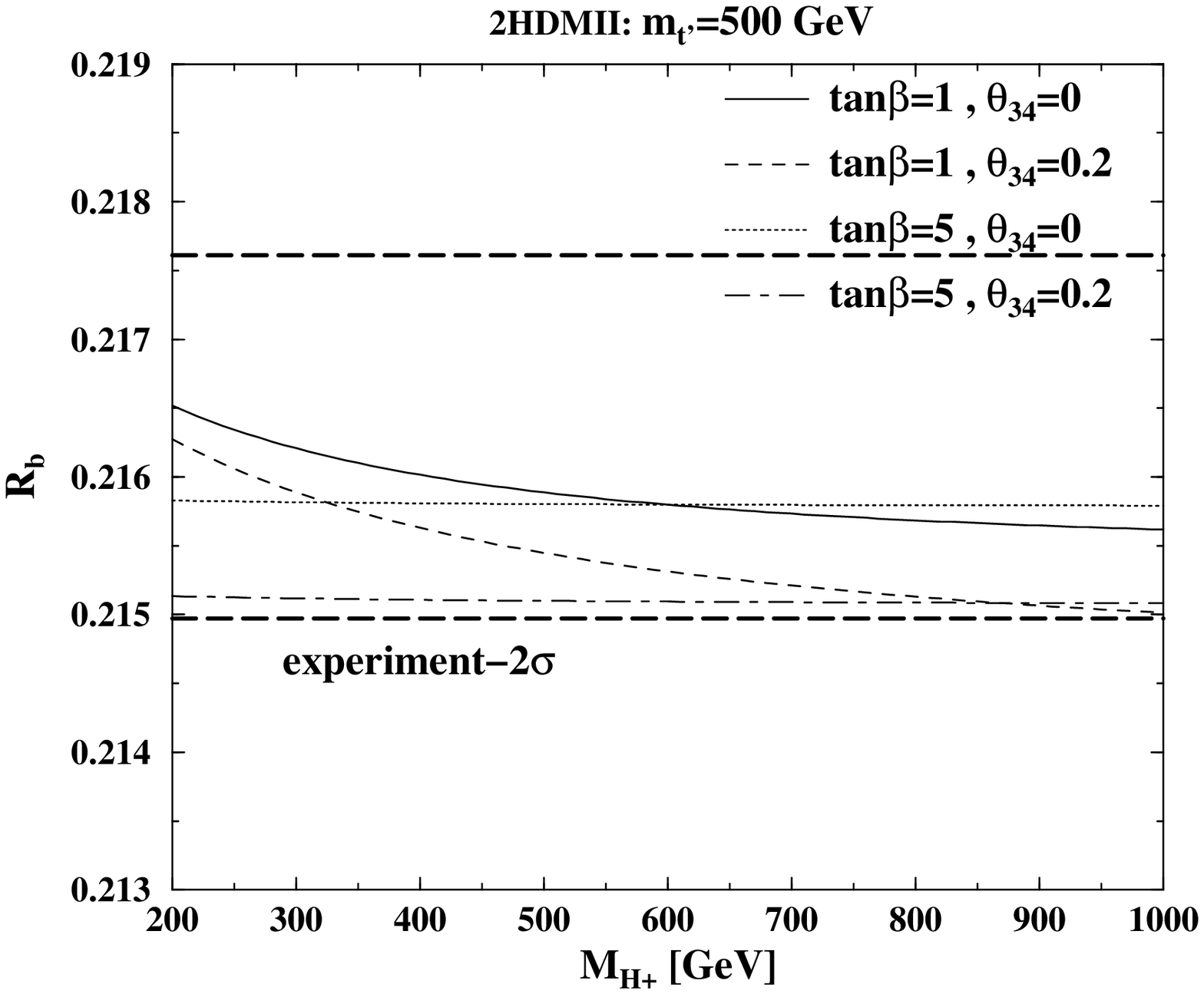,height=8cm,width=8cm,angle=0}
\epsfig{file=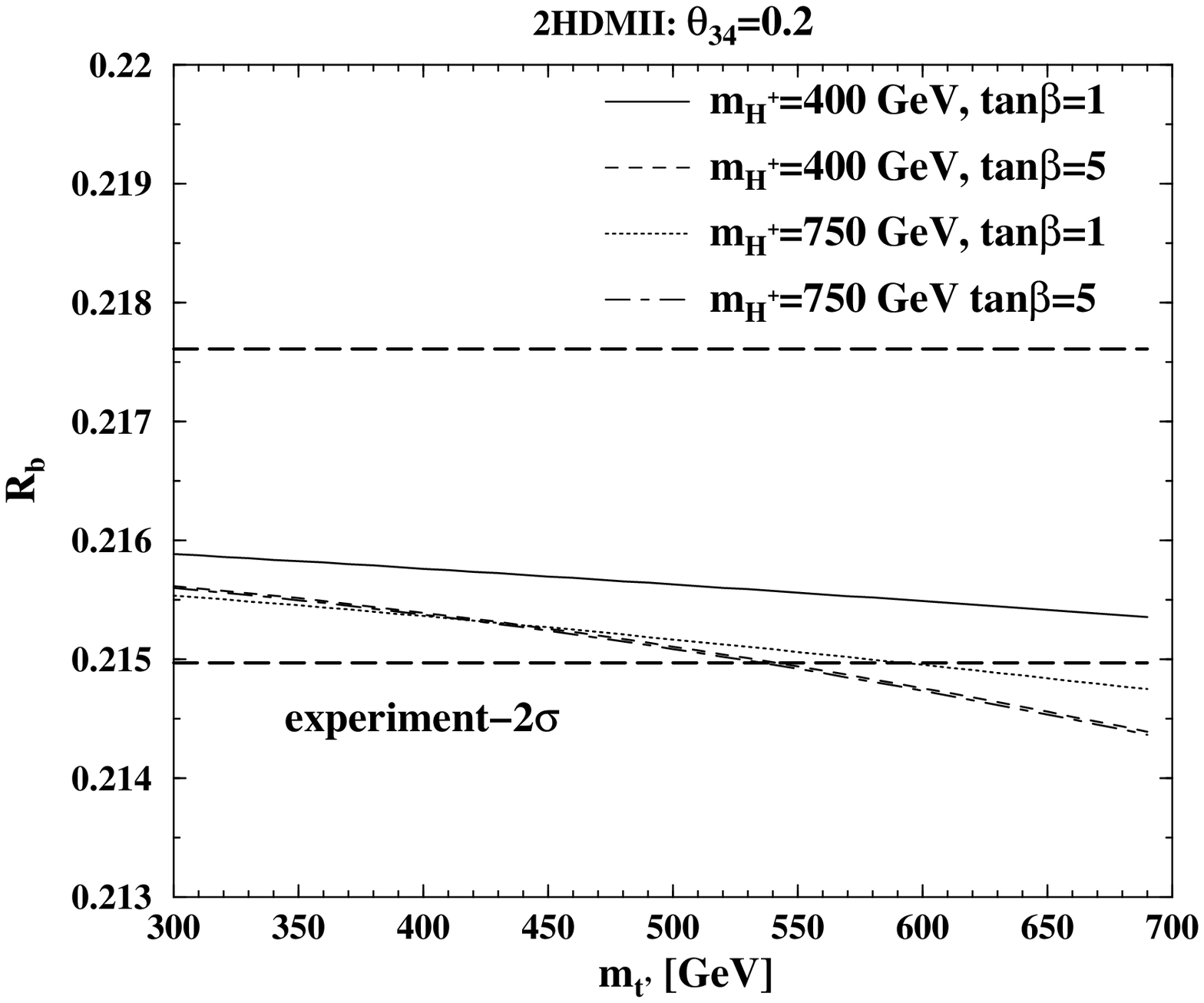,height=8cm,width=8cm,angle=0}
\caption{\emph{$R_b$ in the 4G2HDMI (upper plots, figure taken from \cite{4G2HDM}) and in the
2HDMII (lower plots), as a function of
the charged Higgs mass (left plots) for
$m_{t^\prime}=500$ GeV,
and $(\tan\beta,\theta_{34})=(1,0),(1,0.2),(5,0),(5,0.2)$,
and as a function of $m_{t^\prime}$ (right plots), for $\theta_{34}=0.2$ and
$(m_{H^+}~[{\rm GeV}],\tan\beta)=(400,1),(400,5),(750,1),(750,5)$ (right).
In the 4G2HDMI case we use $\epsilon_t = m_t/m_{t^\prime}$. The long-dashed horizontal
lines represent the upper and lower $2\sigma$ (measured) bounds on $R_b$.
}}
\label{figRb}
\end{center}
\end{figure}
\begin{widetext}
\begin{figure}[htb]
\begin{center}
\epsfig{file=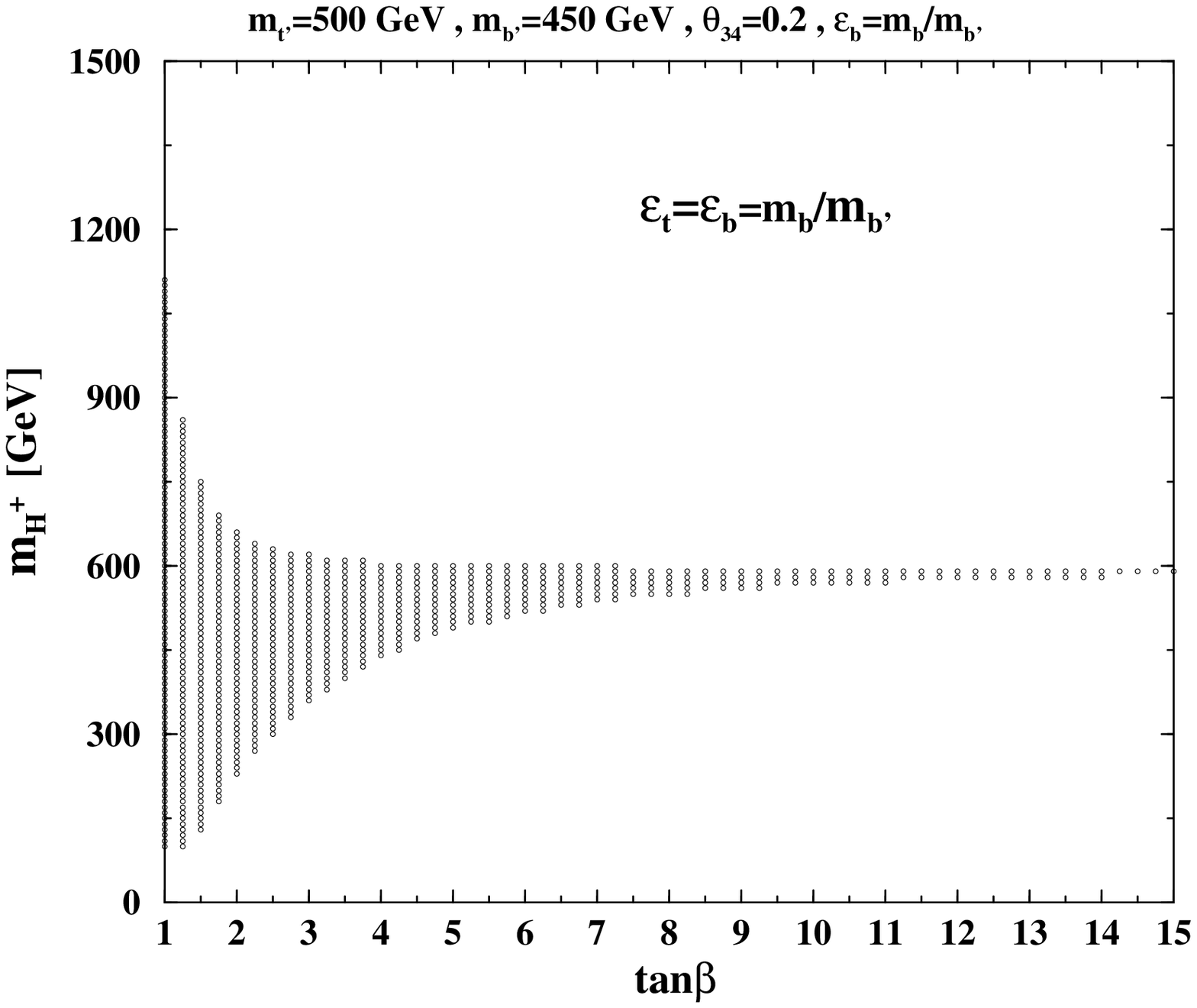,height=5.5cm,width=5.5cm,angle=0}
\epsfig{file=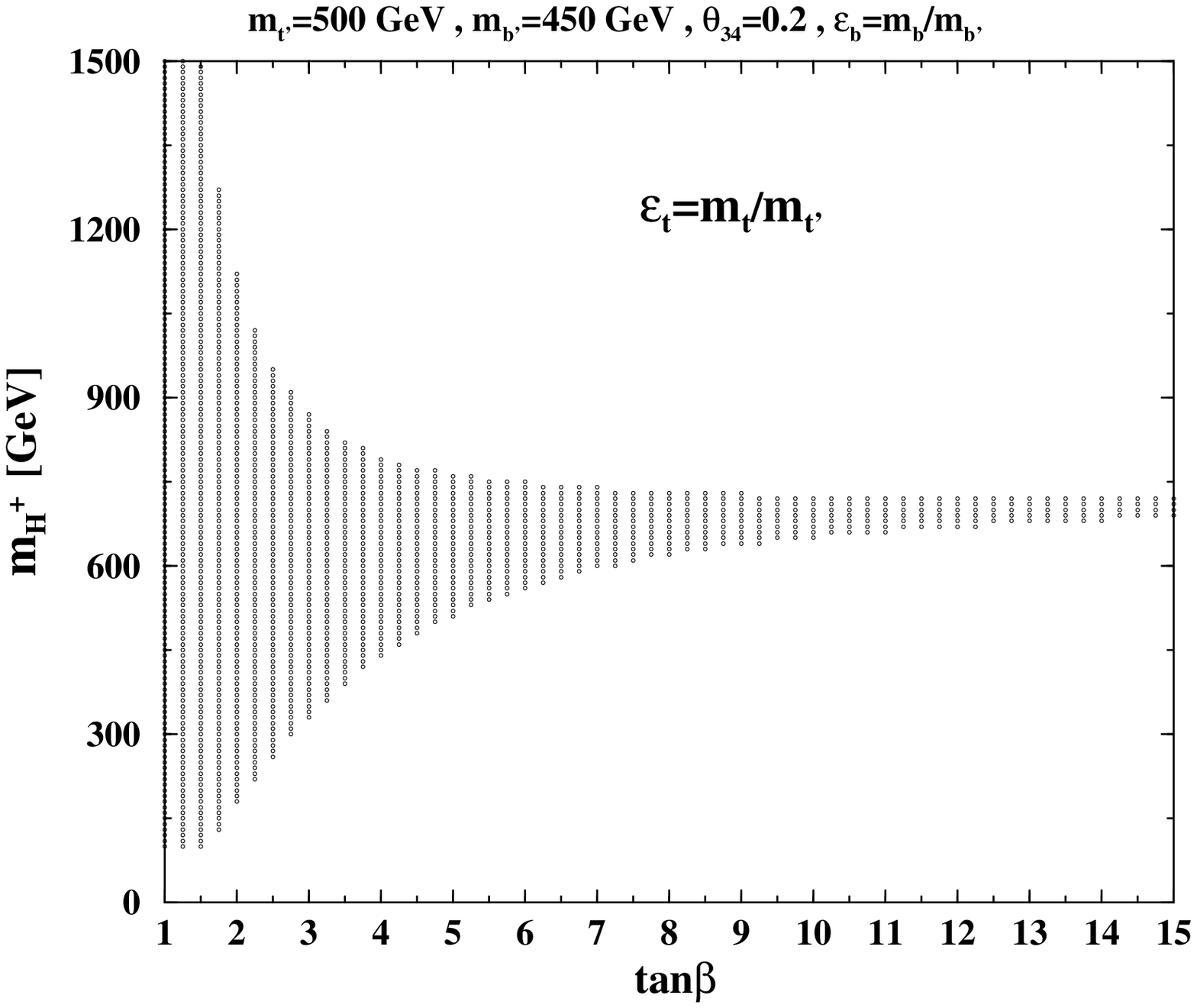,height=5.5cm,width=5.5cm,angle=0}
\epsfig{file=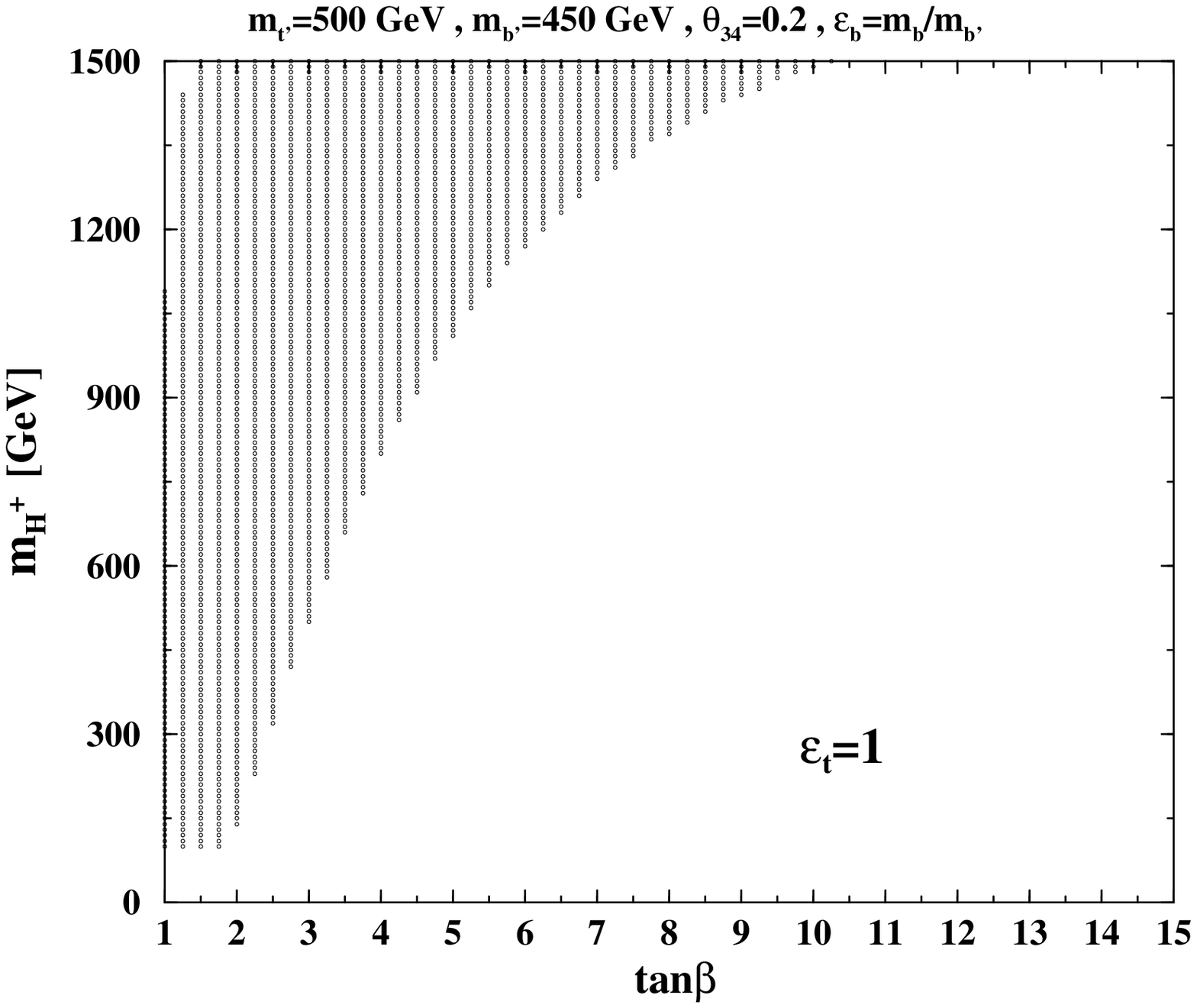,height=5.5cm,width=5.5cm,angle=0}
\caption{\emph{Allowed
area in the $m_{H^+} - \tan\beta$ in the 4G2HDMI, subject to the
$R_b$ measurement (within $2\sigma$),
for $m_{t^\prime}=500$ GeV, $m_{b^\prime}=450$ GeV, $\theta_{34}=0.2$, $\epsilon_b = m_b/m_{b^\prime}$ and for three values of
the $t-t^\prime$ mixing parameter: $\epsilon_t = \epsilon_b \sim 0.01$ (left plot),
$\epsilon_t = m_t/m_{t^\prime} \sim 0.35$ (middle plot) and $\epsilon_t = 1$ (right plot). Figure taken from \cite{4G2HDM}.}}
\label{figmhctb}
\end{center}
\end{figure}
\end{widetext}

\begin{enumerate}

\item {\bf SM4-like corrections:}\\

These are the corrections to $g_{qL}$ due to the 1-loop
$W-t^\prime$ exchanges (denoted here as $g_{qL}^{SM4}$),
which are given by \cite{SAGMN10,chenowitz,RbSM4}:
\begin{eqnarray}
g_{qL}^{SM4} = \frac{g^2}{64 \pi^2 c_W^2}
\left( \frac{m_{t^\prime}^2}{m_Z^2} - \frac{m_{t}^2}{m_Z^2} \right) \sin^2\theta_{34} \label{SM4cor}~,
\end{eqnarray}

\noindent where $\theta_{34}$ is the mixing angle between the 3rd and 4th generation
quarks, i.e., defining $|V_{t^\prime b}| = |V_{t b^\prime}| \equiv \sin\theta_{34}$,
and the 2nd term $\propto - \sin^2\theta_{34} m_t^2/m_Z^2$ is the decrease from the
SM's $tb$ correction to the W-boson vacuum polarization, which
in the 4th generation case,
is
$\propto |V_{tb}|^2 = \cos^2\theta_{34} = 1 - \sin^2\theta_{34}$.

The SM4-like effect on $R_b$ is plotted in Fig.~\ref{figRbSM4}, from which we
can see that $R_b$ puts rather stringent constraints on the
$m_{t^\prime} - \theta_{34}$ plane which is the SM4 subspace of the
parameter space of any 2HDM containing a 4th generation of fermions.
For example, for $m_{t^\prime} \sim 500$ GeV
the $t^\prime - b$ mixing angle is restricted to $\theta_{34} \lsim 0.2$.

\item {\bf $H^+ - t^\prime$ exchanges:}\\

The corrections from the 1-loop $H^+ - t^\prime$ exchanges
are plotted in Fig.~\ref{diagrams}. In the 4G2HDM of types II and III,
these charged Higgs exchange diagrams
are found to have negligible effects on $R_b$ and are,
therefore, not constrained by this quantity. On the other hand,
$R_b$ is rather sensitive to the
charged Higgs 1-loop exchanges within the 4G2HDMI.
This can be seen in Fig.~\ref{figRb}, where
$R_b$ is plotted (for the 4G2HDMI case)
as a function of the charged Higgs and $t^\prime$
masses, fixing $\epsilon_t = m_t/m_{t^\prime}$ and focusing
on the values $\tan\beta=1,5$, $\theta_{34}=0,0.2$ and
$m_{H^+}=400,750$ GeV.

In Fig.~\ref{figmhctb} we further plot the allowed ranges in the $m_{H^+} - \tan\beta$ plane
in the 4G2HDMI, subject to the $R_b$ constraint (at $2\sigma$), for $\tan\beta$ in the range
1-15, fixing $m_{t^\prime} =500$ GeV, $m_{b^\prime} =450$ GeV, $\theta_{34}=0.2$, $\epsilon_b = m_b/m_{b^\prime}$ (which also enters the $t^\prime b H^+$ vertex)
and for three representative values of
the $t-t^\prime$ mixing parameter: $\epsilon_t = \epsilon_b \sim 0.01$, $\epsilon_t = m_t/m_{t^\prime} \sim 0.35$
and $\epsilon_t = 1$.
We see that, as expected, when $\tan\beta$ is lowered,
the constraints on the charged Higgs mass are weakened.
In particular, while there are no constraints from $R_b$
on the charged Higgs and $t^\prime$
masses if $\tan\beta \sim {\cal O}(1)$, for higher values of $\tan\beta$
a more restricted region of the charged Higgs mass is imposed which again
depends on $\theta_{34}$.
We see e.g., that for $\epsilon_t = m_t/m_{t^\prime} \sim 0.35$,
$\tan\beta \sim 1$ is compatible with $m_{H^+}$ values ranging from 200 GeV up to the TeV scale,
while for $\tan\beta \sim 5$ the charged Higgs mass is restricted to be
within the range $ 450 ~{\rm GeV} \lsim m_{H^+} \lsim 750 ~{\rm GeV} $.

For the case of the 2HDMII (i.e., extended with a 4th family of fermions),
which is also plotted in Fig.~\ref{figRb}, we find that
there is essentially no constraint in the $m_{H^+} - \tan\beta$ plane for $m_{t^\prime} \lsim 500$ GeV.

\item {\bf The flavor changing ${\cal H}^0 b b^\prime$ interactions:}\\

The 1-loop Corrections to $R_b$ which involve the
flavor changing (FC) ${\cal H}^0 b b^\prime$ interactions
emanate from the non-diagonal 34 and 43 elements in $\Sigma^d$,
with ${\cal H}^0 = h,H$ or $A$. These corrections are found to be much smaller than
1-loop $H^+$ exchanges, so that they can be safely neglected, in particular for
$\epsilon_b \ll 1$.
\end{enumerate}

\subsection{Constraints from flavor in b-physics}
\subsubsection{${\bar B} \to X_s \gamma$}

Flavor physics plays an important role in discriminating between the various NP models.
In this regard, FCNC decays can provide key information about the SM and its various extensions.

The inclusive radiative decay ${\bar B} \to X_s \gamma$ is indeed known to be a very sensitive probe of NP.
The underlying process is induced by the FC decay
of the $b$-quark into a strange quark and a photon. The Br(${\bar B} \to X_s \gamma$) has already carved out large
regions of the parameter space of most of the NP models \cite{bsgnp,burzumati,CDGG98,DGG00}.
On the other hand, model independent analysis in the effective field theory approach
without \cite{bsgmid1} and with \cite{bsgmid2} the assumption of minimal flavor violation
also show the strong constraining power of the decay ${\bar B} \to X_s \gamma$.
Once more precise data from Super-B factories are available, this decay will undoubtedly
be more efficient in selecting the viable regions of the parameter space in the various
classes of NP models.

\begin{figure*}[htbp]
\vspace{5mm}
\hspace*{-10mm}  $\gamma$ \hspace{36.5mm} $\gamma$
\hspace{42.0mm} $\gamma$ \hspace{36.5mm} $\gamma$ \\[7mm]
\hspace*{-0mm} $u_i$ \hspace{11mm} $u_i$ \hspace{18mm}
            $W^{\pm}$ \hspace{11mm}   $W^{\pm}$ \hspace{21mm}
    $u_i$ \hspace{11mm}  $u_i$ \hspace{18mm} $H^{\pm}$ \hspace{11mm} $H^{\pm}$ \\[-18mm]
\includegraphics[width=75mm,angle=0]{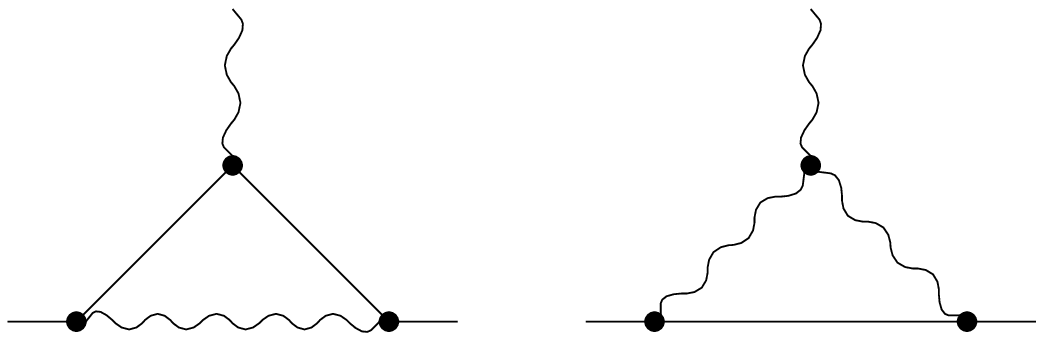}
\hspace{1cm}
\includegraphics[width=75mm,angle=0]{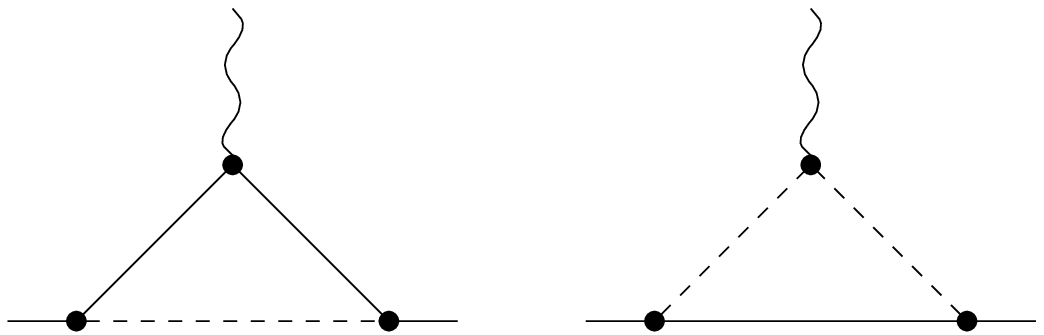}\\[-1mm]
\hspace*{-2mm}
$b$ \hspace{0.5cm}  $W^{\pm}$  \hspace{8mm} $s$ \hspace{12mm}
$b$ \hspace{9mm}   $u_i$   \hspace{7mm} $s$ \hspace{14.5mm}
$b$ \hspace{1cm} $H^{\pm}$ \hspace{8mm} $s$ \hspace{9mm}
$b$ \hspace{12mm}   $u_i$   \hspace{7mm} $s$ \\[-1cm]
\begin{center}
\caption{\emph{Examples of one-loop 1PI diagrams that contribute to $b \to s \gamma$ in a 2HDM framework, with $W$-bosons, charged Higgs and
4th generation quarks exchanges ($u_i=u,c,t,t^\prime$).}}
\label{bsg}
\end{center}
\vspace{-1cm}
\end{figure*}

The calculation of the decay rate of the ${\bar B} \to X_s \gamma$ transition is most conveniently performed after
integrating out the heavy degrees of freedom. The resulting effective theory contains various
FC dimension-five and -six local interactions and the inclusive decay rate is given by
\be
{\Gamma(b \to X_s\gamma)_{E_{\gamma}>E_0} = \frac{G^2_F\, m^5_b\,\alpha_{\it em}}{32\, \pi^4} |V^{\ast}_{ts}V_{tb}|^2 \sum_{i,j=1}^8
{C_{i}(\mu_b)\,C_{j}(\mu_b)} {G_{ij}(E_0,\mu_b)}},
\ee
where the Wilson coefficients, $C_i$, of the effective operators (see below)
are perturbatively calculable
at the relevant renormalization scale and the Renormalization Group Equations (RGE) can
be used to evaluate $C_i$ at the scale $\mu_b \sim m_b/2$. At present, all
the relevant Wilson coefficients $C_i(\mu_b)$ are known at the
Next-to-Next-to-Leading-Order (NNLO) \cite{GHW96,CMM97,AG95,P96,MM95,match2,Czakon:2006ss,Bobeth:1999mk} and
${G_{ij}(E_0,\mu_b)}$ is determined by the matrix elements of the operators $O_1,.....,O_8$ \cite{Chetyrkin:1996vx}:
\bea
O_{1,2} &=& (\bar{s} \Gamma_i c)(\bar{c} \Gamma'_i b), \hspace{1.6cm}
\begin{array}{l} \mbox{\footnotesize (current-current} \\[-1mm]
                 \mbox{\footnotesize ~operators)} \end{array}\nnb\\
O_{3,4,5,6} &=&  (\bar{s} \Gamma_i b) {\textstyle \sum_q} (\bar{q} \Gamma'_i q), \hspace{1cm}
\begin{array}{l} \mbox{\footnotesize (four-quark} \\[-1mm]
                 \mbox{\footnotesize ~penguin operators)} \end{array}\nnb\\
O_7 &=& \f{e m_b}{16 \pi^2}\, \bar{s}_L \sigma^{\mu \nu} b_R F_{\mu \nu}, \hspace{6.5mm}
\begin{array}{l} \mbox{\footnotesize (photonic dipole} \\[-1mm]
                 \mbox{\footnotesize ~operator)} \end{array}\nnb\\
O_8 &=& \f{g m_b}{16 \pi^2}\, \bar{s}_L \sigma^{\mu \nu} T^a b_R G^a_{\mu\nu}. \hspace{2mm}
\begin{array}{l} \mbox{\footnotesize (gluonic dipole} \\[-1mm]
                 \mbox{\footnotesize ~operator)} \end{array} \label{ops},
\eea
which consists of perturbative and non-perturbative corrections.
The perturbative corrections are well under control and are fully known at NLO QCD \cite{misiak08}.
However, quantitative estimates of all the non-perturbative effects are not available,
although they are believed to be $\approx 5\%$ \cite{misiak08}.

The inclusive branching ratio in the SM is given by \cite{smbsg}:
\be
{{\cal B}({\bar B}\to X_s \gamma)^{NNLO}_{E_{\gamma}>1.6\, {\rm GeV}}= (3.15 \pm 0.23 )\times 10^{-4}},
\ee
whereas the current experimental data gives \cite{hfag10}:
\be
{\cal B}({\bar B}\to X_s \gamma)^{exp}_{E_{\gamma}>1.6\, {\rm GeV}}= (3.55 \pm 0.24 \pm 0.09)\times 10^{-4}.
\ee

The SM prediction is, thus, consistent with the experiment (both having a 7\% error) and
is therefore useful for constraining many extensions of the SM.

In the SM4 there are no new operators other than the ones present in the SM. However, there are extra contributions
to the Wilson coefficients corresponding to the operators $O_7$ and $O_8$ from $t'$-loops \cite{SAGMN08,SAGMN10,ajb10B,buras_charm}.
In a 2HDM framework with a 4th generation family, the new ingredient with respect to the SM4 is the
presence of the charged Higgs 1-loop exchanges which contribute
to the Wilson coefficients of the effective theory. In particular, at the parton level within
a 2HDM, ${\bar B} \to X_s \gamma$ proceeds via the penguin diagrams depicted in Fig.~\ref{bsg}.
As was shown in \cite{4G2HDM}, in the 4G2HDMI, 4G2HDMII and 4G2HDMIII frameworks,
the leading effects enter in $C_7$ and $C_8$ from
the 1-loop exchanges of $t^\prime - W$, $t -H^+$ and $t^\prime - H^+$.

\begin{figure}[t]
\begin{center}
\epsfig{file=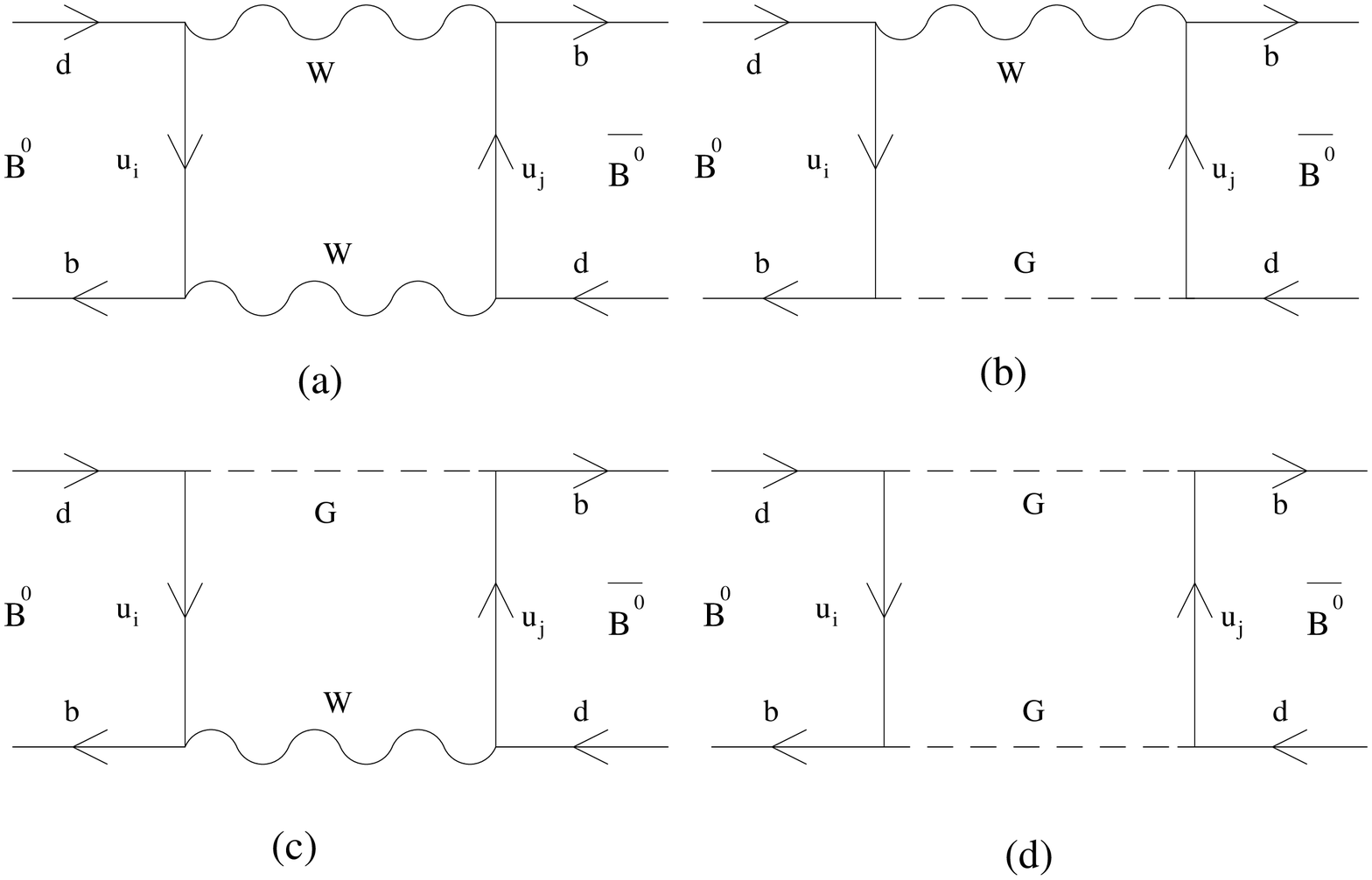,height=6cm,width=15cm,angle=0}
\caption{\emph{The $B^0-{\bar B^0}$ (representative)
box diagrams with different combinations of the gauge bosons
$(W, G)$ and the fermions $(u_i,u_j)$ in the internal lines. The same diagrams contribute to $B^s-{\bar B^s}$ mixing with the $d$-quark in the external lines replaced by the $s$-quark.}}
\label{bbbox}
\end{center}
\end{figure}

\subsubsection{$B_q-\bar B_q$ mixing}

An important role for constraining NP in the b-quark system is also played by $B_q-\bar B_q$ ($q=d,s$) mixing,
the phenomenon of which is described by the dispersive part
$M_{12}^q$ of the $B_q$ mixing amplitude. The current theory precision is limited by lattice results;
the SM prediction still allows NP contributions to $|M^s_{12}|$ of order 20\%~\cite{lenz_uli}.

Within a 2HDM setup, the leading contribution to  $B_q - {\bar B_q}$ ($q=d,s$) mixing comes
from the box diagrams shown in Fig.~\ref{bbbox}, where the $G$-boson is replaced by the
charged Higgs $H^+$, and the fermions $u_{i,j}$ are replaced by $(t, t')$.
Thus, the net contribution to the mass difference $\Delta M_q = 2|M_{{12}_q}|$ is given by \cite{4G2HDM}:
\be
M_{{12}_q} = \frac{G_F^2}{12\pi^2} M_W^2 f_{B_q}^2 B_q M_{B_q} \left[M_{WW} + M_{HH} + M_{HW} \right],
\ee
where
\begin{widetext}
\bea
M_{WW} &=& {\lambda^t_{bq}}^2 \eta_{tt} S_{WW}(x_t) + {\lambda^{t'}_{bq}}^2 \eta_{t't'} S_{WW}(x_{t'}) +
2~ \lambda^t_{bq}  \lambda^{t'}_{bq} \eta_{tt'} S_{WW}(x_t,x_{t'}), \nonumber \\
M_{HH} &=& {\lambda^t_{bq}}^2 S_{HH}(y_t) + {\lambda^{t'}_{bq}}^2 S_{HH}(y_{t'}) +
2~ \lambda^t_{bq}  \lambda^{t'}_{bq} S_{HH}(y_t,y_{t'}), \nonumber \\
M_{HW} &=& {\lambda^t_{bq}}^2 S_{HW}(x_t,z) + {\lambda^{t'}_{bq}}^2 S_{HW}(x_{t'},z) +
2~ \lambda^t_{bq}  \lambda^{t'}_{bq} S_{HW}(x_t,x_{t'},z),
\eea
\end{widetext}
and $z = \frac{m_{H^+}^2}{m_W^2}$, $x_i = \frac{m_i^2}{m_W^2}$, $y_i = \frac{m_i^2}{m_{H^+}^2}$ ($i = t$ or $t'$),
$\lambda^u_{d_i d_j} \equiv V_{u d_i}^\star V_{u d_j}$. Here, $M_{WW}$, $M_{HH}$ and $M_{HW}$
are the contributions from the box diagrams with the combination of the gauge bosons
$(W,W)$, $(W,H)$ and $(H,H)$ in the internal lines ($H$ stands for the charged Higgs), respectively. The detail
expression for the various Inami-Lim functions $S_{i,j}$ are given in Ref.~\cite{4G2HDM}.

\begin{widetext}
\begin{table*}[htbp]
\begin{center}
\begin{tabular}{|c|c|}
\hline
& \\
$f_{bd}\sqrt{B_{bd}} = 0.224 \pm 0.015$\, {GeV} \cite{Gamiz:2009ku,gamizp}& $|V_{ub}| = (32.8 \pm 2.6)\times 10^{-4}$
\footnote{It is the weighted average of
$V_{ub}^{inc}=(40.1\pm 2.7 \pm 4.0)\times 10^{-4}$  and
$V_{ub}^{exc}=(29.7 \pm 3.1)\times 10^{-4}$ for the inclusive and exclusive values, respectively.
In our numerical analysis,
we increase the error on $V_{ub}$ by 50\% and take the total error to be
around 12\% due to the appreciable
disagreement between the two determinations.} \\
$\xi = 1.232 \pm 0.042$ \cite{Gamiz:2009ku,gamizp} & $|V_{cb}| = (40.86 \pm 1.0)\times 10^{-3} $ \\
$\eta_t = 0.5765\pm 0.0065$ \cite{buras1}  & $\gamma = (73.0 \pm 13.0)^{\circ} $\\
$\Delta{M_s} = (17.77 \pm 0.12) ps^{-1}$    & ${\cal{BR}}(B\to X_s \gamma) = (3.55 \pm 0.25)\times 10^{-4}$\\
$\Delta{M_d} = (0.507 \pm 0.005) ps^{-1}$    & $m_b(m_b) = 4.23 \, GeV$ \\
$f_B = (0.208 \pm 0.008)$  GeV  & $\alpha_s(M_Z) = 0.11$ \\
$m_t^{pole} = (170 \pm 4) $ GeV  & $\tau_{B^+} = 1.63\,ps$\\
 & $m_{\tau} = 1.77$  GeV  \\
\hline
\end{tabular}
\caption{\emph{Inputs used for the B-physics parameters in the analysis below. When not
explicitly stated, we take the inputs from Particle Data Group \cite{PDG}.}}
\label{tab1}
\end{center}
\end{table*}
\end{widetext}

For the B-physics parameters we use the inputs given in
Table \ref{tab1} and for the 4th generation quark masses we take
$m_{t'}=500$ GeV and $m_{b'}=450$ GeV.

\begin{widetext}
\begin{figure}[htb]
\begin{center}
\epsfig{file=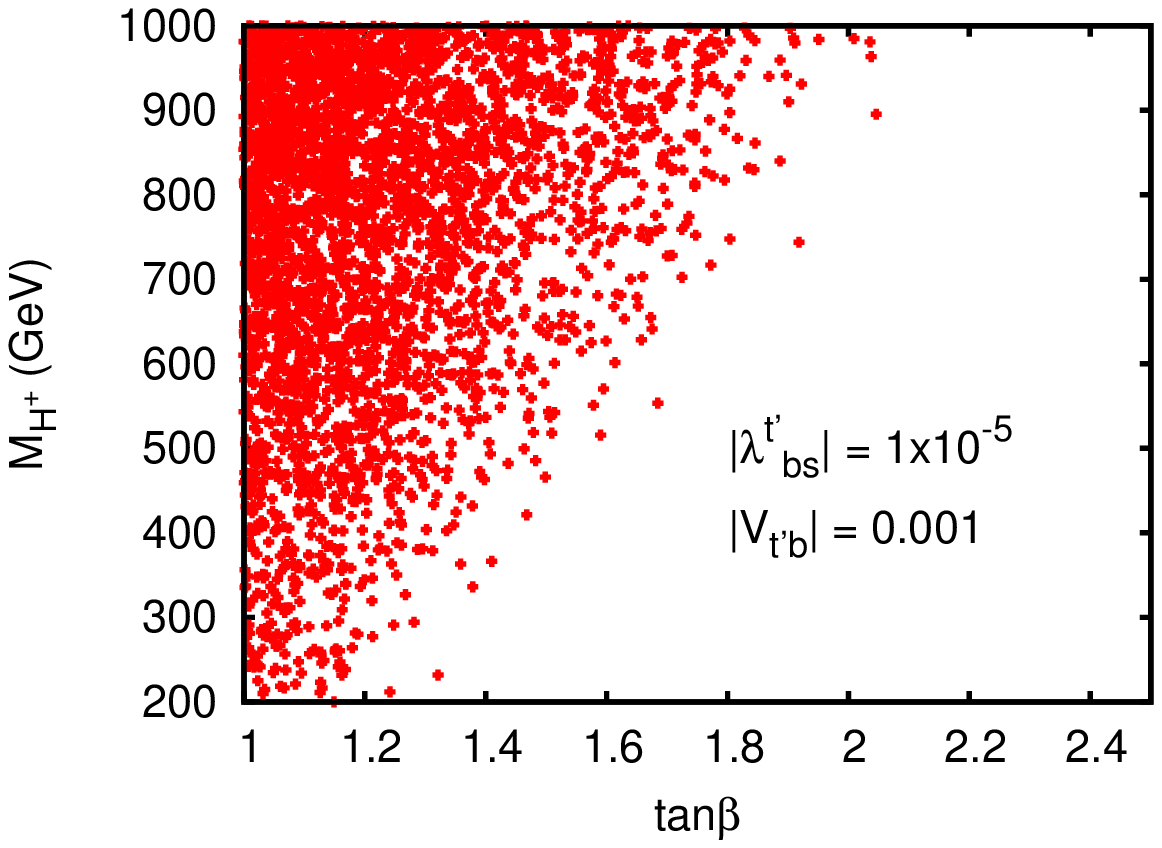,height=5.5cm,width=5.5cm,angle=0}
\epsfig{file=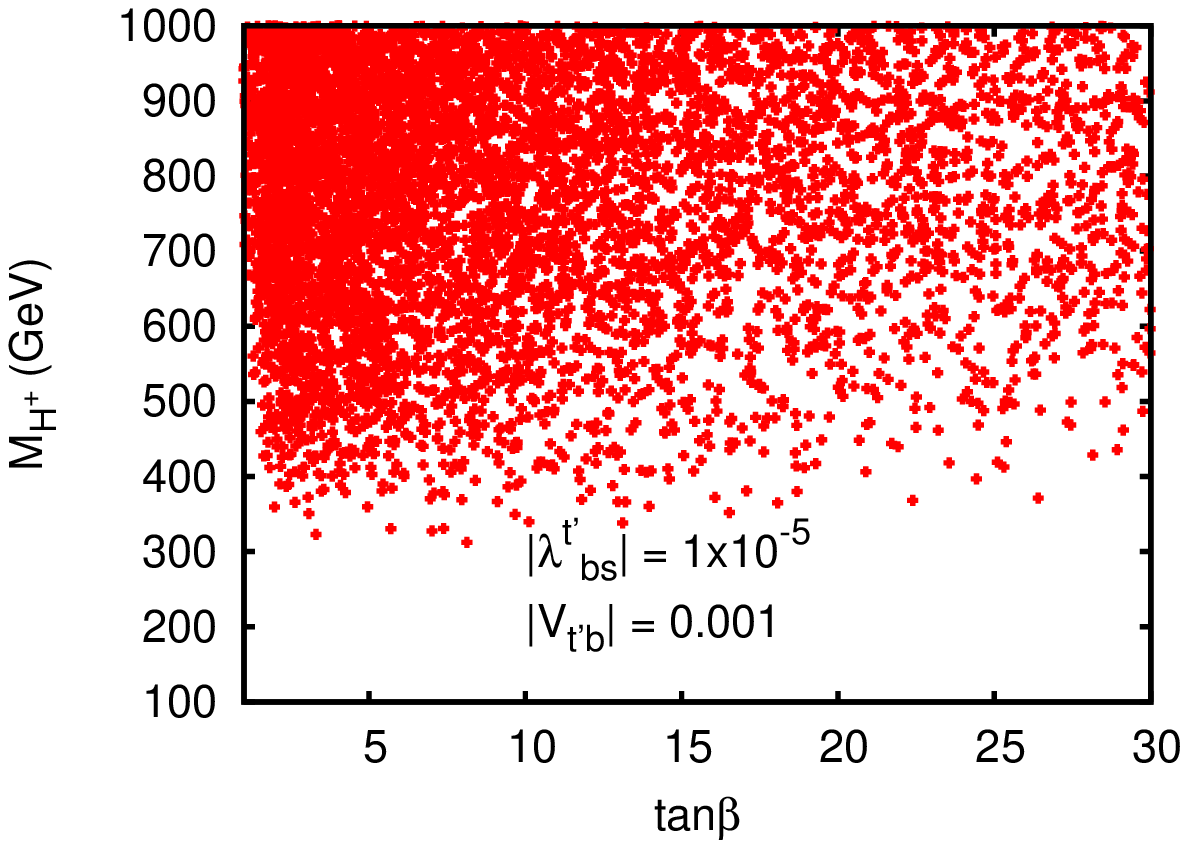,height=5.5cm,width=5.5cm,angle=0}
\epsfig{file=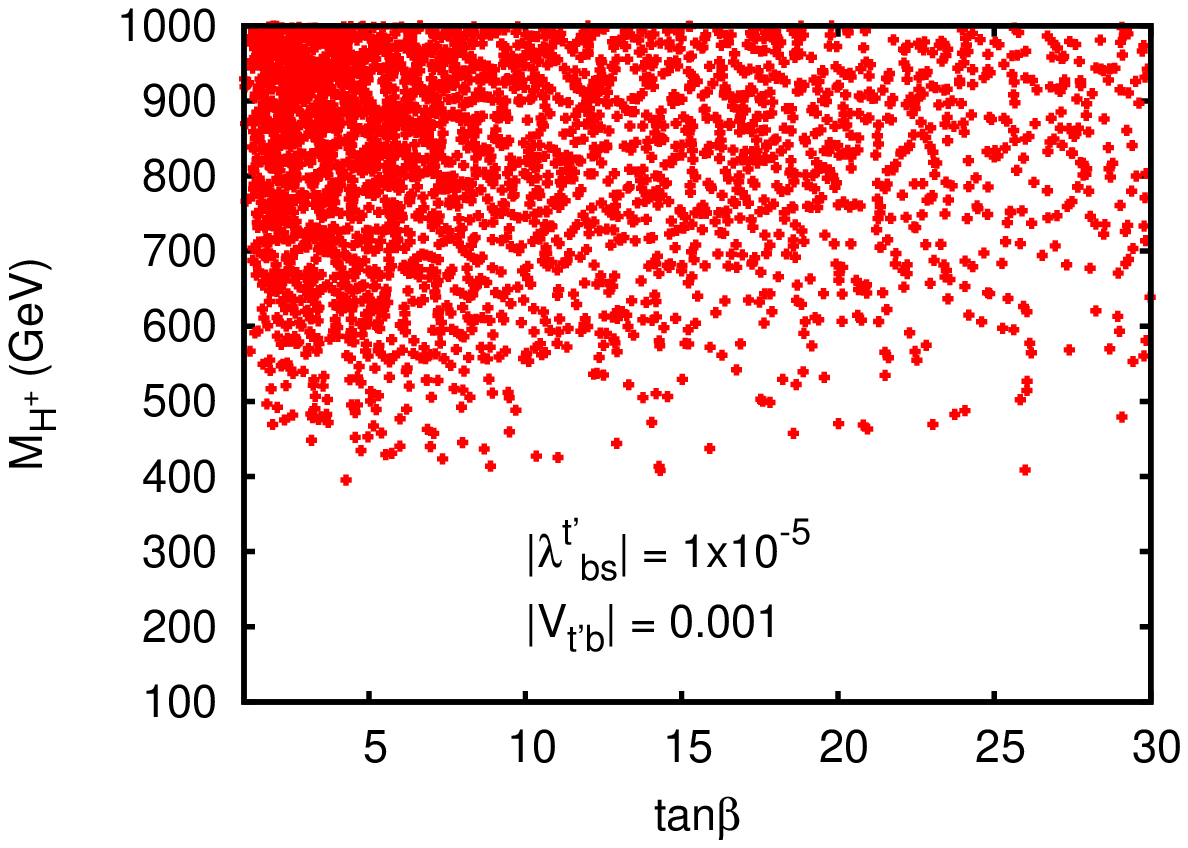,height=5.5cm,width=5.5cm,angle=0}
\caption{\emph{The ``3+1" scenario, $V_{t' b} = 0.001$ ($|\lambda^{t'}_{sb}| = 10^{-5}$):
the allowed parameter space in the
$m_{H^+}-\tan\beta$  plane, following
constraints from $B \to X_s \gamma$ and $B_q$-${\bar B_q}$ mixing,
in the 4G2HDMI (left), the 4G2HDMII (middle) and the 4G2HDMIII (right), for
$m_{t'}=500$ GeV, $m_{b'}=450$ GeV,
$\epsilon_b = m_b/m_{b'}$
and $\epsilon_t = 0.34 (\sim m_t/m_{t'})$. Figure taken from \cite{4G2HDM}.}}
\label{figI-vtb01}
\end{center}
\end{figure}
\end{widetext}
\begin{widetext}
\begin{figure}[htb]
\begin{center}
\epsfig{file=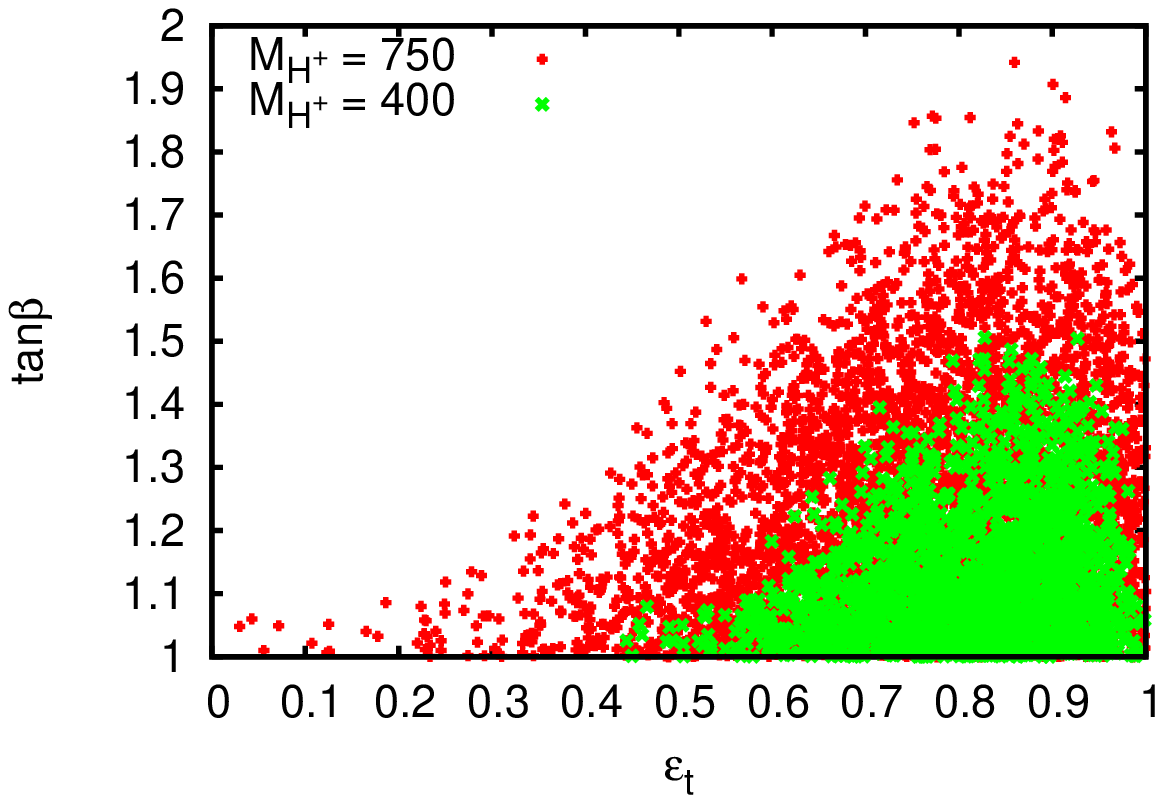,height=5.5cm,width=5.5cm,angle=0}
\epsfig{file=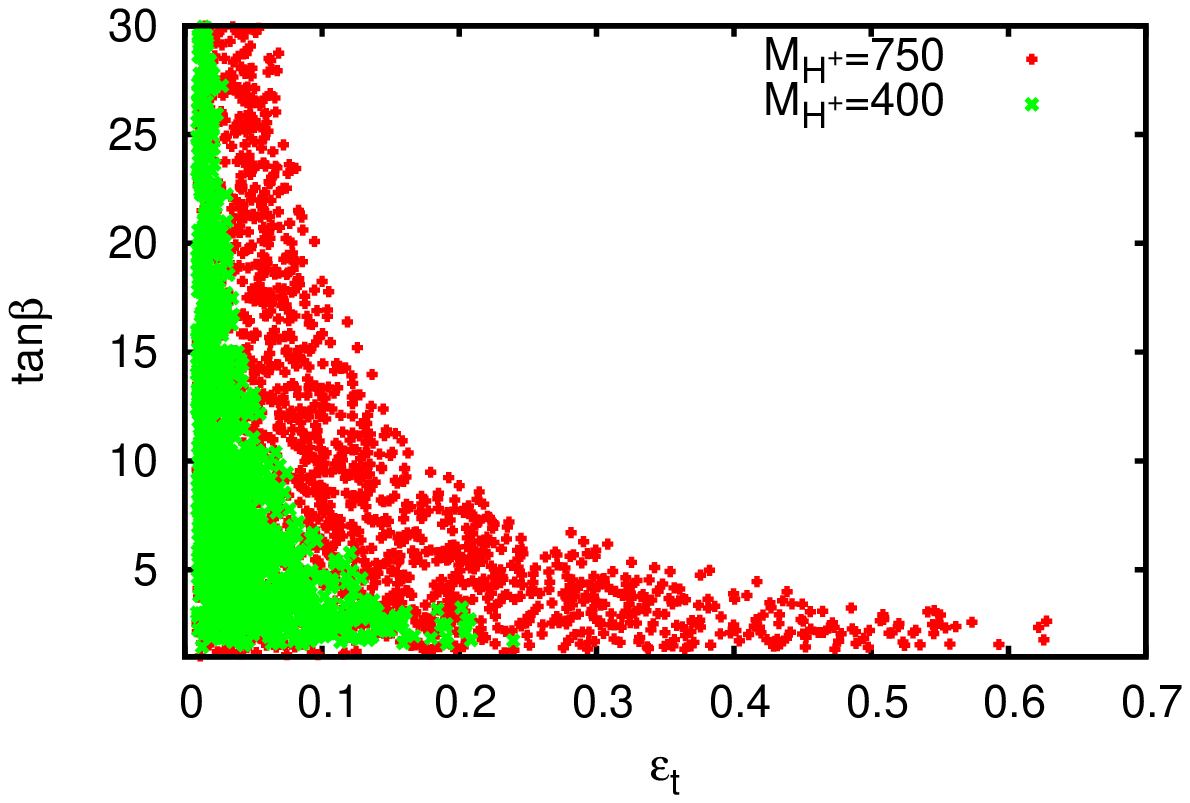,height=5.5cm,width=5.5cm,angle=0}
\epsfig{file=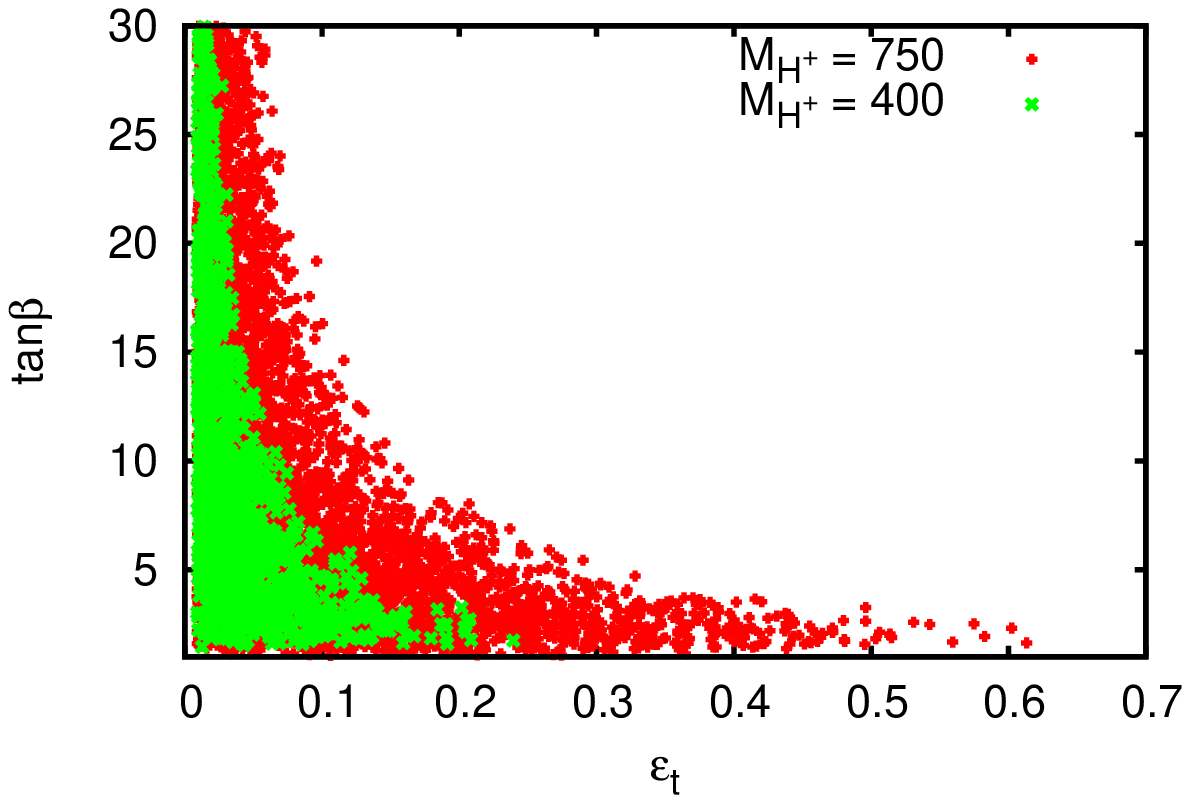,height=5.5cm,width=5.5cm,angle=0}
\caption{
\emph{The ``3+1" scenario, $V_{t' b} = 0.001$ ($|\lambda^{t'}_{sb}| = 10^{-5}$):
the allowed parameter space in the
$\tan\beta - \epsilon_t$ plane, following
constraints from $B \to X_s \gamma$ and $B_q$-${\bar B_q}$ mixing,
in the 4G2HDMI (left), the 4G2HDMII (middle) and the 4G2HDMIII (right), for
$m_{t'}=500$ GeV, $m_{b'}=450$ GeV,
$\epsilon_b = m_b/m_{b'}$
and with $m_{H^+}=400$ and $750$ GeV. Figure taken from \cite{4G2HDM}.}}
\label{figI-vtb02}
\end{center}
\end{figure}
\end{widetext}
\begin{widetext}
\begin{figure}[htb]
\begin{center}
\epsfig{file=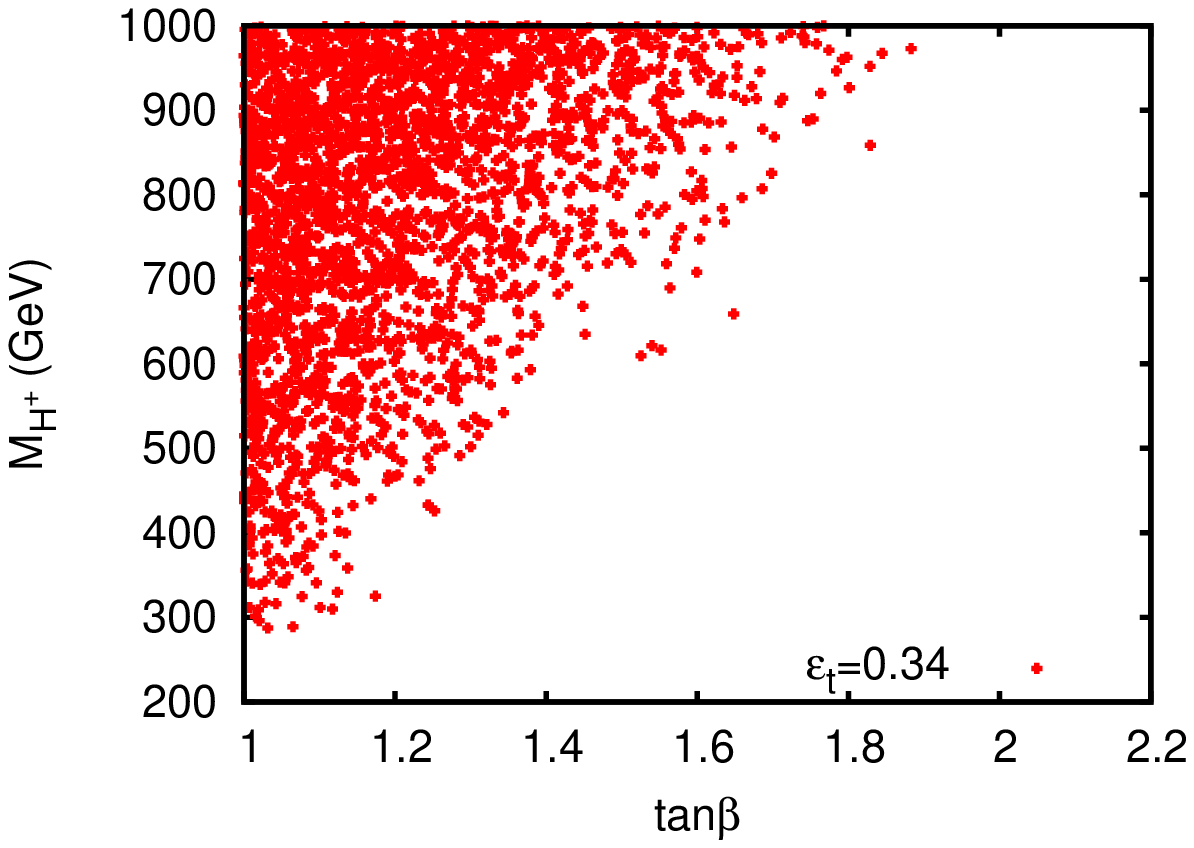,height=5.5cm,width=5.5cm,angle=0}
\epsfig{file=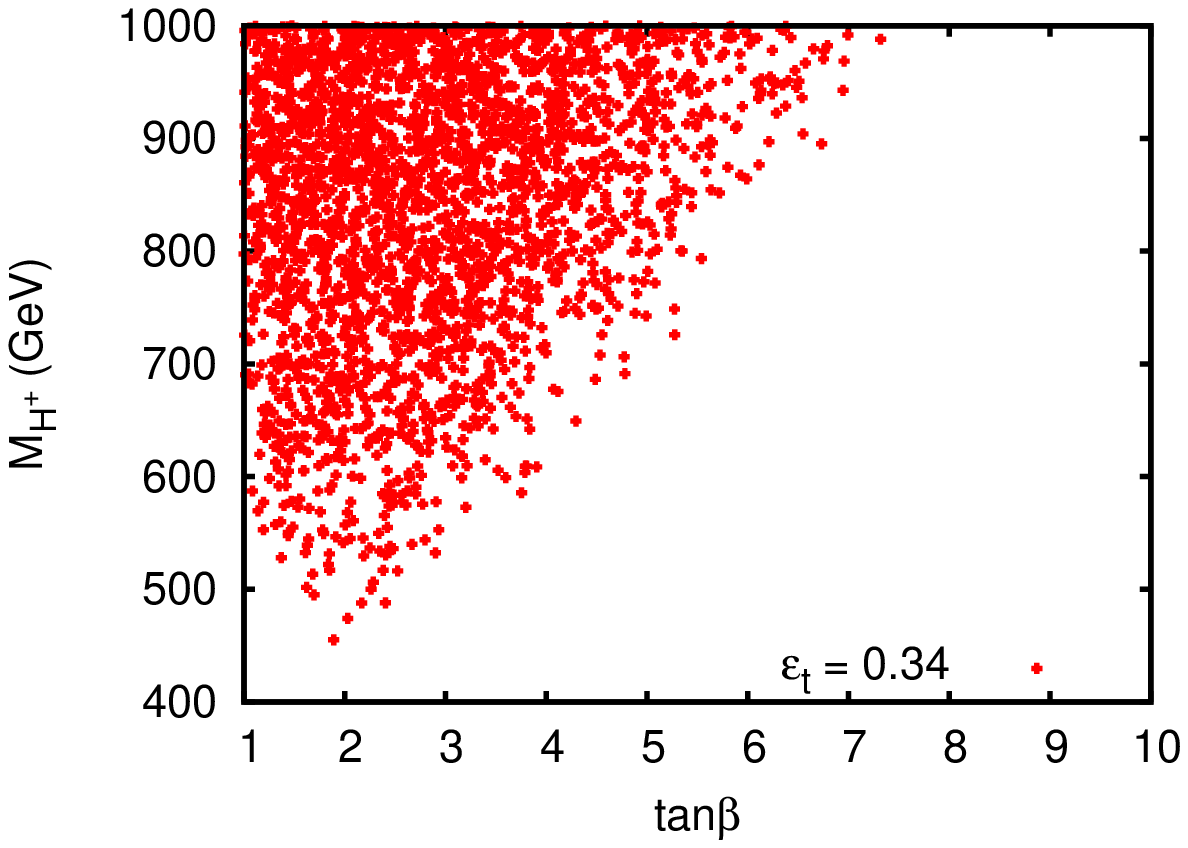,height=5.5cm,width=5.5cm,angle=0}
\epsfig{file=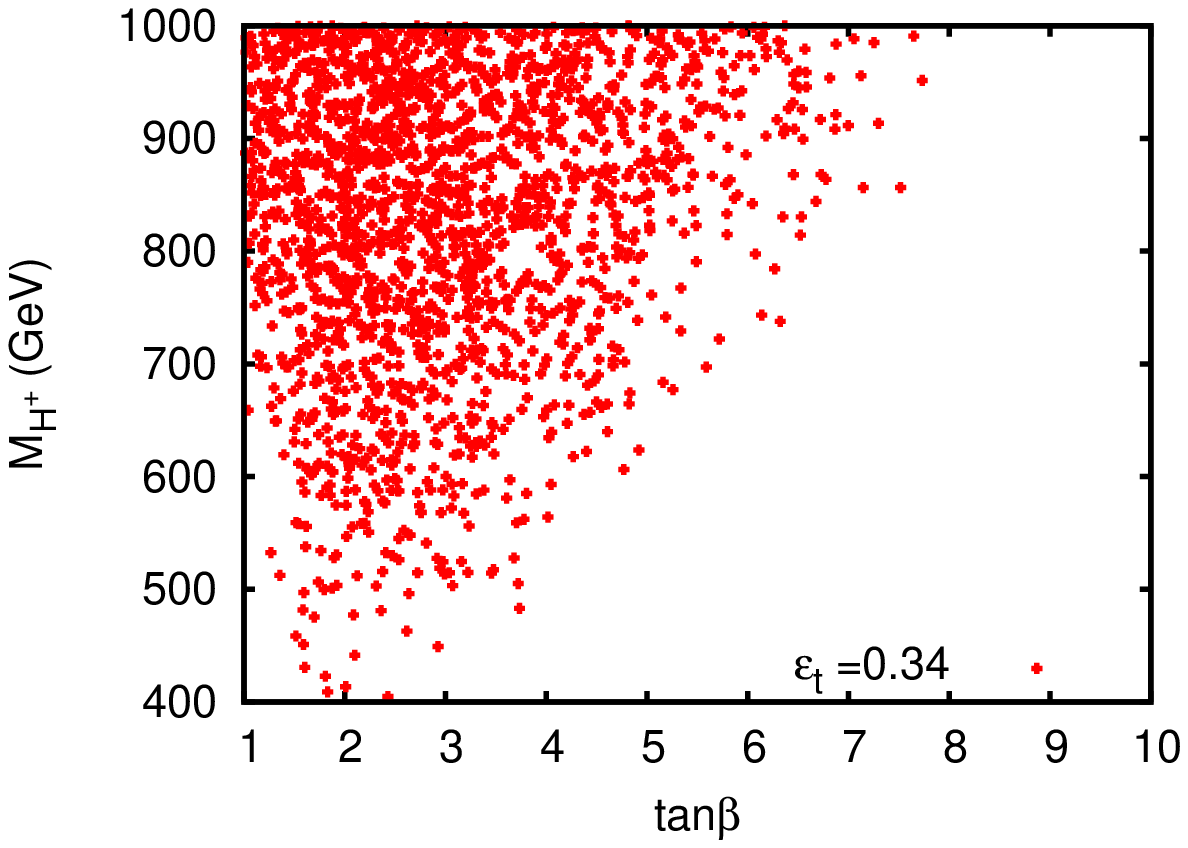,height=5.5cm,width=5.5cm,angle=0}
\caption{\emph{The Cabibbo size mixing case, $V_{t' b} = 0.2$ ($|\lambda^{t'}_{sb}|=0.004$):
the allowed parameter space in the
$m_{H^+} - \tan\beta$ plane, following
constraints from $B \to X_s \gamma$ and $B_q$-${\bar B_q}$ mixing, in the 4G2HDMI (left),
4G2HDMII (middle) and 4G2HDMIII (right), for
$m_{t'}=500$ GeV, $m_{b'}=450$ GeV,
$\epsilon_b = m_b/m_{b'}$ and
$\epsilon_t = 0.34 (\sim m_t/m_{t'})$. Figure taken from \cite{4G2HDM}.}}
\label{figII-vtb}
\end{center}
\end{figure}
\end{widetext}
\begin{widetext}
\begin{figure}[htb]
\begin{center}
\epsfig{file=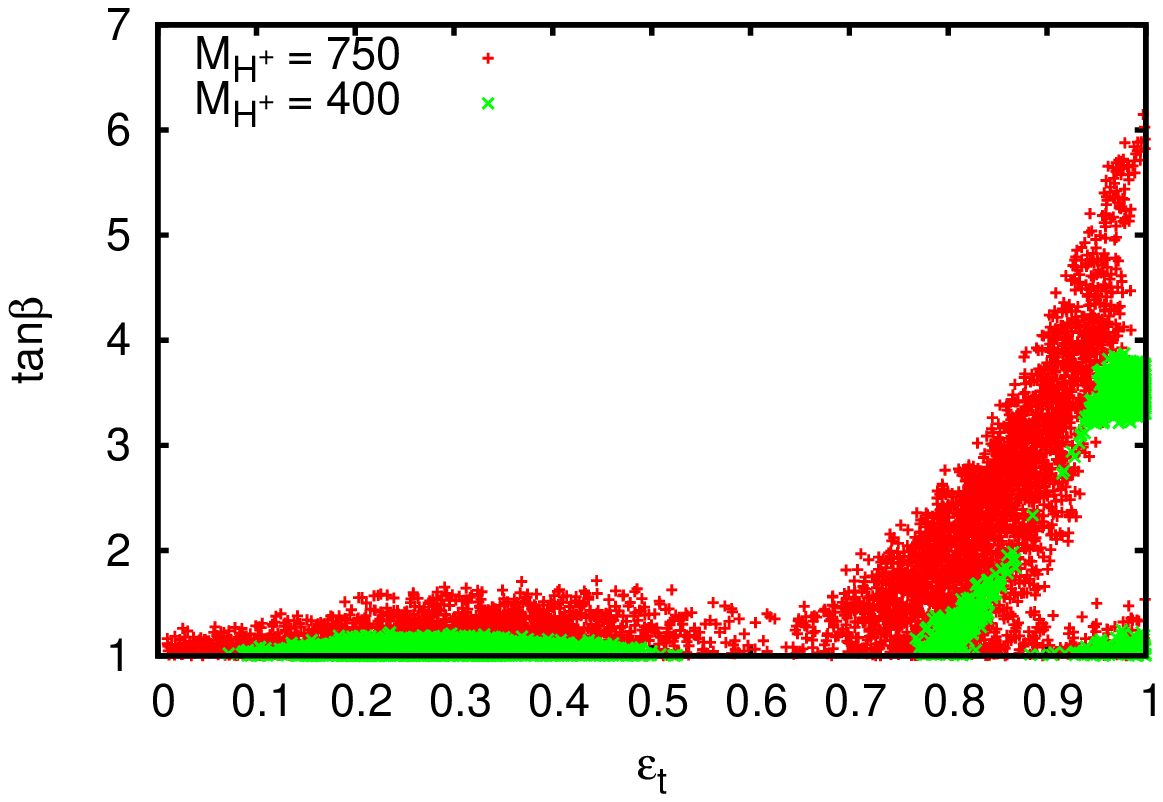,height=5.5cm,width=5.5cm,angle=0}
\epsfig{file=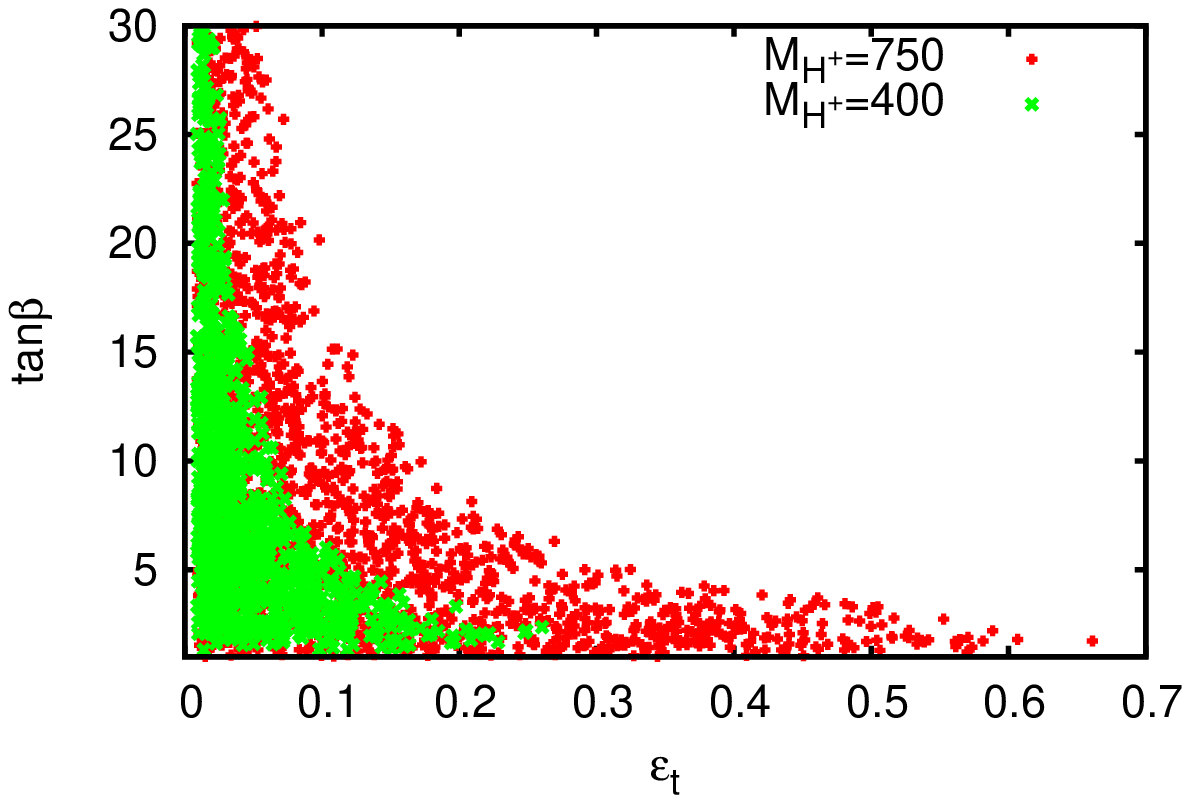,height=5.5cm,width=5.5cm,angle=0}
\epsfig{file=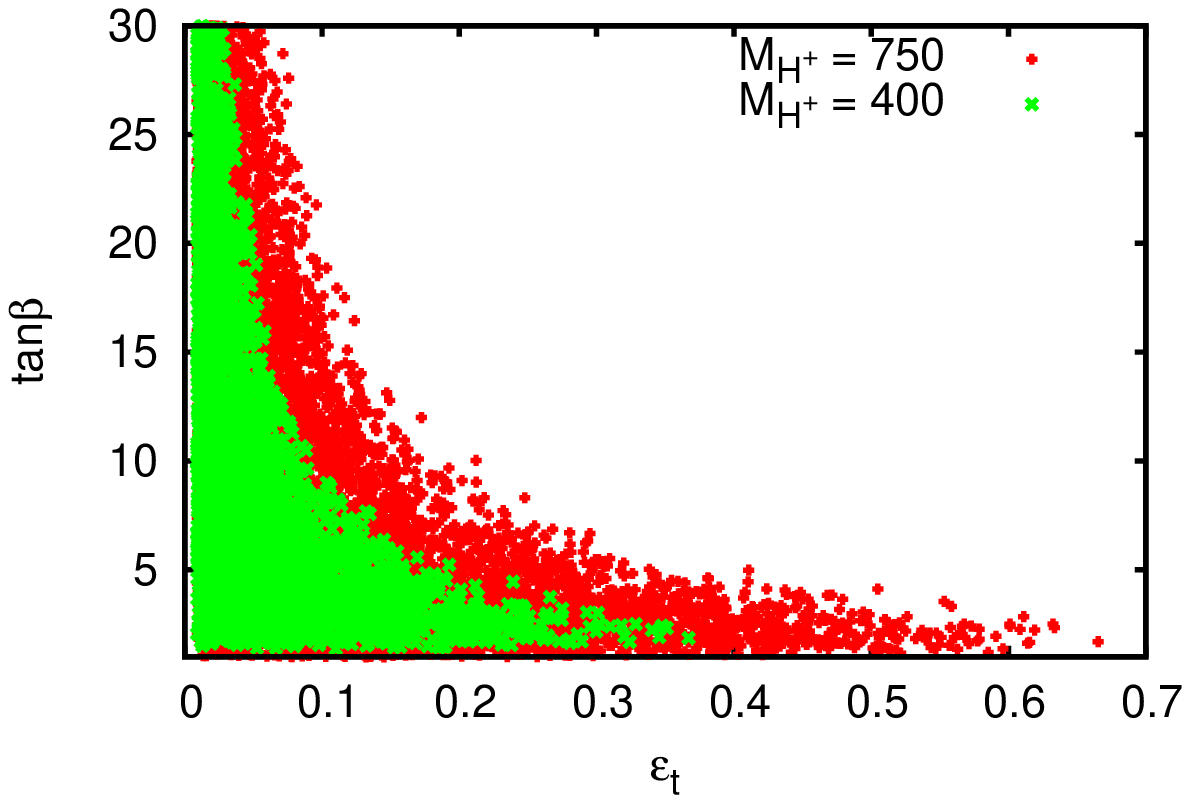,height=5.5cm,width=5.5cm,angle=0}
\caption{\emph{The Cabibbo size mixing case, $V_{t' b} = 0.2$ ($|\lambda^{t'}_{sb}|=0.004$):
the allowed parameter space in the
$\tan\beta - \epsilon_t$ plane, following
constraints from $B \to X_s \gamma$ and $B_q$-${\bar B_q}$ mixing, in the 4G2HDMI (left),
4G2HDMII (middle) and 4G2HDMIII (right), for
$m_{t'}=500$ GeV, $m_{b'}=450$ GeV,
$\epsilon_b = m_b/m_{b'}$ and
with $m_{H^+}=400$ and $750$ GeV. Figure taken from \cite{4G2HDM}.}}
\label{figI-vtb}
\end{center}
\end{figure}
\end{widetext}

\newpage

\subsubsection{Constraints from b-Physics: results}

For the ``standard" 2HDMII with four generations we find that
the constraints from $Br(B\to X_s \gamma)$ and $\Delta M_q$ $(q =d,s)$ have a
simple pattern in the $m_{H^+} - \tan\beta$ plane. In particular,
with $m_{t^\prime} \sim 500$ GeV we find that $M_{H^+} \gsim 600$ GeV for $\tan\beta =1$, while
$M_{H^+} \gsim 500$ GeV for $\tan\beta =5$.

For the 4G2HDM's of types I, II and III, the combined constraints on their parameter space from both
$Br(B\to X_s \gamma)$ and $\Delta M_q$ $(q =d,s)$, are summarized below.
In Figs.~\ref{figI-vtb01} and \ref{figI-vtb02} we show a sample of the results
obtained in \cite{4G2HDM}, where the
allowed ranges are shown in the $m_{H^+} - \tan\beta$ and the
$\tan\beta - \epsilon_t$ planes, respectively.
In these plots we use $|V_{t' b}| = 0.001$ - corresponding to the ``3+1" scenario with a
negligible 4th-3rd generation mixing, i.e., with
$|\lambda^{t'}_{sb}| = 10^{-5}$ correspondingly.
We see e.g., that in the 4G2HDMI, the ``3+1" scenario typically imposes
$\tan\beta \sim 1$ with $\epsilon_t$ typically larger than about 0.4 when $m_{H^+} \lsim 500$ GeV.
In the 4G2HDMII and the 4G2HDMIII one observes a similar correlation
between $\tan\beta$ and $m_{H^+}$, however, larger $\tan\beta$
are allowed for $\epsilon_t \lsim m_t/m_{t'}$
and a charged Higgs mass is typically heavier than 400 GeV.

For the case of a Cabbibo size
mixing between the 4th and 3rd generation quarks,
we set $|V_{t' b}| = |V_{t b'}| =0.2$ and show in
Figs.~\ref{figII-vtb} and \ref{figI-vtb} the allowed parameter space in the
$m_{H^+} - \tan\beta$ and $\tan\beta - \epsilon_t$ planes, in
the 4G2HDM's of types I, II and III, with $m_{t'}=500$ GeV, $m_{b'}=450$ GeV and
$\epsilon_b = m_b/m_{b'}$.
In the 4G2HDMII and the 4G2HDMIII
we see a similar behavior as in the no-mixing case
(i.e., as in the case $V_{t' b} \to 0$), while in the 4G2HDMI we see that ``turning on"
$V_{t' b}$ allows for a slightly larger $\tan\beta$, i.e., up to $\tan\beta \sim 5$ for $\epsilon_t \gsim 0.9$.
Also, similar to the no mixing case, larger values of $\tan\beta$
are allowed in the 4G2HDMII and 4G2HDMIII. Furthermore,
$m_{H^+} \sim 300$ GeV and $\tan\beta \sim 1$ are allowed in the 4G2HDMI.

\subsection{Combined constraints and points of interest}

In Table~\ref{tab2} we give a sample list of interesting points (models) in parameter space of the
4G2HDMI
that ``survive" all constraints from EWPD and flavor physics in the 4G2HDMI, for
$m_h=125$ GeV, $\tan\beta =1$ and $\epsilon_t=m_t/m_{t^\prime}$.
The list includes (see also caption to Table~\ref{tab2}) models with a 4th generation
mass splitting (between the up and down partners of both the 4th family quarks and leptons)
larger than 150 GeV, models where both the 4th generation quarks and leptons are nearly degenerate,
models with a light to intermediate neutral Higgs spectrum, i.e., $m_h=125$ GeV and
$m_A ~ {\rm or} ~ m_H$ in the range
$150 ~{\rm GeV} - 300~{\rm GeV}$,
models with a large inverted mass hierarchy in the quark doublet, i.e.,
$m_{b^\prime} - m_{t^\prime} > 150$ GeV, models with a light charged Higgs with a mass smaller than 400 GeV
and models with a Cabibbo-size as well as an ${\cal O}(0.01)$ size $t^\prime - b$/$t - b^\prime$
mixing angle.

\begin{table}[htb]
\begin{center}
\begin{tabular}{c|c|c|c|c|c|c|c|c|c|c}
\hline \hline
~ & \multicolumn{10}{c}{4G2HDMI: $m_h=125$ GeV, $\tan\beta=1$, $\epsilon_t = m_t/m_{t^\prime}$} \\
\hline \hline
  Point \# & $m_{t^\prime}$ & $m_{b^\prime}$  & $m_{\nu^\prime}$  & $m_{\tau^\prime}$ &
  $m_A$ & $m_H$ & $m_{H^+}$ & $\sin\theta_{34}$ & $\alpha$ & $ \left| \lambda^{t^\prime}_{sb} \right| $ \\
\hline
1  & 570 & 403 & 118 & 184 &  319 & 993 & 806 & 0.02 & $0.46 \pi$ & $ < 0.002$ \\
2  & 596 & 435 & 124 & 277 &  840 & 172 & 595 & 0.09 & $0.32 \pi$ & $ < 0.0005$ \\
3  & 425 & 591 & 1151 & 1085 &  817 & 203 & 646 & 0.08 & $0.46 \pi$ & $ < 0.001$ \\
4  & 441 & 595 & 385 & 556 &    180 & 998 & 661 & 0.21 & $0.69 \pi$ & $ < 0.001$ \\
5  & 429 & 580 & 587 & 759 &    978 & 304 & 454 & 0.13 & $0.95 \pi$ & $ < 0.0005$\\
6  & 555 & 564 & 1185 & 1180 &  501 & 674 & 661 & 0.06 & $0.62 \pi$ & $ < 0.0007$ \\
7  & 409 & 401 & 424 & 429 &  509 & 837 & 472 & 0.1 & $0.68 \pi$ & $ < 0.0006$ \\
8  & 500 & 450 & 1079 & 1005 &  745 & 439 & 750 & 0.05 & $\pi/2$ & $ < 0.0006$ \\
9  & 500 & 450 & 160 & 176 &  733 & 414 & 750 & 0.05 & $\pi/2$ & $ < 0.0006$ \\
10  & 500 & 450 & 786 & 652 &  833 & 308 & 750 & 0.2 & $\pi/2$ & $ < 0.0006$ \\
11  & 500 & 450 & 211 & 268 &  798 & 289 & 750 & 0.2 & $\pi/2$ & $ < 0.0006$ \\
12  & 450 & 500 & 711 & 618 &  500 & 215 & 300 & 0.2 & $\pi/2$ & $ < 0.004$ \\
13\footnote{Points 12 and 13 require $\epsilon_b \lsim m_b/m_{b^\prime}$ in order to have
${\rm BR}(b^\prime \to t H^+) \sim {\cal O}(1)$ (see Fig.~\ref{fig3BR}).}
& 450 & 500 & 108 & 253 &  872 & 295 & 300 & 0.2 & $\pi/2$ & $ < 0.004$ \\
\hline \hline
\end{tabular}
\caption{List of points (models) in parameter space for the 4G2HDM's of types I with $m_h=125$ GeV,
$\tan\beta=1$ and $\epsilon_t=m_t/m_{t^\prime}$,
allowed at 95\%CL by EWPD and B-physics flavor data. Points 1-2 have
$m_{t^\prime} - m_{b^\prime} > 150 $ GeV,
while points 3-5 have a large inverted splitting
$m_{b^\prime} - m_{t^\prime} > 150 $ GeV.
Points 6 and 7  have nearly degenerate 4th generation quark
and lepton doublets. Points 8-11 give
${\rm BR}(t^\prime \to t h) \sim {\cal O}(1)$ (see Fig.~\ref{fig1BR} in section \ref{pheno}),
while points 12 and 13 give ${\rm BR}(b^\prime \to t H^+) \sim {\cal O}(1)$ (see Fig.~\ref{fig3BR} in section \ref{pheno}).
Point 4 has a light 180 GeV pseudoscalar Higgs (A) and points 12 and 13 have
a light 300 GeV charged Higgs. Points 1,8 and 9 have a small $t^\prime - b/t-b^\prime$
mixing angle ($\theta_{34} \leq 0.05$), while points 4 and 10-13 have a Cabibbo-size
$t^\prime - b/t-b^\prime$ mixing angle ($\theta_{34} \sim 0.2$).}
\label{tab2}
\end{center}
\end{table}

\newpage

\section{Other useful effects in flavor physics \label{sec4}}

We discuss below some important low energy observables which are potentially
sensitive to the 4th generation dynamics within the multi-Higgs framework,
and have shown some degree of discrepancy between their measured values and the
SM predictions.

\subsection{Muon $(g-2)$ and lepton flavor violation}

The muon anomalous magnetic moment ($\mu$AMM), $a_{\mu} = (g_\mu - 2)/2$, is well known to
plays an important role in the search for NP. In the SM, the total contributions
to the $\mu$AMM, $a_{\mu}^{SM}$, can be
divided into three parts: the QED, the electroweak (EW) and the hadronic contributions.
While the QED \cite{qed} and EW \cite{ew} contributions are well understood, the
main theoretical uncertainty lies with the hadronic part which is difficult to control \cite{qcd}.

Since the first precision measurement of $a_{\mu}$, there has been a discrepancy between
its experimental value and the SM prediction. This discrepancy has been slowly growing due
to recent impressive theoretical and experimental progress.
Comparing theory and experiment, the deviation
amounts to \cite{gminus2}:
\be
a_{\mu}^{exp} - a_{\mu}^{SM} = (255 \pm 80) \cdot 10^{-11}
\label{amuexp}
\ee
which corresponds to a $\sim 3\sigma$ effect. In order to confirm this
result the uncertainties have to be further reduced.

It is interesting to interpret the difference as a contribution from loop exchanges of new particles. A number of groups have studied the contribution to $a_{\mu}$ in
various extensions of the SM to constrain their parameters space (for reviews
see Ref.~\cite{nyffeler}). In most extensions of the SM, new charged or neutral states
can contribute to the $\mu$AMM at the one-loop (lowest) level.
In Ref.~\cite{4G2HDM2}, we have shown that the $\sim 3\sigma$ excess in $a_{\mu}$ (with respect
to the SM prediction) can be accounted for by one-loop exchanges of the
heavy 4th generation neutrino ($\nu^\prime$) in the 4G2HDMI setup
when applied to the leptonic sector (i.e.,
where the ``heavy" Higgs doublet couples only to the 4th generation lepton doublet and the
``light" Higgs doublet couples to leptons of the lighter 1st-3rd generations, see \cite{4G2HDM2}).

\bigskip

\begin{figure}[htbp]
\hspace*{-2mm}  $\gamma$ \hspace{50.5mm} $\gamma$ \\ [10mm]
\hspace*{3mm}
            $\tau'$ \hspace{15mm}   $\tau'$ \hspace{22mm}
              $\phi^{\pm}$ \hspace{15mm}     $\phi^{\pm}$ \\[-30mm]
\vspace{6mm}
\begin{center}
\includegraphics[width=90mm,angle=0]{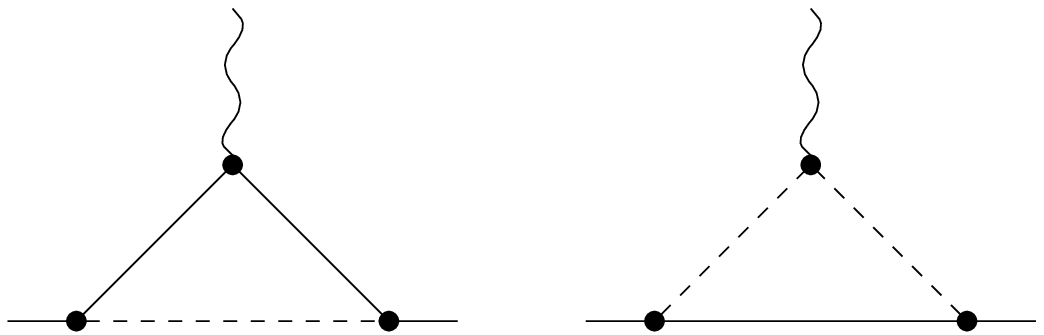}
\end{center}
\vspace{-5mm}
\hspace{0mm}
$l_i$ \hspace{10mm}   $\phi^0$   \hspace{10mm} $l_j$ \hspace{14mm}
$l_i$ \hspace{1cm} $\nu^{\prime}$ \hspace{12mm} $l_j$ \hspace{6.5mm}
\begin{center}
\caption{\emph{One-loop diagrams for $l_i \to l_j \gamma$ with charged and neutral scalar exchanges.}}
\label{fig:one-loop}
\end{center}
\vspace{-1cm}
\end{figure}
\bigskip

The effective vertex of a photon with a charged fermion can in general be written as
\begin{equation}
\bar{u}(p') e \Gamma_\mu u(p) = \bar{u}(p')  e \left[ \gamma_\mu F_1(q^2)
+ \frac{i  \sigma_{\mu\nu} q^\nu}{2m_f} F_2(q^2) \right] u(p) \, ,
\end{equation}
where, to lowest order, $F_1(0) =1$ and $F_2(0) =0$. While $F_1(0)$ remains unity at
all orders due to charge conservation, quantum
corrections yield $F_2(0) \ne 0$. Thus, since
$g_\mu \equiv 2 \cdot \left(F_1(0) +  F_2(0)\right)$, it follows that $a_\mu \equiv (g_\mu - 2)/2 = F_2(0)$.

In the 4G2HDMI \cite{4G2HDM,4G2HDM2} the one-loop contribution to the muon anomaly can be subdivided as
\begin{equation}
a_{\mu} = [a_{\mu}]^{SM4}_W + [a_{\mu}]^{4G2HDMI}_{\cal H},
\end{equation}
where $[a_{\mu}]^{4G2HDMI}_{\cal H}$ contains the charged and neutral Higgs contributions coming from
the one-loop diagrams in Fig.~\ref{fig:one-loop}, where the diagrams with
$\tau^\prime$ and $\nu^\prime$ in the loop dominate.
The SM4-like contribution, $[a_{\mu}]^{SM4}_W$, comes from the one-loop diagram with $W^{\pm} - \nu^\prime$
in the loop and is given by \cite{jlev}:
\be
\frac{[a_{\mu}]^{SM4}_W}{|U_{24}|^2} = \frac{G_F m^2_{\mu}}{4\sqrt{2}\pi^2} A(x_{\nu'})~,
\label{ag2sm4}
\ee
where $U_{24}$ is the 24 element of the CKM-like PMNS leptonic matrix,
$x_i = m_i^2/m_W^2$. For values of $m_{\nu'}$ in the range $100 ~{\rm GeV} \lsim m_{\nu^\prime} \lsim 1000~{\rm GeV}$,
one finds $1.5\times 10^{-9} \lsim [a_{\mu}]^{SM4}_W/|U_{24}|^2 \lsim 3.0\times 10^{-9}$,
so that for $|U_{24}|^2 << 1$ (as expected) the simple SM4 cannot accommodate the observed discrepancy in $a_\mu$.
The detail expression for $[a_{\mu}]^{4G2HDMI}_{\cal H}$ has been given in~\cite{4G2HDM2}.
It is interesting to note that the dominant contribution to $[a_{\mu}]^{4G2HDMI}_{\cal H}$, or
for that matter to $a_{\mu}$, comes from the charged Higgs loops and the contribution from
diagrams with the neutral Higgs exchanges are subleading \cite{4G2HDM2}.
In addition,
$a_\mu$ was found to be sensitive only to the product $\delta_{\Sigma_2} \cdot \delta_{U_2}$, where
\begin{eqnarray}
\delta_{U_i} \equiv \frac{U_{i4}^*}{U_{44}^*} ~,~
\delta_{\Sigma_i} \equiv \frac{\Sigma_{4i}^{e *}}{\Sigma_{44}^\nu}
\label{delta12}~,
\end{eqnarray}
 and $\Sigma^e(\Sigma^\nu)$ are the new  mixing matrices (i.e., in the 4G2HDMI) in the charged(neutral)-leptonic sectors. That is, similar
to the quark sector (see Eq.~\ref{sigma}), these matrices are obtained
after diagonalizing the lepton mass matrices
\begin{widetext}
\begin{eqnarray}
\Sigma_{ij}^e = L_{R,4i}^\star L_{R,4j} ~,~
\Sigma_{ij}^\nu = N_{R,4i}^\star N_{R,4j} ~, \label{sigma2}
\end{eqnarray}
\end{widetext}
where $L_R,N_R$ are the rotation (unitary) matrices of the right-handed
charged and neutral leptons, respectively.$^{\footnotemark[3]}$\footnotetext[3]{Note that
since $N_{R,4i}$ and $L_{R,4j}$ parameterize mixings among the
4th generation and the 1st-3rd generations leptons, one expects
$\Sigma^\ell_{ij} \ll \Sigma^\ell_{4k}$ for $i,j,k=1,2,3$, see Eq.~\ref{sigma2}.}

\begin{widetext}
\begin{figure}[htb]
\begin{center}
\epsfig{file=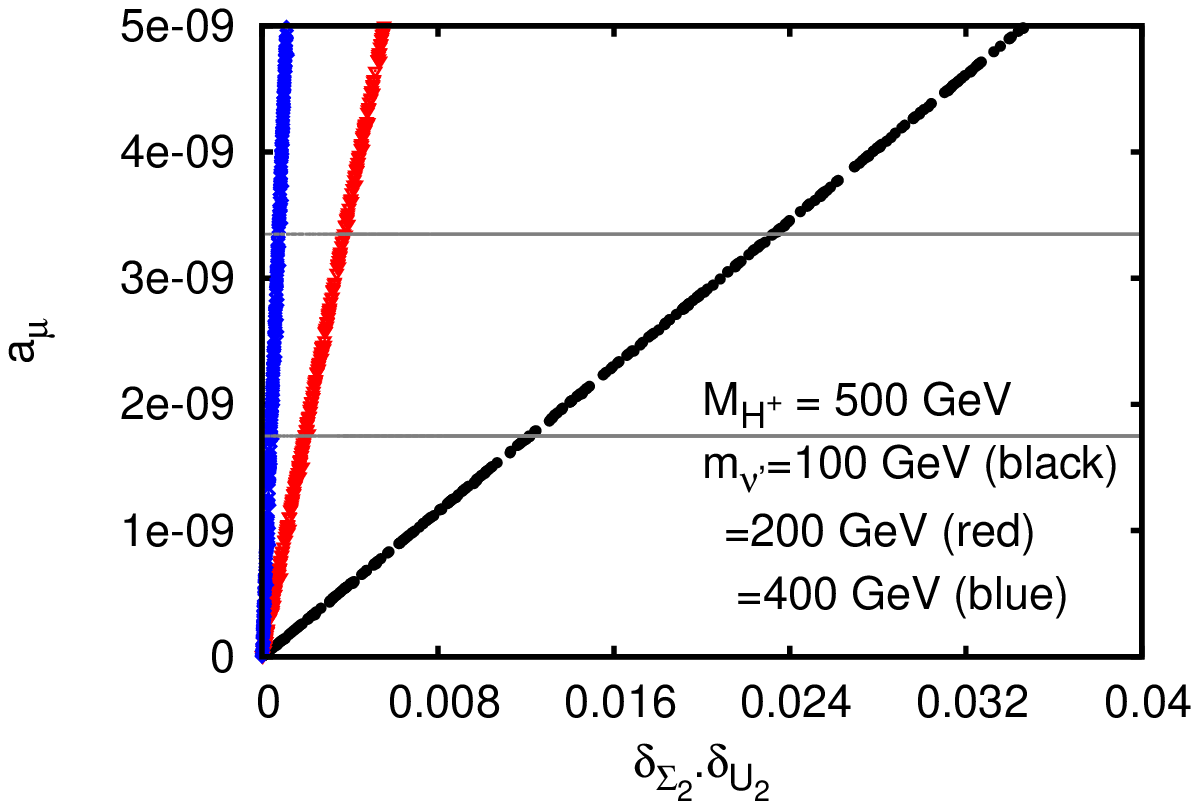,height=7cm,width=7cm,angle=0}
\epsfig{file=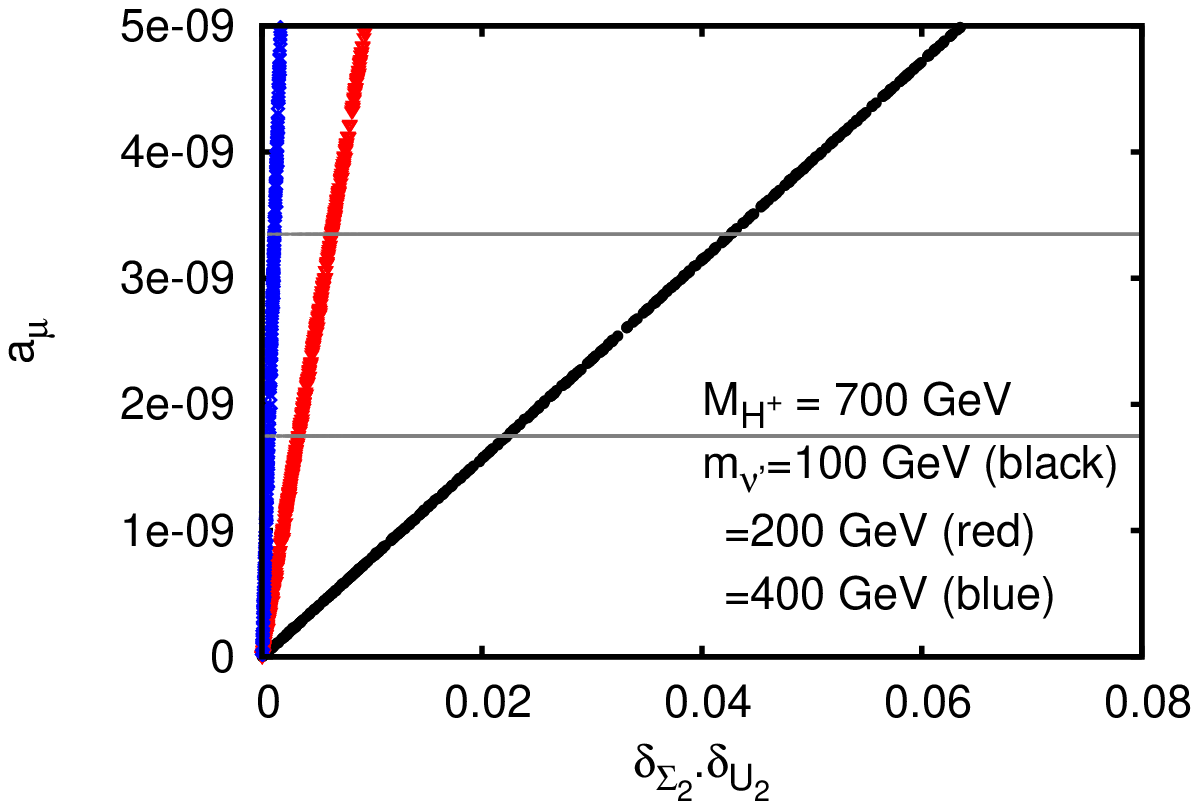,height=7cm,width=7cm,angle=0}
\caption{\emph{The muon $g-2$ as a function of the product
$\delta_{\Sigma_2} \cdot \delta_{U_2}$, for
$m_{\nu^\prime} = 100,~200,~400$ GeV, $m_{\tau^\prime} = m_{\nu^\prime}$ and with $m_{H^+}=500$ GeV (left)
and $m_{H^+}=700$ GeV (right). The horizontal lines are the measured 1-$\sigma$ bounds on $a_\mu$ (see Eq.~\ref{amuexp}).
Figure taken from \cite{4G2HDM2}.}}
\label{g-2}
\end{center}
\end{figure}
\end{widetext}

In Fig.~\ref{g-2} we plot $a_\mu$ as a function of the product
$\delta_{\Sigma_2} \cdot \delta_{U_2}$ (assuming its real) for several values of
$m_{\nu^\prime}$ and $m_{H^+}$ and fixing
$m_{\tau^\prime} = m_{\nu^\prime}$. Depending on the mass $m_{\nu'}$, we find that
$\delta_{U_2} \cdot \delta_{\Sigma_2} \sim 10^{-3} - 10^{-2}$ is
typically required to accommodate the measured value of $a_\mu$.

The constraint on the 4G2HDMI parameters and in particular on the quantities $\delta_{\Sigma_2}$ and $\delta_{U_2}$ which
control the $\mu$AMM were studied in \cite{4G2HDM2}, by analyzing the lepton flavor violating (LFV) decays $\tau \to \mu\gamma$ and $\mu \to e \gamma$.
These decays are absent in the SM, and are useful for constraining NP models
that can potentially contribute to the muon anomaly.

The current experimental 90\%CL upper bounds on these LFV decays are \cite{PDG,meg}
\be
Br(\tau \to \mu \gamma) < 4.4 \times 10^{-8} ~~ , ~~ Br(\mu \to e \gamma) < 2.4\times 10^{-12} \label{lfvbounds}~.
\ee

The amplitude for the transition $\ell_i \to \ell_j \gamma$ can be defined as
\begin{equation}
{\cal M}(\ell_i \to \ell_j \gamma) =  \bar{u}_{\ell_j}(p')
\left[i\sigma_{\mu\nu} q^\nu \left( A  +  B \gamma_5 \right) \right] u_{\ell_i}(p)  \epsilon^{\mu\ast} \, ,
\label{liljg}
\end{equation}
where $\epsilon^{\mu\ast}$ is the photon polarization. The decay width is then given by
\be
\Gamma(\ell_i \to \ell_j \gamma) = \frac{m^3_{\ell_i}}{8 \pi}
\left( 1 - \frac{m_{\ell_j}^2}{m_{\ell_i}^2} \right) \left[
\left( 1 + \frac{m_{\ell_j}^2}{m_{\ell_i}^2} \right) \left( |A|^2 + |B|^2 \right)
+4 \frac{m_{\ell_j}}{m_{\ell_i}} \left( |A|^2 - |B|^2 \right) \right] ~.
\ee

Here also, the new 4G2HDMI contribution to the amplitude, ${\cal M}(\ell_i \to \ell_j \gamma)^{4G2HDMI}$,
can be divided as
\be
{\cal M}(\ell_i \to \ell_j \gamma)^{4G2HDMI} \equiv {\cal M}^{SM4}_W(\ell_i \to \ell_j \gamma) +
{\cal M}^{4G2HDMI}_{H^+}(\ell_i \to \ell_j \gamma) + {\cal M}^{4G2HDMI}_{{\cal H}^0}(\ell_i \to \ell_j \gamma) ~,
\ee
where ${\cal M}^{SM4}_W(\ell_i \to \ell_j \gamma)$ is the SM4-like W-exchange contribution which is much smaller than
the charged and neutral Higgs amplitudes, ${\cal M}^{4G2HDMI}_{H^+}(\ell_i \to \ell_j \gamma)$
and ${\cal M}^{4G2HDMI}_{{\cal H}^0}(\ell_i \to \ell_j \gamma)$ (calculated from the diagrams in Fig.~\ref{fig:one-loop}).
As in the $\mu$AMM case, the dominant contribution to LFV decays
was found the be from the charged Higgs exchange diagrams \cite{4G2HDM2}.
In addition, the decays $\mu \to e \gamma$ and $\tau \to \mu \gamma$ are sensitive
to $\delta_{U_2}$ and $\delta_{\Sigma_2}$ through the products
$(\delta_{U_2} \delta_{\Sigma_1},\delta_{U_1} \delta_{\Sigma_2})$ and
$(\delta_{U_3} \delta_{\Sigma_2},\delta_{U_2} \delta_{\Sigma_3})$, respectively, so that,
in principle, one can avoid constraints on the quantities $\delta_{U_2}$ and $\delta_{\Sigma_2}$
if $\delta_{U_1},\delta_{U_3},\delta_{\Sigma_1}$ and $\delta_{\Sigma_3}$ are sufficiently small.

In Ref.~\cite{4G2HDM2}, we have shown that it is possible to address both the
$BR(\mu \to e \gamma)$ and the muon anomaly $a_{\mu}$ within the 4G2HDMI framework,
if $\delta_{U_1} \ll \delta_{U_2}$ and $\delta_{\Sigma_1} \ll \delta_{\Sigma_2}$, which is indeed expected
if we consider the observed hierarchical pattern of the
quark's CKM matrix as a guide.
However, in order to account also for the measured upper limit on
$BR(\tau \to \mu \gamma)$ (see Eq.~\ref{lfvbounds}), one requires
$\delta_{U_3} < \delta_{U_2}$ and $\delta_{\Sigma_3} < \delta_{\Sigma_2}$. Therefore, the
typical benchmark texture for the 4th generation elements
of the matrices $U_{i4}$ $\Sigma^e_{4i}$ that can account for the observed muon anomaly and still be consistent with the
current constraints from the LFV decays $\tau \to \mu\gamma$ and $\mu \to e \gamma$ is
\begin{eqnarray}
U_{i4} \sim (\Sigma^e_{4i})^T \simeq
\left( \begin{array}{c} \epsilon^5  \\ \epsilon  \\ \epsilon^2 \\ 1
\end{array} \right)
 \label{texture}~,
\end{eqnarray}
where e.g., $\epsilon \sim 0.1$ for $m_{\nu^\prime} = 100$ GeV.

The above texture implies a hierarchical pattern which is
different than one would expect from the observed hierarchical pattern
of the quark's CKM matrix. Nonetheless, without a fundamental theory of flavor,
our insight for flavor should be data driven also in the leptonic sector.
Besides, the above texture is sensitive to the current precision in the measurement
of the muon g-2 which can change e.g., if
more accurate calculations end up showing that part of the hadronic
contributions cannot be ignored.

\subsection{Insight from $B$ physics}

\subsubsection{$B_s \to \mu^+\mu^-$}

\begin{figure}[t]
\begin{center}
\resizebox{12cm}{10cm}
{\includegraphics*[-43mm,100mm][163mm,244mm]{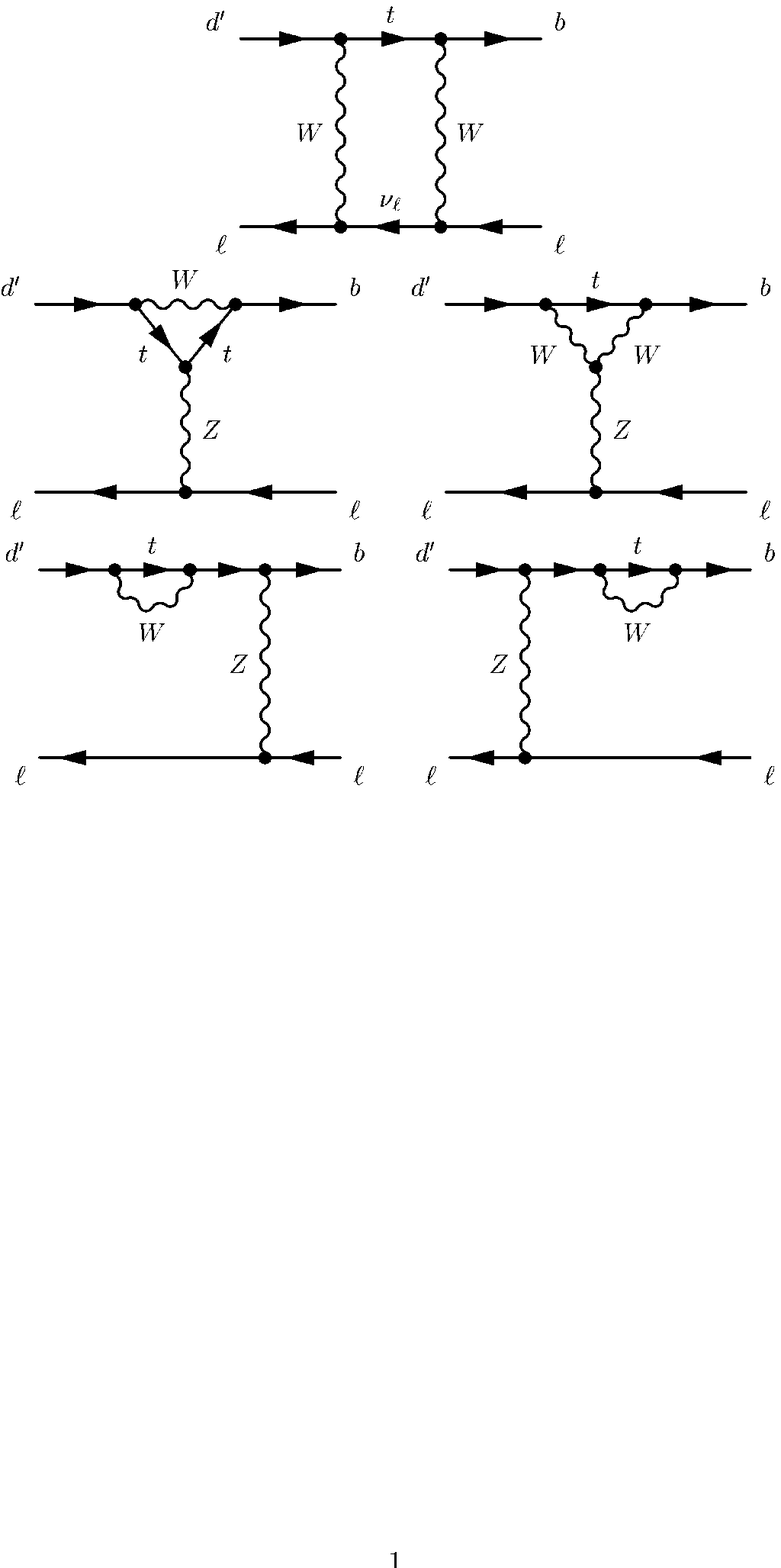}}
\end{center}
\caption{\emph{Dominant SM diagrams for the decay $B_{d'}\to \ell^+\ell^-$, $d'=d$ or $s$.}}
\label{smdiags}
\end{figure}

Among the various $B_q$ rare decays, the purely leptonic $B_{d/s} \to \mu^+\mu^-$ decays are highly sensitive to indirect effects of
NP,
since the quark level decays are based on the FCNC $b\to d,s$ transitions which are severely (loop) suppressed in the SM.
In particular, the decay $B_s\to \mu^+\mu^-$ has received special attention in the past decade, since
its branching fraction, $Br(B_s\to \mu^+\mu^-)$, can be significantly enhanced by
loop exchanges of new particles predicted by various NP scenarios. For example,
$Br(B_s\to \mu^+\mu^-)$ imposes restrictive constraints
on the SUSY parameter space (see e.g., \cite{uli_bsmumu,mahmoudi}),
where in some scenarios better limits than those obtained from direct
searches have been claimed. However, the excluded SUSY parameter space depends strongly on the
choice of $\tan\beta$ since the $B_s \to \mu^+ \mu^-$ rate typically varies as $(\tan\beta)^6$.

\begin{widetext}
\begin{figure}[htb]
\begin{center}
\epsfig{file=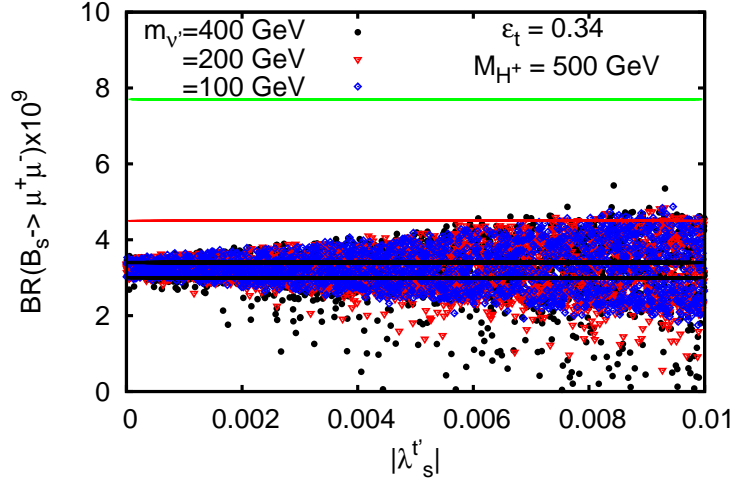,height=7cm,width=10cm,angle=0}
\caption{\emph{$BR(B_s \to \mu \mu)$ as a function of $\lambda_{bs}^{t^\prime} \equiv V_{t^\prime b} V_{t^\prime s}^\ast$,
from box diagrams with $H^+$ and $(t, \nu^\prime)$, $(t^\prime, \nu^\prime)$ exchanges in the 4G2HDMI.
The parameters
$\delta_{U_2}$ and $\delta_{\Sigma_2}$ are varied within the constraints imposed by $a_{\mu}$ (see previous section),
keeping both of them $\lsim 0.2$.
Also shown are the experimental 95\% CL upper bounds from LHCb (red horizontal line) and from CMS
(green horizontal line). The SM predicted range of values (at $1 \sigma$) is shown within the black horizontal lines.
Figure taken from \cite{4G2HDM2}.}}
\label{bsmumu}
\end{center}
\end{figure}
\end{widetext}

In the LHC era the current limit on $Br(B_s \to \mu^+ \mu^-)$ has been improved.
The two different experiments LHCb and CMS, using 1$fb^{-1}$ and 5$fb^{-1}$ data sample, respectively,
yield \cite{lhcb,cms}

\begin{align}
Br(B_s \to \mu^+ \mu^-)  &< 4.5 \times 10^{-9}\;, \hskip 30pt LHCb @ 95\% CL \nonumber \\
                         &< 7.7 \times 10^{-9}\;, \hskip 30pt CMS @ 95\% C
\end{align}
whereas the SM prediction for this decay is \cite{ajb10B}:
\be
Br(B_s \to \mu^+ \mu^-) = (3.2 \pm 0.2) \cdot 10^{-9} ~.
\ee

In fact, LHCb has the sensitivity to measure the $Br(B_s \to \mu^+ \mu^-)$ down to  $ \sim 2 \times 10^{-9}$, which is
about $5\sigma$ smaller than the SM prediction.

In general, the matrix element for the decay ${\bar B_s}\to \ell^+\ell^-$ can be written as \cite{mBs}
\be
{\cal M} = \frac{G_F \alpha}{2 \sqrt{2} \pi \sin\theta_W^2} \left[ F_{S}\,{\bar \ell} \ell + F_{P}\,{\bar \ell}\gamma_5 \ell
+  F_{A}\,P^{\mu} {\bar \ell} \gamma_{\mu}\gamma_5 \ell \right], \,
\label{ampbsll}
\ee
where $P^{\mu}$ is the four momentum of the initial $B_s$ meson and $F_{i}$'s are functions of Lorentz invariant quantities.
Squaring the matrix and summing over the lepton spins, we obtain the branching fraction
\be
Br({\bar B_s}\to \ell^+ \ell^-) = \frac{G_F^2 \alpha^2 M_{B_s} \tau_{B_s}}{64 \pi^3} \sqrt{1- \frac{4 m_{\ell}^2}{M_{B_s}^2}}
\left[\left(1- \frac{4 m_{\ell}^2}{M_{B_s}^2}\right) |F_S|^2 + |F_P + 2 m_{\ell} F_A|^2\right].
\ee
In the SM, the dominant effect in ${\bar B_s}\to \ell^+ \ell^-$ arise from the diagrams shown in
Fig.~\ref{smdiags}, which contribute only to $F_A$ in Eq.~\ref{ampbsll}.

As in other NP models, in the 4G2HDMI
there will be contributions to $F_S$, $F_P$ and $F_A$ coming from the
charged-Higgs exchange penguin and box diagrams (replacing $W^+ \to H^+$ in Fig.~\ref{smdiags}).
In Ref.~\cite{4G2HDM}, constraint on the 4G2HDMI parameter spaces were estimated, using the recent data on
$Br(B_s \to \mu^+ \mu^-)$. This was done in the context of the muon $(g-2)$, in the sense that only those interactions
(in the leptonic
vertex) which are associated with $a_{\mu}$ have been considered. In particular,
considering only the $\ell^{\pm} \nu' H^{\pm}$ vertex, the only diagrams that contribute to
${\bar B_s}\to \ell^+ \ell^-$ are the Higgs-exchange box diagrams in Fig.~\ref{smdiags}, where one or
two $W$-bosons are replaced by $H^+$ and $(t, \nu_{\ell})$ are being replaced by both $(t, \nu^\prime)$ and $(t^\prime, \nu^\prime)$.
It was then found that the contribution from the new box diagrams in the 4G2HDMI that
involve the heavy 4th generation neutrino is consistent with the current experimental bound
on $BR(B_s \to \mu \mu)$ for values of $\delta_{U_2}$ and $\delta_{\Sigma_2}$ that
reproduce the observed muon $g-2$, see Fig.~\ref{bsmumu}.

It is also interesting to note that the Br$(B_s \to \mu^+\mu^-)$,
in both the SM4 and the 4G2HDMI, can differ from the SM value
by at-most a factor of \cal{O}(3) in either direction (for a detail discussions see \cite{4G2HDM2}).

\subsubsection{$B^+ \to \tau^+\nu$ and $B \to D^{(*)} \tau \nu$}

Other purely leptonic and semileptonic decays of the $B$ meson, such as
$B\to \tau$ decays, can also provide useful tests of the SM and its extensions.
Of particular interest are the purely leptonic $B\to \tau\nu$ and the semileptonic $B^+ \to D^{(*)} \tau \nu$
decays. The SM contribution to the branching ratios of these decays arise at the tree-level from
the charged weak interactions. An important NP contribution to these decays is
the tree-level exchange of a charged Higgs in multi-Higgs models, so that these decays offer interesting
probes of the Higgs sector and, particularly, of its Yukawa interactions.

The SM expression for the decay rate of $B \to \tau\nu$ is given by
\be
{Br(B\to \tau\nu)}_{SM} = \frac{G_F^2 m_{\tau}^2 m_B }{8 \pi}(1-\f{m_{\tau^2}}{m_B^2})^2f_B^2 |V_{ub}|^2 \tau_{B}.
\label{brfsm}
\ee
where $f_B$ is the decay constant and $\tau_B$ is the $B^+$ life time.
The SM prediction for $Br(B^+\to \tau^+\nu)$ is, therefore, sensitive to the decay constant $f_B$ and to the CKM
element $|V_{ub}|$ and is thus limited by the uncertainty in the determination of these quantities.
Using the available constraints on $f_B$ and the inclusive determination of $V_{ub}$:
$f_B = 200 \pm 20$ MeV and
$V_{ub} = (39.9 \pm 1.5 \pm 4.0)\cdot 10^{-4}$ \cite{utfit}, the SM prediction for the decay rate is
\be
{Br(B\to \tau\nu)}_{SM} = (0.86 \pm 0.12 )\cdot 10^{-4}~.
\ee

Furthermore, the SM prediction on Br$(B\to \tau \nu)$,
obtained directly from a fit to various other observables (i.e., without using $V_{ub}$ and the lattice results
for $f_B$) is \cite{utfit}
\be
{Br(B\to \tau\nu)}_{SM} = (0.73 \pm 0.12 )\cdot 10^{-4}~.
\ee

Both results show some degree of discrepancy with the
current world average on $BR(B\to \tau\nu)$ which is \cite{hfag10},
\be
Br(B^+\to \tau^+ \nu_{\tau}) = (1.67 \pm 0.3 )\cdot 10^{-4}.
\label{btaunuexp}
\ee

We want to indicate here how the 4G2HDM can address this if the discrepancy is confirmed.

From the theoretical point of view, several models of NP
predict large deviations from the SM for processes involving third
generation fermions. For instance, in a ``standard" 2HDM where
the two Higgs doublets are coupled separately to up- and down-
type quarks (i.e., the 2HDMII setup described in section \ref{sec2}),
the $B\to \tau\nu$ amplitude receives an additional
tree-level contribution from the heavy charged-Higgs exchange, leading to
\be
\frac{Br(B\to \tau\nu)^{2HDMII}}{Br(B\to \tau\nu)^{SM}} = \left[1 - \frac{m_B^2 \tan^2\beta}{M^2_H}\right] ~,
\label{btaununp}
\ee
so that for large $\tan\beta$, the r.h.s of Eq.~\ref{btaununp} can be significantly different from ``1". However, in this
particular case (of the 2HDMII),
the charged-Higgs contribution reduces the SM value for the branching ratio, thus further worsening the
situation with respect to the experimentally measured value.

In the 4G2HDMI, the effective tree-level interactions that will contribute to $B\to\tau\nu$
can be written as
\bea
{\cal H}_{\it eff} &=& \f{G_F V_{ub}}{\sqrt{2}} \left[{\bar u}\gamma_{\mu}(1-\gamma_5)b \, {\bar \tau}\gamma^{\mu}(1-\gamma_5)\nu
- \f{m_{\tau} m_b A_{bu} }{M_H^2} \left\{ A^{\ell}_u {\bar u}(1+\gamma_5)b\,{\bar \tau}(1-\gamma_5)\nu \right.\right.  \nonumber \\
&{}& \left. \left. + A^{\ell}_d {\bar u}(1+\gamma_5)b\, {\bar \tau}(1+\gamma_5)\nu \right\} \right],
\label{4G2HDM_eff}
\eea
where the second term represents the tree-level charged-Higgs  exchange and the first term results from the diagram with $W$-boson exchange. Also, $A_{bu}$, $A^{\ell}_u$ and $A^{\ell}_d$ are factors coming from the $b\to u H$ and $\tau \to \nu_{\tau} H$ vertices, respectively, given by
\bea
A_{bu} &=& \tan\beta - (\tan\beta+\cot\beta)(\Sigma_{bb} + \f{m_{b'}}{m_b}\f{V_{ub'}}{V_{ub}} \Sigma_{b'b}), \nonumber \\
A^{\ell}_{u} &=& -\tan\beta + (\tan\beta+\cot\beta) \left\{\Sigma^{\ell}_{33} U_{33} + \frac{m_{\tau'}}{m_{\tau}} \Sigma^{\ell}_{43} U_{43}\right\}, \nonumber \\
A^{\ell}_{d} &=&  - \frac{m_{\nu_{\tau'}}}{m_{\tau}}(\tan\beta+\cot\beta) \Sigma^{\nu}_{43} U_{34}.
\label{abu}
\eea
 A simple calculation, using Eqs.~\ref{4G2HDM_eff} and \ref{abu}, yields
\bea
Br(B \to \tau \nu) &=& {Br(B\to \tau\nu)}_{SM} \left[{\left|1 - \f{m_B^2}{M_H^2} A_{bu} A^{\ell}_u\right|^2 + \left|\f{m_B^2}{M_H^2}
A_{bu} A^{\ell}_d\right|^2}\right] ~.
\label{brft}
\eea

Thus, taking for example $\Sigma_{ij} \approx m_j/m_i$, only a moderate enhancement to
$BR(B\to \tau\nu)$ is possible at large $\tan\beta$. If, on the other hand, $\Sigma_{ij} \gg m_j/m_i$, then
the BR$(B\to \tau\nu)$ can be significantly enhanced compared to the SM prediction.
Of course, the experimental deviations at the moment
are only a few sigmas, but, if they get confirmed, then
we have indicated here how we may be able to address them.

Semileptonic $B$ decays such as $B \to D^{(*)} \tau \nu$  are more complicated to handle than the pure leptonic ones, since
the theoretical predictions for these decays to exclusive final states require knowledge
of the form factors involved.  There are, however, several other observables (besides the branching fraction), such as
the decay distributions and the $\tau$ polarization, which can be useful in this cases for probing NP.

As in the case of
$B \to \tau\nu$, the semileptonic decay
$B \to D^{(*)} \tau \nu$ is also known to be the a sensitive mode to the tree-level charged-Higgs exchange.
Furthermore, the precise measurement of $B \to D^{(*)} \ell \nu$ at the B-factories and the theoretical developments
of heavy-quark effective theory (HQET) has improved our understanding of exclusive semileptonic decays \cite{PDG,bdtaunu_exp}.

In particular, the ratios  $R(D^{(*)}) \equiv {\rm BR}(B \to D^{(*)} \tau \nu)/{\rm BR}(B \to D^{(*)} \ell \nu)$ reduces considerably
the main theoretical uncertainties, hence, turns out to be a more useful observable \cite{soni_kiers}.
The updated SM predictions of these rates, averaged over electron and muons, are given by
\cite{bdtaunu_exp2},
\be
R(D)_{SM} = 0.297 \pm 0.017, \hskip 3cm R(D^*)_{SM} = 0.252 \pm 0.003,
\ee
so that at this level of precision the experimental uncertainties are expected to dominate.

The most recently measured values of these observables are given by \cite{bdtaunu_exp2},
\be
R(D)_{exp} = 0.440 \pm 0.058 \pm 0.042, \hskip 3cm R(D^*)_{exp} = 0.332 \pm 0.024 \pm 0.018
\label{eq51} ~.
\ee

The measured values, therefore, exceed the SM predictions for $R(D)_{SM}$ and $R(D^*)_{SM}$ by 2.0$\sigma$ and 2.7$\sigma$,
respectively, so it is argued that the possibility of both the measured values agreeing with the SM is excluded
at the 3.4$\sigma$ level. In addition, the combined analysis of $R(D)$ and $R(D^*)$ rules out
the 2HDMII charged Higgs boson with 99.8\% confidence level for any value of $\tan\beta/M_H$
when combined with Br$(B\to X_s\gamma)$, see \cite{bdtaunu_exp2}. Once again, it is not clear to us how serious
to take the indications of the deviations in Eq.~\ref{eq51}.
Nonetheless, we briefly indicate here
how this discrepancy (if experimentally confirmed) can be addressed in the 4G2HDMI, for which
the effective tree-level interactions that contribute to $B\to D^{(*)} \tau\nu$
are given in Eq.~\ref{4G2HDM_eff} with the $u$-quark replaced by the $c$ quark. Thus, similar to the case of
$B\to \tau\nu$, we expect a moderate enhancement to both $R(D)$ and $R(D^*)$ in
the 4G2HDMI if $\Sigma_{ij} \approx m_j/m_i$, and a larger effect for larger values of $\Sigma_{ij}$.

\section{New aspects of the Phenomenology of the 4G2HDMI \label{pheno}}

In the 4G2HDMI (i.e., the 4G2HDM with $\beta_d = \beta_u=0$, see Eq.~\ref{eq:LY4G}) one obtains
(see Eq.~\ref{sigsimple}):
\begin{eqnarray}
\Sigma^d \simeq \left(\begin{array}{cccc}
0 & 0 & 0 & 0 \\
0 & 0 & 0 & 0 \\
0 & 0 &  |\epsilon_b|^2 & \epsilon_b^\star \\
0 & 0 &  \epsilon_b  & \left( 1- \frac{|\epsilon_b|^2}{2} \right)
\end{array}\right)~,~
\Sigma^u \simeq \left(\begin{array}{cccc}
0 & 0 & 0 & 0 \\
0 & 0 & 0 & 0 \\
0 & 0 &  |\epsilon_t|^2 & \epsilon_t^\star \\
0 & 0 &  \epsilon_t  & \left( 1- \frac{|\epsilon_t|^2}{2} \right)
\end{array}\right)
 \label{sigmaI}~,
\end{eqnarray}
which leads to new interesting patterns (in flavor space) in the both the neutral and charged Higgs sectors.
For example, the
${\cal H}^0q_iq_j$ Yukawa interactions of Eqs.~\ref{Sff1}-\ref{Sff2} (${\cal H}^0=h,H,A$), give rise to
potentially enhanced tree-level $t^\prime \to t$ and $b^\prime \to b$ FC transitions, and absence of ``dangerous" tree-level
FCNC transitions between the 4th and the 1st and 2nd generations quarks as well as among the 1st-2nd and 3rd generation quarks.
In particular, the
FC ${\cal H^0}t^\prime t$
interactions in this case are (taking $\alpha \to \pi/2$):
\begin{figure}[htb]
\begin{center}
\epsfig{file=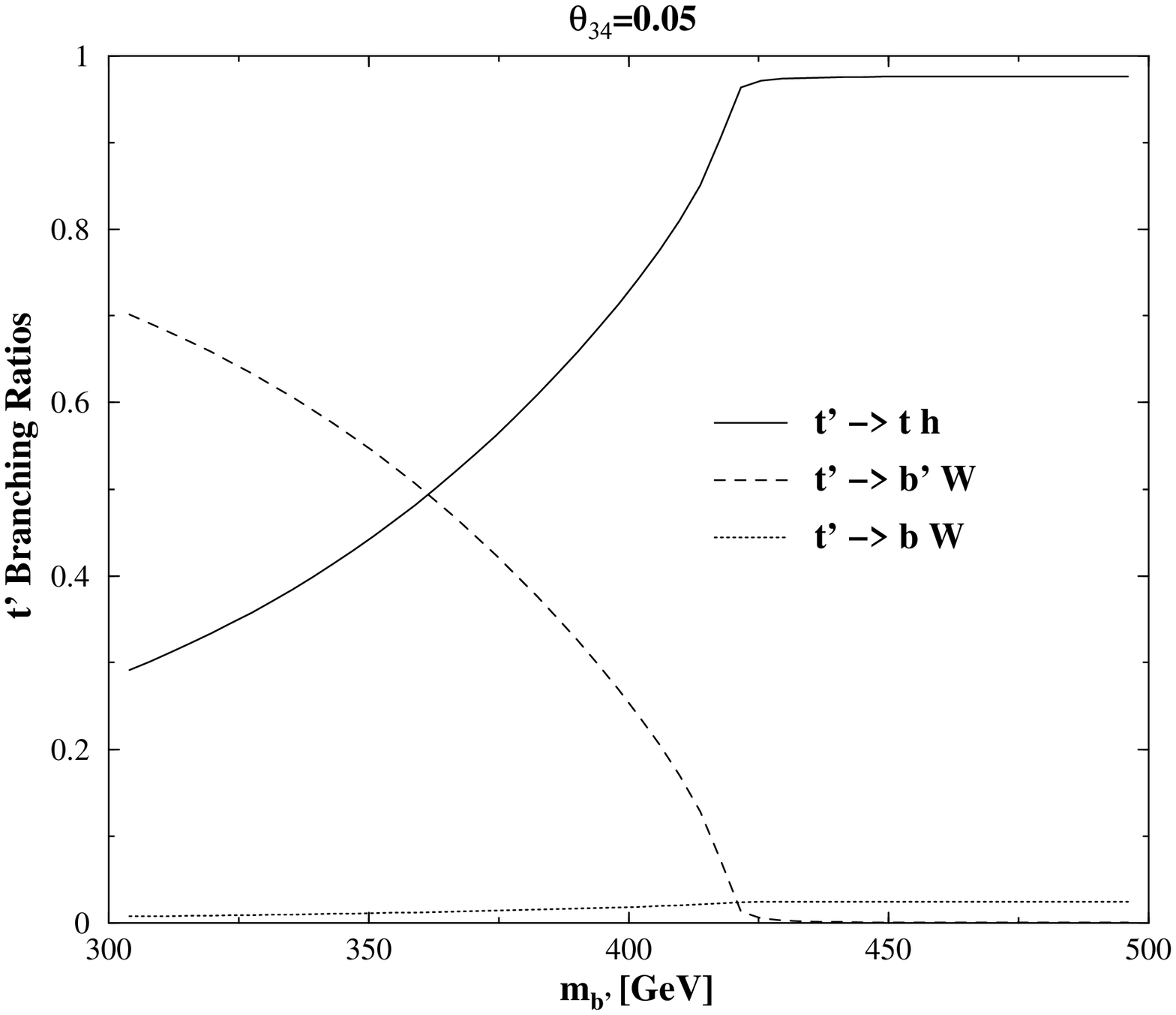,height=8cm,width=8cm,angle=270}
\epsfig{file=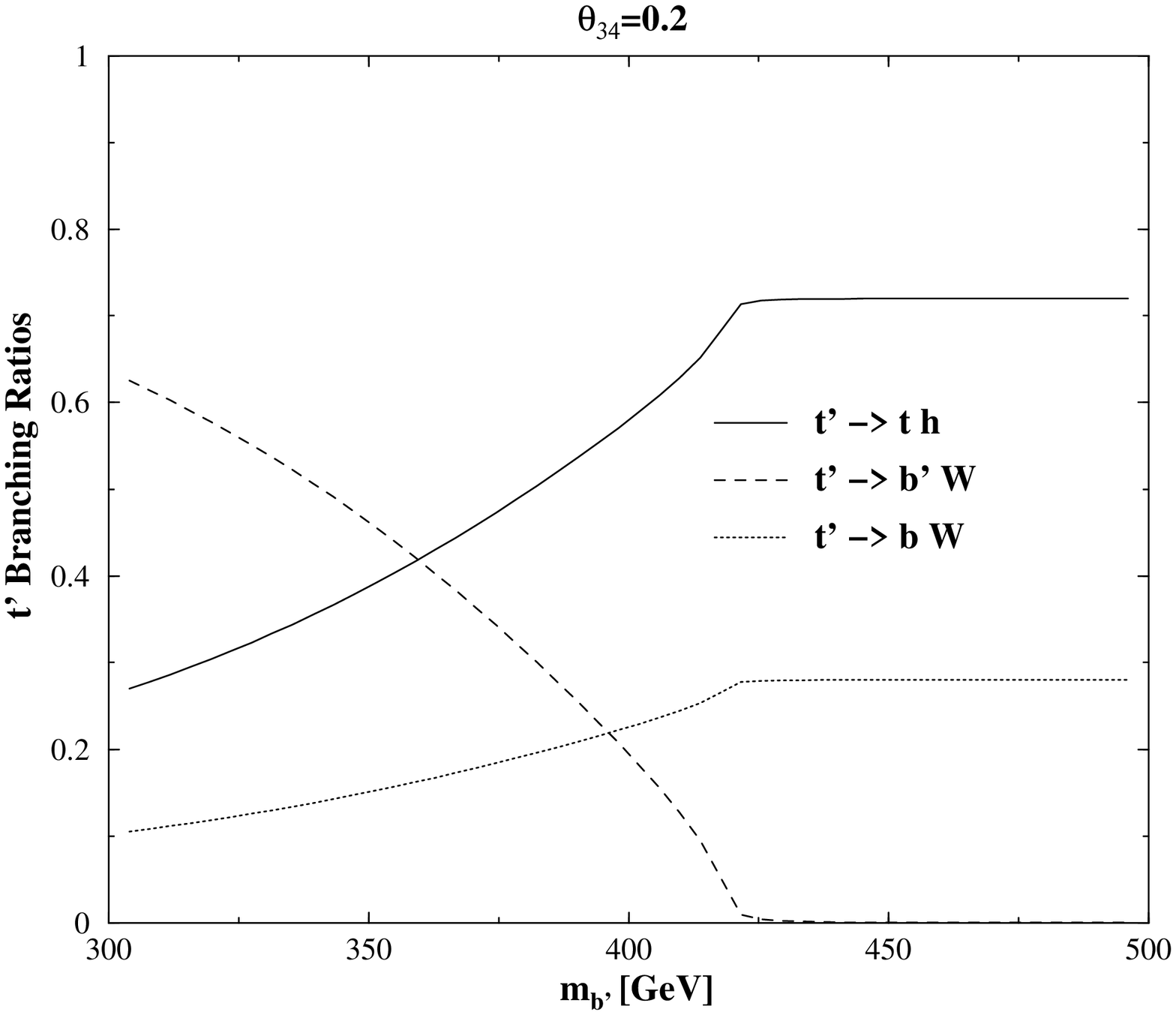,height=8cm,width=8cm,angle=270}
\caption{\emph{The branching ratios for the $t^\prime$ decay channels
$t^\prime \to t h$, $t^\prime \to bW$ and $t^\prime \to b^\prime W^{(\star)}$ ($W^{(\star)}$
is either on-shell or off-shell depending on the $b^\prime$ mass) in the 4G2HDMI,
as a function of $m_{b^\prime}$, for $m_h=125$ GeV, $m_{t^\prime}=500$ GeV, $\epsilon_t =m_t/m_{t^\prime}$,
$\tan\beta=1$, $\theta_{34}=0.05$ (left) and $\theta_{34}=0.2$ (right).
Also, $\alpha=\pi/2$ and $m_{H^+} > m_{t^\prime}$, $m_{A} > m_{t^\prime}$ is assumed.}}
\label{fig1BR}
\end{center}
\end{figure}
\begin{figure}[htb]
\begin{center}
\vspace{1.0cm}
\epsfig{file=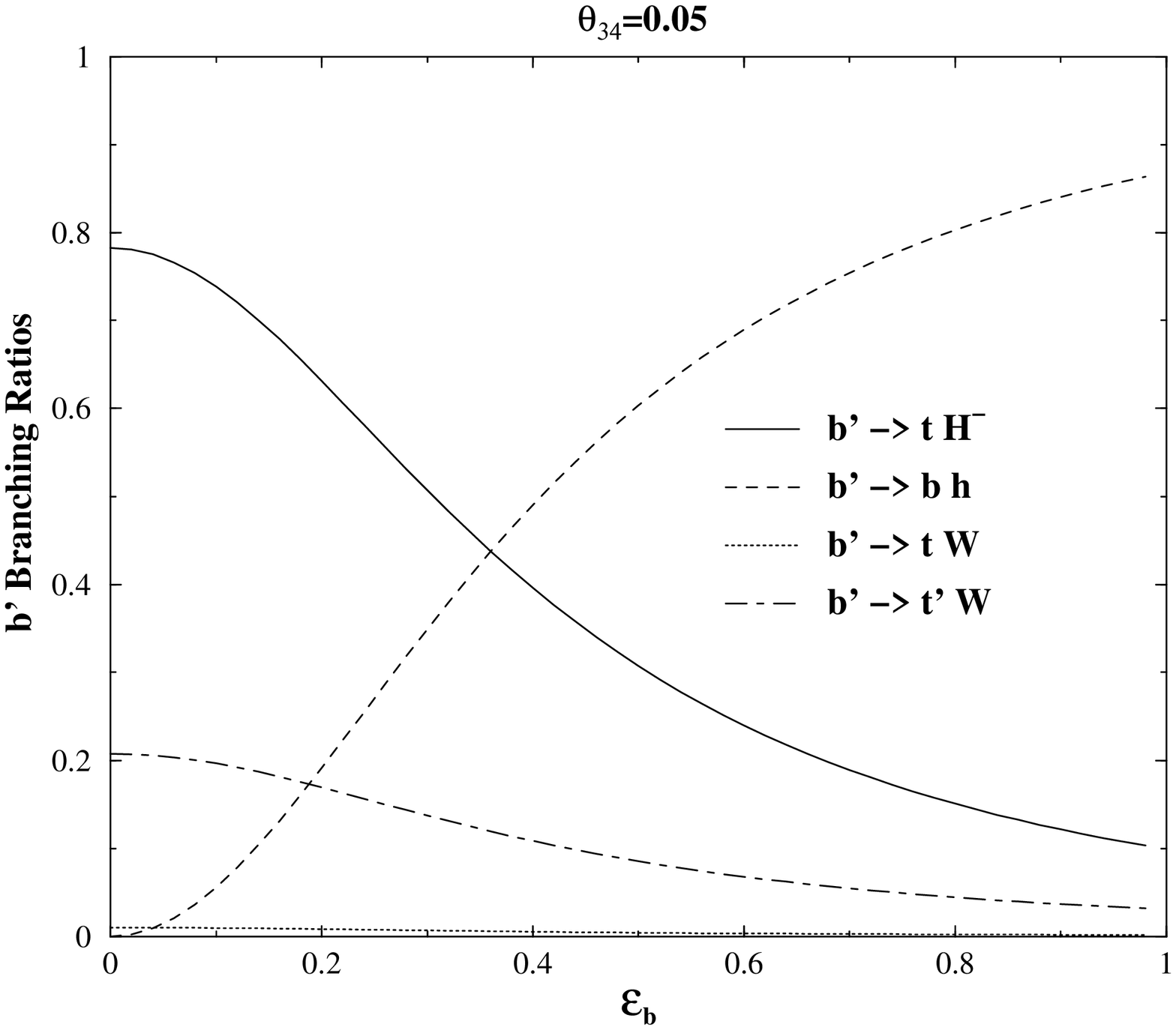,height=8cm,width=8cm,angle=270}
\epsfig{file=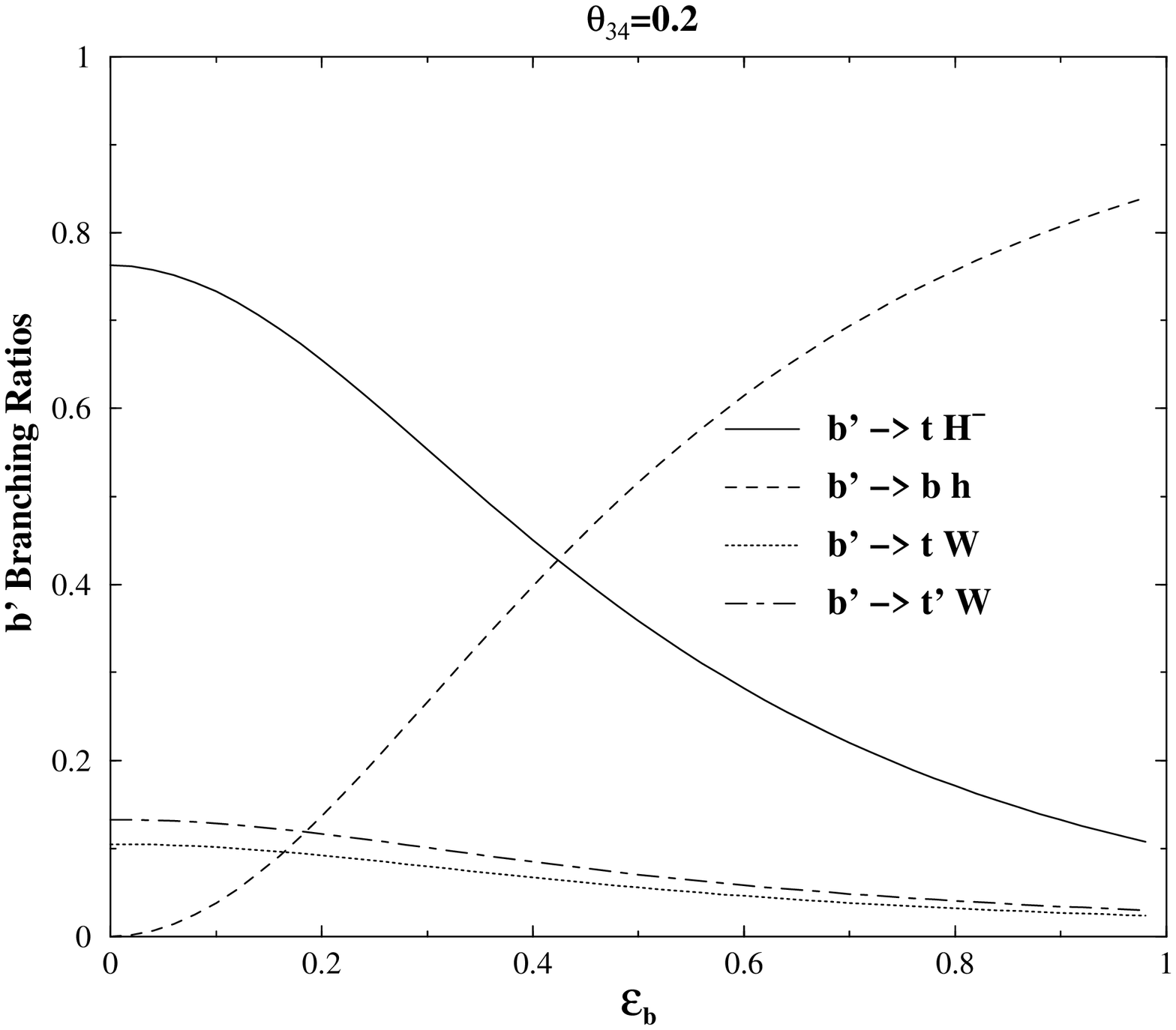,height=8cm,width=8cm,angle=270}
\caption{\emph{The branching ratios for the $b^\prime$ decay channels
$b^\prime \to t H^+$, $b^\prime \to b h$, $b^\prime \to tW$ and $b^\prime \to t^\prime W$
in the 4G2HDMI,
as a function of $\epsilon_b$ for $m_h=125$ GeV, $m_{b^\prime}=500$ GeV, $m_{t^\prime}=400$ GeV,
$m_{H^+}=300$ GeV, $\tan\beta=1$, $\epsilon_t=m_t/m_{t^\prime}$
and , $\theta_{34}=0.05$ (left) and $\theta_{34}=0.2$ (right).
Also, $\alpha=\pi/2$ and $m_{A} > m_{b^\prime}$ is assumed.}}
\label{fig3BR}
\end{center}
\end{figure}
\begin{widetext}
\begin{eqnarray}
{\cal L}(h t^\prime t) &=&
-\frac{g}{2}\frac{m_{t^\prime}}{m_W} \epsilon_t \sqrt{1+ t_\beta^2}
~ \bar t^\prime \left( R + \frac{m_t}{m_{t^\prime}} L \right) t h
\label{htpt}~, \\
{\cal L}(H t^\prime t) &=&
-\frac{g}{2}\frac{m_{t^\prime}}{m_W} \epsilon_t \frac{\sqrt{1+ t_\beta^2}}{t_\beta}
~ \bar t^\prime \left( R + \frac{m_t}{m_{t^\prime}} L \right) t H
\label{hhtpt}~, \\
{\cal L}(A t^\prime t) &=&
i \frac{g}{2}\frac{m_{t^\prime}}{m_W} \epsilon_t \frac{1+ t_\beta^2}{t_\beta}
~ \bar t^\prime \left( R - \frac{m_t}{m_{t^\prime}} L \right) t A
\label{Atpt}~,
\end{eqnarray}
\end{widetext}
and similarly for the ${\cal H}^0 b^\prime b$ vertices
by changing $\epsilon_t \to \epsilon_b$ (and an extra minus sign
in the $A b^\prime b$ coupling).

If $\epsilon_t \sim m_t/m_{t^\prime}$, then the above ${\cal H^0}t^\prime t$
couplings can become sizable, to the level that it might
dominate the decay pattern of the $t^\prime$ (see below).
In fact, large FC effects are also expected in $b^\prime \to b$ transitions
since, even for a very small $\epsilon_b \sim m_b / m_{b^\prime}$,
the FC $h b^\prime b$ and $A b^\prime b$ Yukawa couplings can become sizable
if e.g., $\tan\beta \sim 5 $ for which case they are $\propto \frac{5 m_b}{m_W}$. Therefore, such new
FCNC $t^\prime \to t$ and $b^\prime \to b$  transitions can have drastic phenomenological
consequences for high-energy collider searches of the 4th generation fermions, as we be discussed below.

Furthermore, the flavor diagonal interactions
of the Higgs species with the up-quarks of the 1st, 2nd and 3rd generations
are proportional to $\tan\beta$ in this model,
thus being a factor of $\tan^2\beta$ larger than
the corresponding ``conventional" 2HDMII (i.e., the type II 2HDM) couplings (which
are $\propto \cot\beta$).
For example, this gives rise to an enhanced flavor diagonal
$ht \bar t$ interactions, while
suppressing the $ht^\prime \bar t^\prime$ one,
\begin{widetext}
\begin{eqnarray}
{\cal L}(h t t) & \approx & \frac{g}{2}\frac{m_t}{m_W} \sqrt{1+ t_\beta^2} \left(1 -|\epsilon_t|^2 \right)
~ \bar t t h
\stackrel{|\epsilon_t|^2 \ll 1}{\longrightarrow}
\frac{g}{2}\frac{m_t}{m_W} \sqrt{1+ t_\beta^2}
~ \bar t t h
\label{htt}~,\\
{\cal L}(h t^\prime t^\prime) &\approx& \frac{g}{4}\frac{m_{t^\prime}}{m_W}
\sqrt{1+ t_\beta^2} |\epsilon_t|^2
~ \bar t^\prime t^\prime h \label{htptp}~,
\end{eqnarray}
\end{widetext}
when $|\epsilon_t|^2 \to 0$.

Another important new feature of this model occurs
in the charged Higgs couplings involving the 3rd and 4th generation quarks,
which are completely altered by the presence of the $\Sigma_d$
and $\Sigma_u$ matrices and can thus lead to interesting new effects in both the leptonic (see previous section)
and quark sectors.
For example, taking $V_{t'b},V_{tb'} \ll V_{tb},V_{t^\prime b^\prime}$, the
$H^+ t^\prime b$ and $H^+ t b^\prime$ Yukawa couplings are given in the 4G2HDMI by:
\begin{eqnarray}
&&{\cal L}(H^+ t^\prime b) \approx \frac{g}{\sqrt{2}m_W} t_\beta
\left(1 + t_\beta^{-2} \right)
\bar t^\prime \left( m_t \epsilon_t V_{tb} L - m_{b^\prime} \epsilon_b V_{t^\prime b^\prime}
R \right) b H^+ ~, \\
&&{\cal L}(H^+ t b^\prime) \approx
 \frac{g}{\sqrt{2}m_W} t_\beta \left(1 + t_\beta^{-2} \right)
\bar t \left( m_t^\prime \epsilon_t^\star V_{t^\prime b^\prime}  L - m_{b} \epsilon_b^\star V_{tb}
R \right) b^\prime H^+
\label{hpcoup}~.
\end{eqnarray}

Recalling that in the ``standard" 2HDMII (which would underly a supersymmetric four generation model)
the $\bar t_R b^\prime_L H^+$ would be $\propto m_{t} V_{t b^\prime} / t_\beta$, we find
that in the 4G2HDMI the $\bar t_R b^\prime_L H^+$ coupling is potentially
enhanced by a factor of:
\begin{eqnarray}
\frac{\bar t_R b^\prime_L H^+({\rm 4G2HDMI})}
{\bar t_R b^\prime_L H^+({\rm 2HDMII})} \sim
\epsilon_t \cdot t_\beta^2 \cdot \frac{m_t^\prime}{m_{t}} \cdot
\frac{V_{tb}}{V_{t b^\prime}} ~,
\end{eqnarray}
so that
if e.g., $t_\beta \sim 1$, and $\epsilon_t \sim m_t/m_{t^\prime}$,
there is a factor of $V_{tb}/V_{t^\prime b}$ enhancement to the
$\bar t_R b^\prime_L H^+$ interaction.

These new aspects of phenomenology in the
Yukawa interactions sector can have far reaching implications
for collider searches of the heavy 4th generation quarks and leptons, as
will be discussed in more detail in the next sections.
To see that, one can study the new decay patterns of $t^\prime$ and $b^\prime$
that follow from the above new Yukawa terms.
In particular, in Fig.~\ref{fig1BR} we plot the branching ratios of the
leading $t^\prime$ decay channels (assuming $m_{H^+},m_A > m_{t^\prime}$):
$t^\prime \to t h, ~ bW ,~b^\prime W^{(\star)}$
[$W^{(\star)}$ stands for either on-shell or off-shell $W$ depending
on $m_{b^\prime}$], as a function of
the $b^\prime$ mass.
We use $m_h=125$ GeV, $m_{t^\prime}=500$ GeV, $\tan\beta=1$, $\epsilon_t = m_t / m_{t'}$ and
$\theta_{34}=0.05$ and $0.2$.
We see that the $BR(t^\prime \to t h)$ can easily reach
${\cal O}(1)$ (even for a rather large $\theta_{34} \sim 0.2$ for which $t^\prime \to b W$ becomes sizable),
in particular when $m_{t^\prime} - m_{b^\prime} < m_W$;
see e.g., points 8-11 in Table \ref{tab2}
for which $BR(t^\prime \to t h) \sim {\cal O}(1)$.

In Fig.~\ref{fig3BR} we plot the branching ratios of the leading
$b^\prime$ decay channels $b^\prime \to t H^-,~b h, ~tW ,~t^\prime W$,
as a function
of $\epsilon_b$ for $m_{b^\prime}=500$ GeV, $m_h=125$ GeV,
$\tan\beta=1$, $m_{H^+}=300$ GeV,
$\epsilon_t = m_t / m_{t'}$
and
$\theta_{34}=0.05$ and $0.2$.
We see that in the $b^\prime$ case the dominance of $b^\prime \to t H^-$ (if kinematically allowed)
should be much more pronounced due to the expected smallness of the $b - b^\prime$ mixing parameter,
$\epsilon_b$, which controls the FC decay $b^\prime \to b h$; see e.g., points 12 and 13 in Table \ref{tab2}
for which $BR(b^\prime \to b H^-) \sim {\cal O}(1)$. On the other hand, if $\epsilon_b$ is larger than
about 0.4, then $b^\prime \to b h$ dominates.

\section{implications of the 4G2HDMI for direct searches of 4th generation quarks}

The direct searches of the 4th generation quarks at the LHC currently provide the most stringent limits on their masses.
In particular, CMS reported a 450 GeV lower limit \cite{CMS_limit} on the $t'$ mass in the semileptonic
channel ($pp\rightarrow t^\prime  \overline{t^\prime  }\rightarrow \left[W^+\right]_{hadronic} b
\left[W^-\right]_{leptonic} \overline{b}\rightarrow \ell \nu bq\overline{q} \overline{b} $) and a 557 GeV lower
limit \cite{CMS_limit_dilepton} in the dilepton channel, ($pp\rightarrow t^\prime\overline{t^\prime}\rightarrow
\left[W^+\right]_{leptonic} b \left[W^-\right]_{leptonic} \overline{b}\rightarrow \ell^+ \ell^-
\nu \overline{\nu} b\overline{b}$). The most recent lower bound on the $b'$ mass are
480 GeV \cite{Atlas_bprime} (ATLAS) and 611 GeV \cite{CMS_bprime} (CMS).
\begin{figure}[htb]
\begin{center}
\includegraphics[scale=0.4] {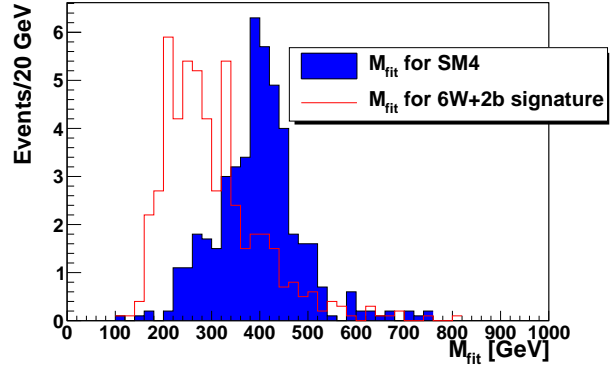}
\end{center}
\caption{\emph{$M_{fit}$ distribution for the SM4 $2W+2b \to l\nu \overline{b} b \overline{q} q$ signature (blue) and for the 4G2HDMI 6$W$+2$b$ signature (red),  for a set of 7 TeV LHC events with $\int Ldt=1$ fb$^{-1}$. For both signatures
$m_{t^\prime  }=450$ GeV is assumed. The peak of the distribution of $M_{fit}$ for the SM4 signature is around $m_{t^\prime  }$, while for the new signature the peak is shifted to a significantly lower value coinciding with the peak of the $\overline{t}t$ background.
Figure taken from \cite{4G2HDM_Geller}.}
\label{Mfit}}
\end{figure}

These searches assumed $Br\left(t^\prime   \rightarrow b W^+\right) \sim {\cal O} \left(1 \right)$, as expected
within the SM4 framework. As was argued above, this is quite unlikely to be the case in models with more that one Higgs doublet,
for which new decay patterns can emerge from the interaction of the heavy quarks with the extended
Higgs sector, e.g. $t'\to ht$ ($b' \to hb$), $t' \to H^+ b$  ($t' \to H^+ b$). In addition, the SM4
forbidden channels $t'\to b' W $ and $b' \to t' W$, depending on the mass hierarchy in the fourth generation doublet,
may no longer be in contradiction with the EWPD if there are more Higgs doublets (see \cite{4G2HDM} and section \ref{sec3}), and may be
kinematically open as well. Taking into account such possible new decay modes to the neutral and charged scalars, one can
define the generic signature \cite{4G2HDM_Geller}: $t' \bar t'/b' \bar b' \to n_W W + n_b b$, with $n_W$ and
$n_b$ being the number of $W$ and $b$ and $\bar b$ jets in the event,
respectively.

\begin{figure}[htb]
\begin{center}
	\includegraphics[scale=0.5]
{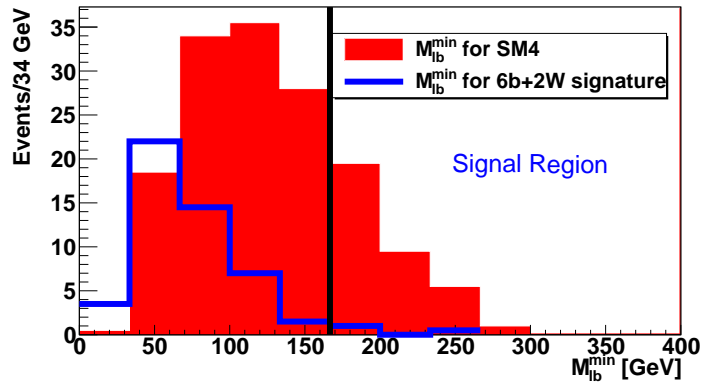}
\end{center}
\caption{\emph{$M^{min}_{lb}$ for the SM4-like $pp\rightarrow t^\prime \overline{t^\prime }\rightarrow 2W+2b$ signature (red) and for the 4G2HDMI $pp\rightarrow t^\prime \overline{t^\prime }\rightarrow 2W+6b$ signature(blue) with
$m_{t^\prime }=350$ GeV for a set of 7 TeV LHC events with $\int Ldt=5$ fb$^{-1}$ in the dilepton channel. The black line is
plotted at the top mass and the region to the right of this line is the ``signal region". Figure taken from \cite{4G2HDM_Geller}. \label{Mlb_dilepton}}.}
\end{figure}

Focusing on the $t'$ case, \cite{Whiteson2012} have reinterpreted the ATLAS $b^\prime$
search (reported in \cite{Atlas_bprime})
to extract limits on $t^\prime$ if it decays via non-SM4 channels such
$t^\prime \to th$ and $t^\prime \to tZ$, whereas
\cite{4G2HDM_Geller} have considered, more specifically, the decay channels $t' \bar t' \to 6W+2b$ and $t' \bar t' \to 2W+6b$, as representatives of such new signatures beyond the SM4.
As was indeed demonstrated in both \cite{4G2HDM_Geller} and \cite{Whiteson2012},
when $t' \to bW$ and $b' \to t W$ are no longer the leading decay
channels, the attempts to impose the SM4-motivated dynamics on processes with a completely different topology
result in a relaxed limit on the fourth generation quarks with respect to the SM4 case.
Specifically, for the $t'$, the CMS analysis in the semileptonic channel was based on the
complete reconstruction of each $\ell \nu bq\overline{q} \overline{b} $ event
(including the reconstruction of the hadronic $W$). The total distribution of
$M_{fit}$ (the reconstructed mass of the $t'$) and $H_T$ (the scalar sum of all
transverse momenta in the event) was used to set a bound on the $t'$ mass.
On the other hand, for the new signatures (e.g., $t' \to t h$), the number of jets in
each event is higher (for example, in the $2W+6b$ signature, there are 8 jets in the
semileptonic channel) and the reconstruction will miss a large part of them, resulting in
$H_T$ and $M_{fit}$ being substantially lower - peaking around the main
$t\overline{t}$ background. An example of this effect is plotted in Figure~\ref{Mfit}.
\begin{figure}[htb]
\begin{center}
	\includegraphics[scale=0.5]
{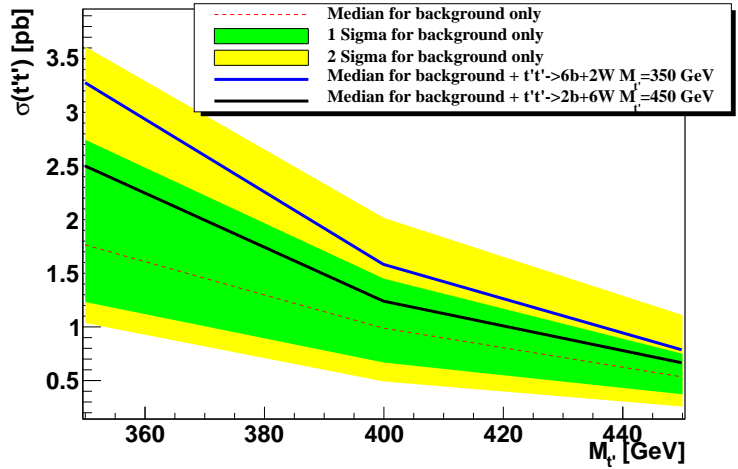}
\end{center}
\caption{\emph{The 95\% CL exclusion plot distribution on the $t^\prime $ mass assuming the SM4 signature in the semileptonic channel ($1\ell+nj+\rlap{ /}{E}_T$) . For the case of background only, the red doted line is the median and the yellow and green bands are the $\pm1$ and $\pm2$ standard deviations accordingly. The black line is the median for background + $t^\prime  t^\prime  \to 2b+6W$ with $m_{t^\prime }=450$ GeV and the blue line is the median for background + $t^\prime  t^\prime  \to 6b+2W$ with $m_{t^\prime }=350$ GeV. The curves for the 4G2HDMI signatures with $m_{t^\prime }=350-450$ GeV lie between these two lines. Figure taken from \cite{4G2HDM_Geller}. \label{Limit}}}
\end{figure}

The analysis in the dilepton channel relies on the fact that $M_{lb}$, which is the invariant mass of a
pair of any lepton and a $b$-jet in the event, is much higher in the underlying $t^\prime  \bar t^\prime $ signal
with respect to the leading $t\overline{t}$ background. In particular, in the case of
$t\overline{t}$, $M_{lb}$ has an upper bound that corresponds to the mass of the top quark,
and therefore in the region above $ \sim 170$ GeV (the ``signal region") $M_{lb}$ is a clean signal
of the SM4-like $t^\prime t^\prime$ production. However, this dilepton search strategy will fail for signatures with more
than 2 leptons or $b$-jets, as in the case of the 4G2HDMI 2$W$+6$b$ and 6$W$+2$b$ signatures,
since the combinatorial background will lower $M_{lb}$, resulting in much less
events in the signal region. An example for this effect is plotted in Figure~\ref{Mlb_dilepton}.

Assuming now that the physics which underlies the 4th generation dynamics goes beyond
the SM4, one can estimate
the extent to which the new signatures are already excluded by the current LHC searches
\cite{4G2HDM_Geller,Whiteson2012}.
Here we will briefly recapitulate the analysis performed
in \cite{4G2HDM_Geller} for both the semileptonic and dilepton channels mentioned above.
For the semileptonic channel, \cite{4G2HDM_Geller} demonstrated, using a naive
simulation of the new beyond SM4 signals in question, what the exclusion plot would be
(using the CMS search strategy which is based on the SM4 $t^\prime \to bW$ decay topology)
if the
data contains in it the 4G2HDMI signals. This was done by ``injecting"
$t^\prime  t^\prime  \to 6b+2W$ events with $m_{t^\prime}=350$ GeV and $t^\prime  t^\prime  \to 2b+6W$ events
with $m_{t^\prime}=450$ GeV.
The results are shown in Figure~\ref{Limit}, which shows that the expected
exclusion curves for the background + $t^\prime  t^\prime  \to 6b+2W$ and
background + $t^\prime  t^\prime  \to 2b+6W$ cases are
less than $2\sigma$ apart from the background only curve.
The curves for the 4G2HDMI signatures with $m_{t^\prime }=350-450$ GeV lie between
the two signal curves shown in the figure.
Thus, using the CMS analysis one would not be able to differentiate between
the no-signal and the 4G2HDMI signal scenarios within $2 \sigma$, so that
we expect the bound on the $t'$ mass within the 4G2HDMI framework
to be no larger than about 400 GeV in the semileptonic channel.
This result is consistent with the most stringent existing limit,
$m_{t'}>423$ GeV,
calculated in \cite{Whiteson2012} by
using templates from the $b'$ search at ATLAS \cite{Atlas_bprime} and
assuming that ${\rm BR}(t^\prime \to t h) \sim 1$.

For the dilepton channel,
the number of events with $M_{lb}^{min}$ in the signal region is negligible for $m_{t'}=350$ GeV
(the lowest mass considered in the CMS analysis) and even less than that for higher $m_{t'}$
(see Figure~\ref{Mlb_dilepton}). One can, therefore, conclude that the CMS  dilepton  analysis is completely irrelevant
for the 4G2HDMI signatures.

\begin{figure}[t]
\begin{center}
	\includegraphics[scale=0.5]
{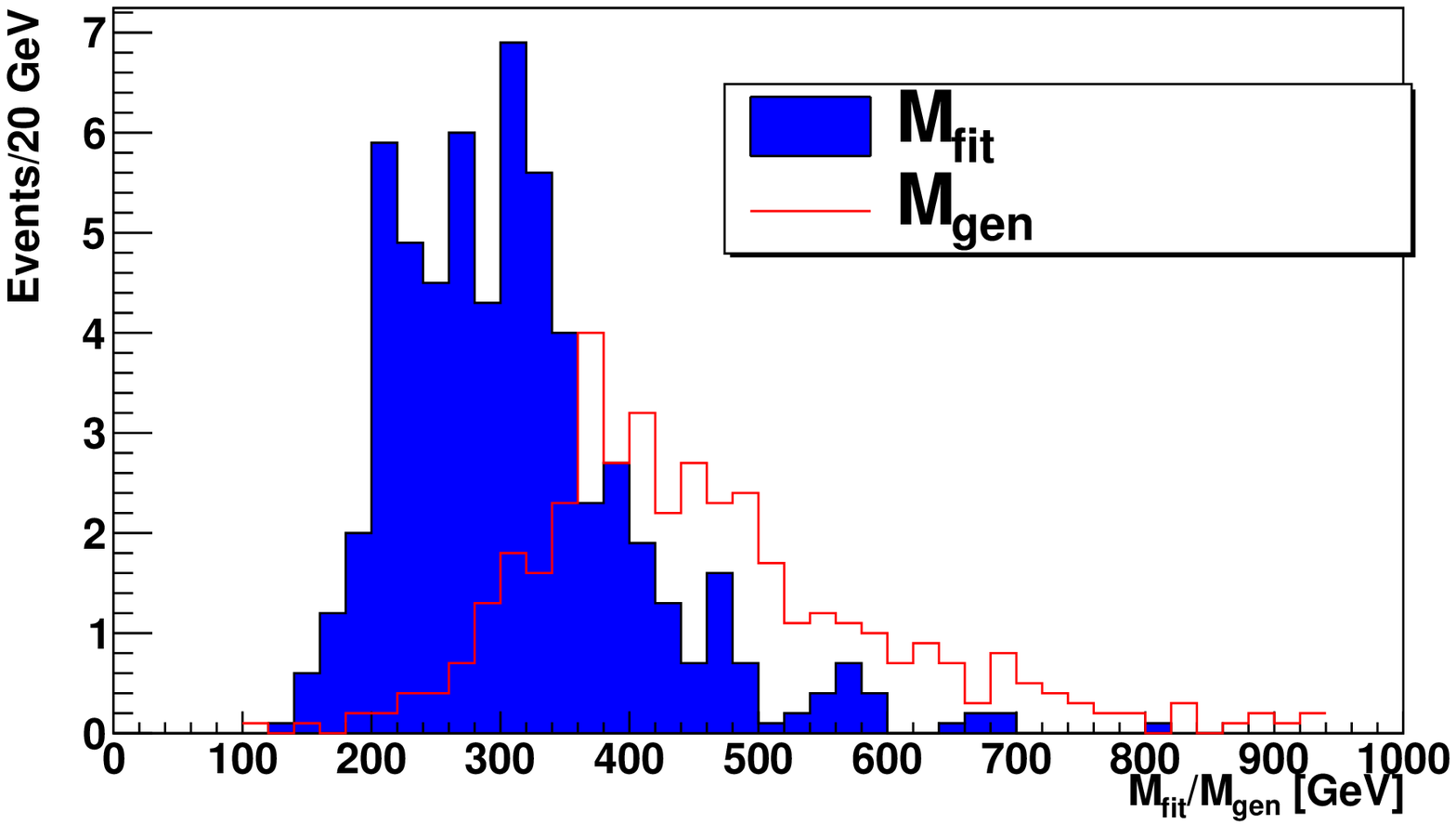}
\end{center}
\caption{\emph{Comparison between $M_{fit}=m\left(l\nu b\right)=m\left(q\overline{q}b\right)$ - the reconstructed $t^\prime $ mass using the CMS method - (in blue) and $M_{gen}=m\left(Left\;  Side \right)=m\left(Right\; Side \right)$ - the reconstructed $t^\prime $ mass using the method suggested in \cite{4G2HDM_Geller} - (in red), for the $pp\rightarrow t^\prime \overline{t^\prime }\rightarrow 2W+6b$ signature with $m_{t^\prime }=450$ GeV at the LHC with a c.m. of 7 Tev and $\int Ldt=1$ fb$^{-1}$, in the semileptonic channel ($1\ell+nj+\rlap{ /}{E}_T$). See also text. Figure taken from \cite{4G2HDM_Geller}. \label{gen_fit}}}
\end{figure}

As was suggested in \cite{4G2HDM_Geller}, an analysis that uses a more general reconstruction
method could avoid the kinematic
misrepresentation of the beyond SM4 events in both the semileptonic and dilepton channels,
and thus yield a higher sensitivity to NP (beyond the SM4) events containing the 4th generation fermions.  An example of that is plotted in Figure~\ref{gen_fit} for the semileptonic channel, which shows how the misconstruction of the $t'$ mass can be surmounted.

\section{implications for direct searches of the Higgs \label{sechiggs}}

The recently observed new Higgs-like particle with a mass of $\sim 125$ GeV (at the level of $\sim 5 \sigma$, see \cite{CMSHiggs,ATLASHiggs}) is
the first potential evidence for a Higgs boson which can be consistent with the SM picture.
Furthermore, a study of the combined Tevatron data has revealed a smaller broad excess in the $b\overline{b}W$ channel,
which can be related to the production of $hW$
with a Higgs mass between 115 GeV and 135 GeV \cite{TevatronHiggs}.
These searches further exclude a SM Higgs with masses between $ \sim 130 - 600$ GeV.

The quantity that is usually being used for comparison between the LHC and Tevatron
results and the expected signals in
various models is the ratio:
 \begin{eqnarray}
 R^{Model(Obs)}_{XX}=\frac{\sigma \left(pp/p\overline{p} \to h \to XX\right)_{Model(Obs)}}{\sigma \left(pp/p\overline{p} \to h \to XX\right)_{SM}} ~,
 \end{eqnarray}
which is the observed ratio of cross-sections, i.e., the signal strengths $R^{Obs}_{XX}$,
and the errors in the different channels are \cite{CMSHiggs,ATLASHiggs,TevatronHiggs}:$^{[1]}$\footnotetext[1]{We combine the results
from the CMS and ATLAS experiments (for $pp/p\overline{p} \to hW \to b\overline{b} W $
we combine the results from CMS and Tevatron), where in cases where the measured value
was not explicitly given we estimate it from the published plots.}
\begin{itemize}
 \item $VV \to h \to \gamma \gamma$: $2.2\pm 1.4$ (taken from $\gamma \gamma+2j)$
 \item $gg \to h \to \gamma \gamma$: $1.68\pm 0.42$
 \item $gg \to h \to W W^*$: $0.78\pm 0.3$
 \item $gg \to h \to Z Z^*$: $0.83\pm 0.3$
 \item $gg \to h \to \tau \tau$: $0.2\pm 0.85$
 \item $pp/p\overline{p} \to hW \to b\overline{b} W $: $1.8\pm 1.5$
\end{itemize}

One can easily notice that the channels which have the highest sensitivity to the Higgs signals and contributed the most to the recent 125 GeV Higgs discovery are
$h\to \gamma \gamma$ and $h\to ZZ^*,WW^*$. In all other channels the results are not
conclusive, and at this time, they are consistent with the background only
hypothesis at the level of less than $2\sigma$.

As was shown recently in \cite{12091101},
the above reported measurements are not compatible with the SM4 at the level of $5 \sigma$.
In particular, light Higgs production through gluon fusion is
enhanced by a factor of $\sim 10$ in the SM4 due to the contribution of
diagrams with $t'$ and $b'$ in the loops, which
in general leads to larger signals (than what was observed at the LHC)
in the $h \to ZZ/WW/\tau\tau$ channels. For
a light Higgs with a mass $m_h<150$ GeV and 4th generation masses of ${\cal O}(600)$ GeV, $h\to ZZ/WW$ is
in fact suppressed by a factor of $\sim 0.2$ due to NLO corrections \cite{Passarino,Passarino2}, and the exclusion is based mainly on the $\tau\tau$ channel.
In the $h\to \gamma\gamma$ channel there is also a substantial suppression of ${\cal O}(0.1)$
due to (accidental) destructive interference in the loop \cite{Kribs_EWPT,He3} and another ${\cal O}(0.1)$ factor due to NLO corrections \cite{Passarino,Passarino2}.
If $\nu_4$ is taken to be light enough, then $Br(h\to \nu_4 \nu_4)$ becomes ${\cal O}\left(1\right)$,
suppressing all the other channels and the exclusion gets eased. This, however, further suppresses the
$\gamma\gamma $ channel to the level that the observed excess can no longer be accounted for \cite{Lenz2}.
Therefore, as was also noted in \cite{Lenz,Lenz2,Nir}, the SM4 is strongly disfavored for any
$m_{\nu_4}$, even without considering the $\tau\tau$ channel.
\begin{figure}[htb]
\begin{center}
\epsfig{file=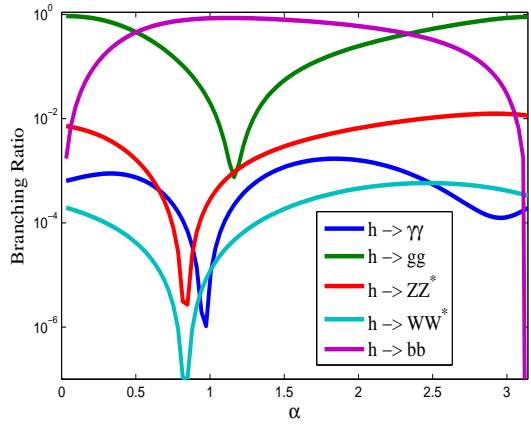,height=6cm,width=8cm,angle=0}
\vspace{-0.8cm}
\end{center}
\caption{\emph{The relevant branching ratios of $h$ in the 4G2HDMI,
as a function of $\alpha$,
with $m_h=125$ GeV, $M_{4G}=400$ GeV, $\epsilon_t=0.5$ and $\tan\beta=1$. Figure taken from \cite{GellerHiggs}.}
\label{Br_H}}
\end{figure}

The comparison to any given model can be performed using a $\chi^2$ fit defined as:
\begin{equation}
\chi^2=\sum_X{\frac{\left(R^{Model}_{XX}-R^{Obs}_{XX}\right)^2}{\sigma_{XX}^2}}~,
\end{equation}
 where $\sigma_{XX}$ are the errors on the observed cross-sections and
 $R^{Model}_{XX}$ is calculated using the program Hdecay \cite{Hdecay} with recent
 NLO contributions (which also
 include the heavy 4th generation fermions for the 4th generation scenarios).
 One can take advantage of the fact that $\frac{\sigma\left(YY \to h \right)_{Model}}{\sigma\left(YY \to h \right)_{SM}}=\frac{\Gamma\left(h\to YY \right)_{Model}}{\Gamma\left(h\to YY \right)_{SM}}$, and calculate $R^{Model}_{XX}$ using
 \begin{eqnarray}
 R^{Model}_{XX}=\frac{\Gamma \left(h\to YY \right)_{Model}}{\Gamma \left(h\to YY \right)_{SM}} \cdot \frac{Br \left(h\to XX \right)_{Model}}{Br \left(h\to XX \right)_{SM}}~,
  \end{eqnarray}
 where $YY\to h$ is the Higgs production mechanism, i.e., either by gluon fusion $gg\to h$, vector boson fusion $WW/ZZ \to h$ or associated
  Higgs-W production, $W^*\to hW$ at Tevatron.

 In multi-Higgs 4th generation frameworks, the picture becomes more complicated,
 since there are new scalar states with new Yukawa couplings depending on
 $\tan{\beta}$ and $\alpha$ ($\alpha$ is the mixing angle in the neutral Higgs sector),
 as well as couplings to the $W$ and the $Z$ bosons which are proportional to
 $\sin \left(\alpha-\beta \right) $
and $\cos \left(\alpha-\beta \right)$ (with the exception of the pseudoscalar
$A$ which does not couple at tree-level
to the $W$ and the $Z$).
Furthermore, the specific
Yukawa structure can vary depending on the type of the multi-Higgs model, e.g., for the 4G2HDMI case considered below there is an additional parameter, $\epsilon_t$,
which parameterizes the $t_R - t^\prime_R$
mixing (see section \ref{sec2} and \cite{4G2HDM}).
In Fig.~\ref{Br_H} we plot the branching ratios of $h$ as a function of $\alpha$ in the 4G2HDMI, for $m_h=125$ GeV,
$\tan \beta=1$, $\epsilon_t=0.5$ and $M_{4G}=m_{t'}$=$m_{b'}=m_{l_4}=m_{\nu_4}=400$ GeV.

 Let us now examine how well the the 2HDM scenarios with a 4th generation of fermions
 fit the measured Higgs mediated cross-sections listed above with $m_h=125$ GeV.
 The simplest case to study is the ``standard" 2HDMII (i.e., the 2HDM of type II extended to include a fourth fermion family)
 with the pseudoscalar $A$ being the lightest scalar,
 since its couplings do not depend on $\alpha$ \cite{Gunion,Bern}.
 However, as was already noted in \cite{Gunion}, for the ``standard" 2HDMII
 the case of a light $A$ decaying to the $\gamma \gamma$ mode is excluded when all
 4th generation fermions are heavy.
With the new results, in particular, the signals of the 125 GeV Higgs decaying into a pair of vector bosons, the case of the $A$ being the lightest
scalar is excluded irrespective of the 4th generation fermion masses.

\begin{figure*}[htb]
\begin{center}$
\begin{array}{cc}
\epsfig{file=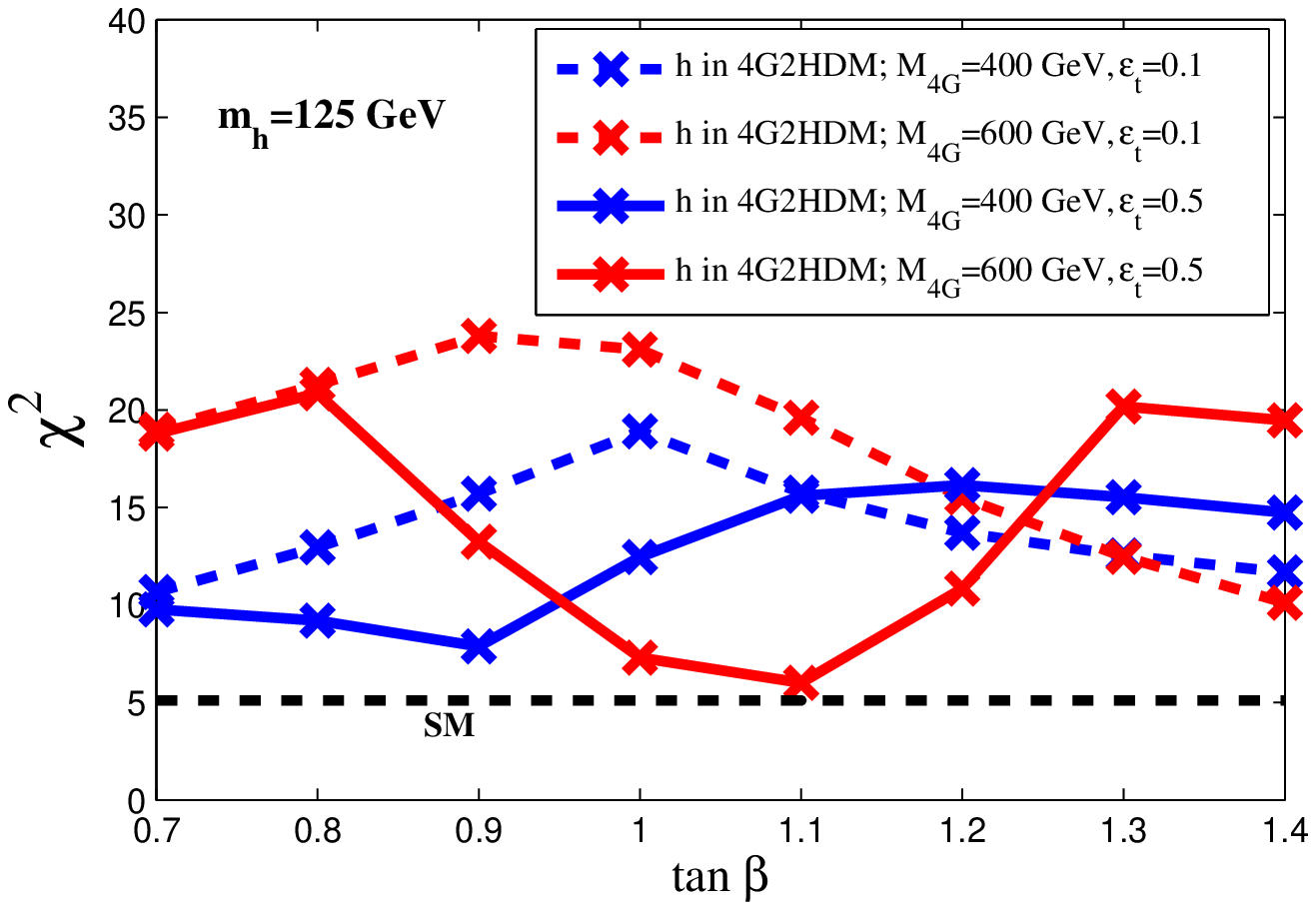,height=6cm,width=8.25cm,angle=0}
\epsfig{file=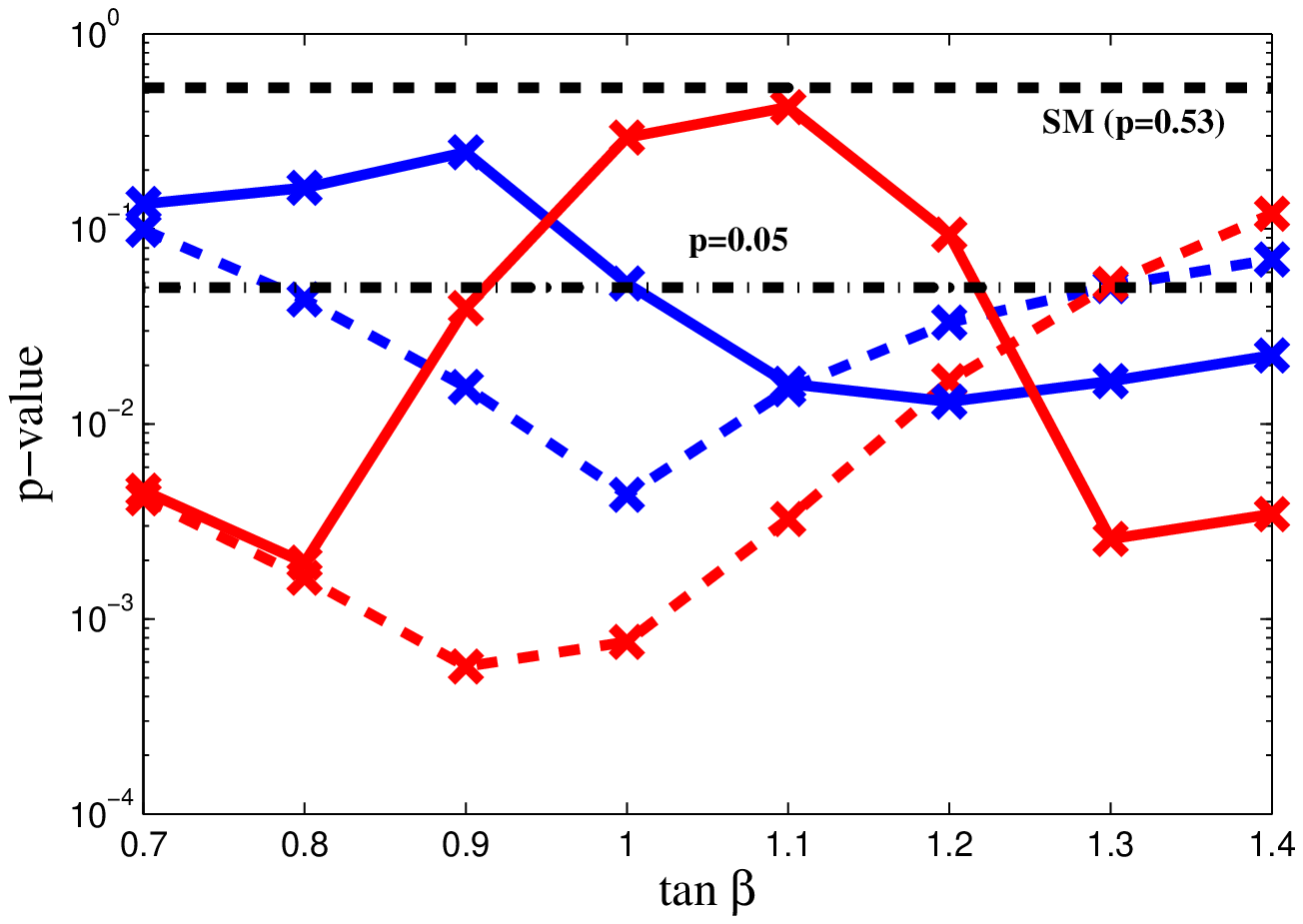,height=6cm,width=8.25cm,angle=0}
\vspace{-0.8cm}
\end{array}$
\end{center}
\caption{\emph{$\chi^2$ (left plot) and p-values (right plot), as a function of $\tan\beta$,
for the lightest 4G2HDMI CP-even scalar $h$, with $m_h=125$ GeV, $\epsilon_t =0.1$ and $0.5$ and
$M_{4G} \equiv m_{t'}=m_{b'}=m_{l_4}=m_{\nu_4}=400$ and $600$ GeV.
The value of the Higgs mixing angle $\alpha$ is
the one which minimizes $\chi^2$ for each value of $\tan\beta$.
The SM best fit is shown by the horizontal dashed-line and the dash-doted line in the right plot
corresponds to $p=0.05$ and serves as a reference line. Figure taken from \cite{GellerHiggs}. \label{ChiSqAll1}}}
\end{figure*}
\begin{figure*}[htb]
\begin{center}$
\begin{array}{cc}
\epsfig{file=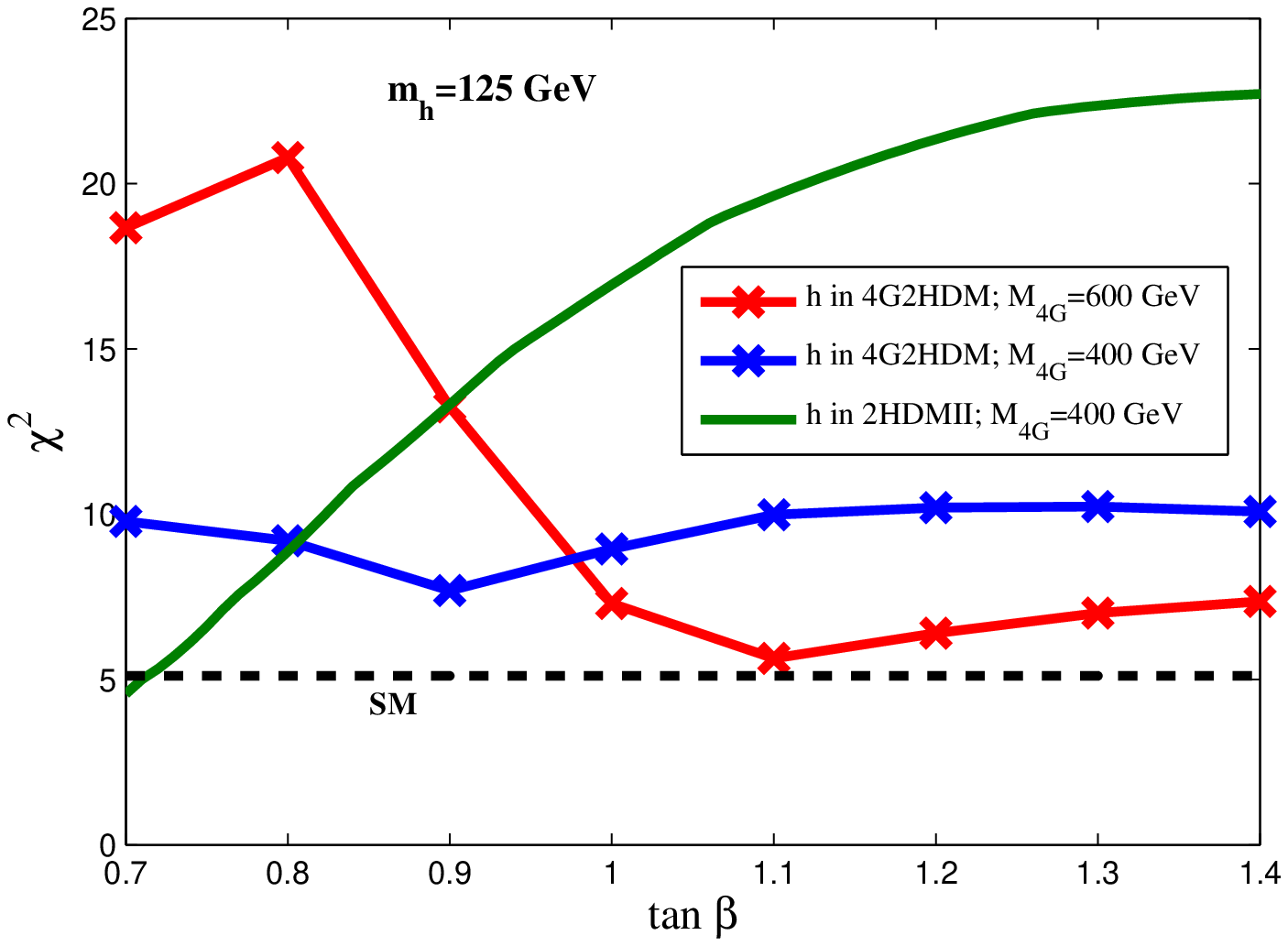,height=6cm,width=8.25cm,angle=0}
\epsfig{file=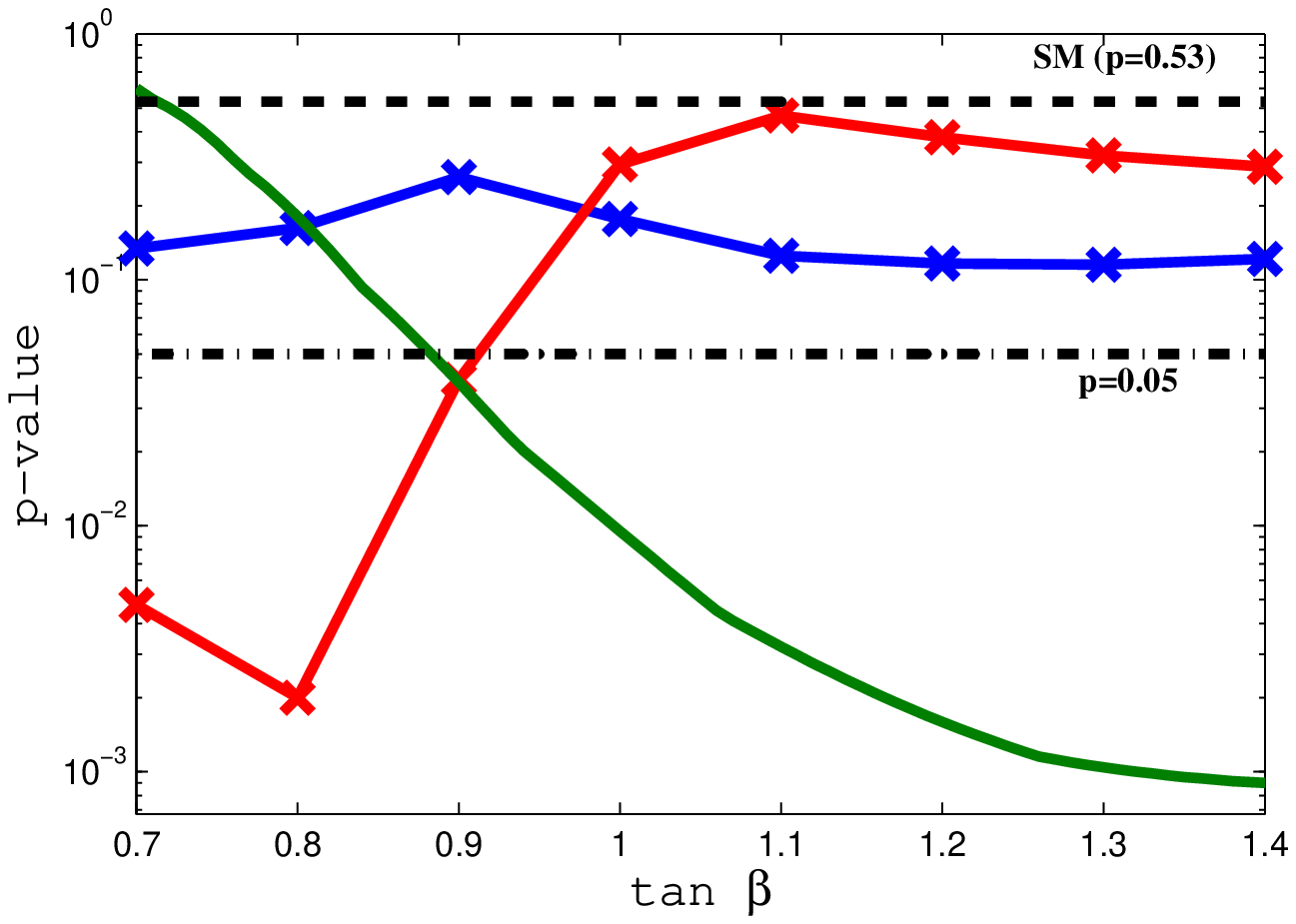,height=6cm,width=8.25cm,angle=0}
\vspace{-0.8cm}
\end{array}$
\end{center}
\caption{\emph{Same as Figure \ref{ChiSqAll1},
where here we minimize with respect to both $\epsilon_t$ and $\alpha$
for each value of $\tan\beta$.
Also shown are the $\chi^2$ and p-values
for a $125$ GeV Higgs in the SM and in
the type II 2HDM with a 4th generation of fermions (denoted by 2HDMII). Figure taken from \cite{GellerHiggs}.}
\label{ChiSqAll2}}
\end{figure*}

Here we wish to extend the previous analysis made for the 2HDMII scenario by calculating
the $\chi^2$ for the light Higgs with a mass $m_h = 125$ GeV,
both for the 4G2HDMI of \cite{4G2HDM} and for the 2HDMII with a 4th generation of fermions, and to
compare it to the SM.
We follow the analysis in \cite{GellerHiggs}, which used the latest version of
Hdecay \cite{Hdecay}, where
all the relevant couplings for the 4G2HDMI and for the 2HDMII
frameworks were inserted. For the treatment of the NLO corrections to $h\to VV$, \cite{GellerHiggs} used the approximation of a degenerate 4th generation spectrum, where
two cases were studied: $m_{t'}=m_{b'}=m_{\ell_4}=m_{\nu_4} \equiv M_{4G}= 400$ and $600$ GeV
 (while the first case, i.e., $M_{4G}= 400$, is excluded for the SM4, it is
 not necessarily excluded for the 2HDM setups,
 as discussed in the previous section).
Note that the 4th
generation neutrino is taken to be heavy enough,
so that the decays of the light Higgs into a pair of $\nu^\prime$
are not considered, thus limiting the discussion to the effects of the
altered Higgs couplings in the 2HDM
frameworks with respect to the SM4.

Indeed, \cite{GellerHiggs} found that the best fit is obtained for the light CP-even Higgs, $h$,
whereas the other neutral Higgs particles of the 2HDM setups, i.e., $H$ and $A$,
cannot account for the observed data.

The resulting $\chi^2$ and p values
in the 4G2HDMI case
(combining all the six reported Higgs decay channels above),
with $m_h=125$ GeV, $M_{4G}=400$ and 600 GeV, $\epsilon_t = 0.1$ and 0.5 and
for $0.7<\tan \beta<1.4$ (this range is roughly the EWPD and flavor physics allowed range in these 2HDM setups, see section \ref{sec3}) are shown
in Fig.~\ref{ChiSqAll1}.
The value of the Higgs mixing angle $\alpha$ is the one
which minimizes the $\chi^2$ for each value of $\tan\beta$.
The SM best fit is also shown in the plot.
In Fig.~\ref{ChiSqAll2} we further show the resulting
$\chi^2$ and p-values as a function of $\tan\beta$,
this time minimizing for each
value of $\tan\beta$ with respect to both $\alpha$ and $\epsilon_t$ (in the 4G2HDMI case).
For comparison, we also show in Fig.~\ref{ChiSqAll2} the $\chi^2$ and p-values for a
125 GeV $h$ in the 2HDMII with a 4th generation
and in the SM.

Looking at the p-values in Figs.~\ref{ChiSqAll1} and \ref{ChiSqAll2}
(which ``measure" the extent to which
a given model can be successfully used to interpret the Higgs data
in all the measured decay channels)
we see that, $h$ of the 4G2HDMI with $\tan\beta \sim {\cal O}(1)$ and
$M_{4G}=400-600$ GeV is a
good candidate for the recently observed 125 GeV Higgs,
giving a fit comparable to the SM fit. This conclusion is not changed by
explicitly adding the EWPD as an additional
constraint to the above analysis (i.e., the p-values stay roughly the
same, see \cite{GellerHiggs}).
The ``standard" 2HDMII setup with $M_{4G} = 400$ GeV
is also found to be consistent with the Higgs data
in a narrower range of $\tan\beta \lsim 0.9$.
Also, the fit favors a large $t - t^\prime$ mixing
parameter $\epsilon_t$, implying ${\rm BR}(t^\prime \to th) \sim {\cal O}(1)$
which completely changes the $t^\prime$ decay pattern
\cite{4G2HDM} and, therefore, significantly
relaxing the current bounds on $m_{t^\prime}$ (see previous section).

\begin{figure}[ht]
\begin{center}
$
\begin{array}{cc}
\epsfig{file=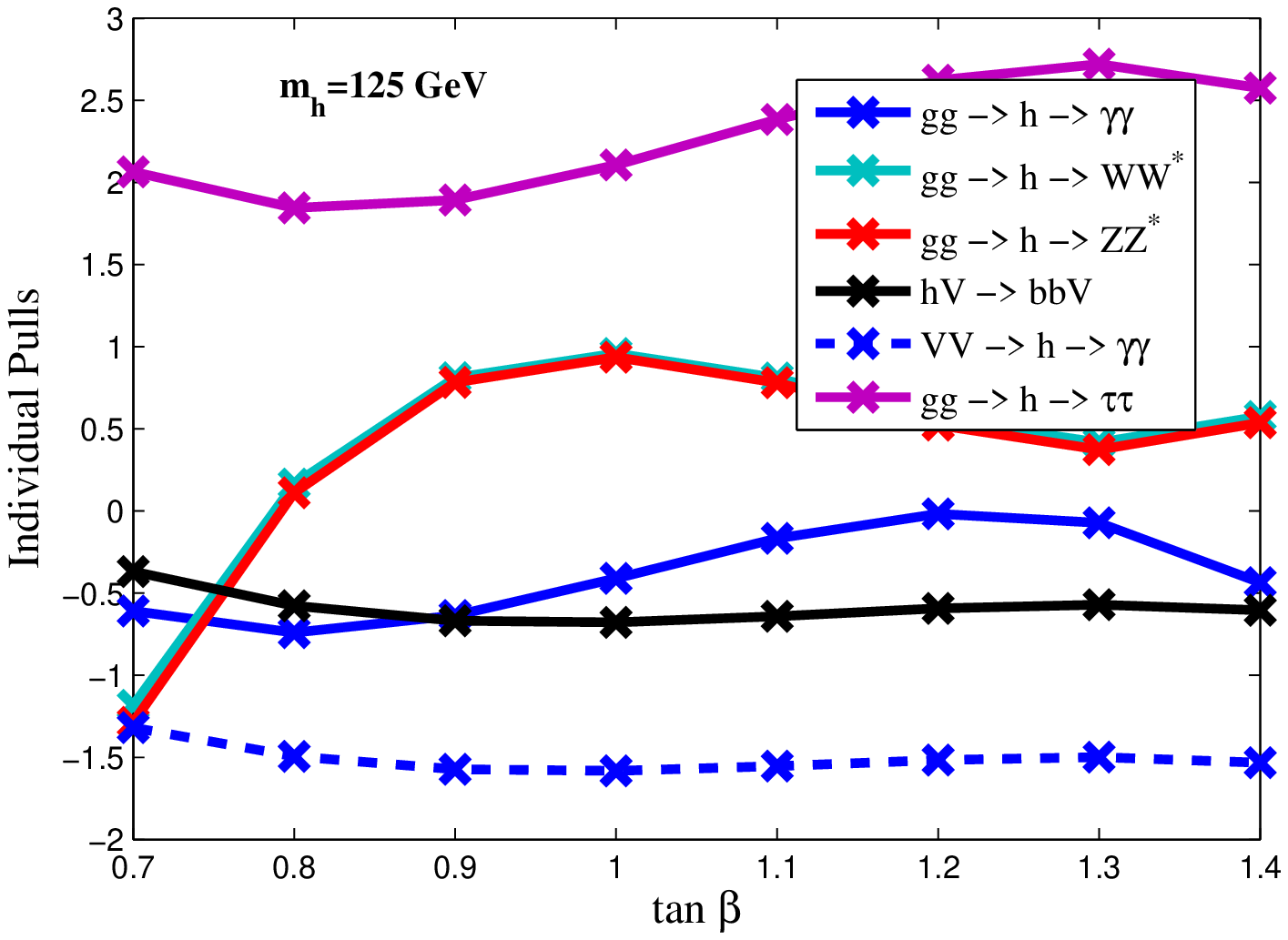,height=6cm,width=8.25cm,angle=0}
\epsfig{file=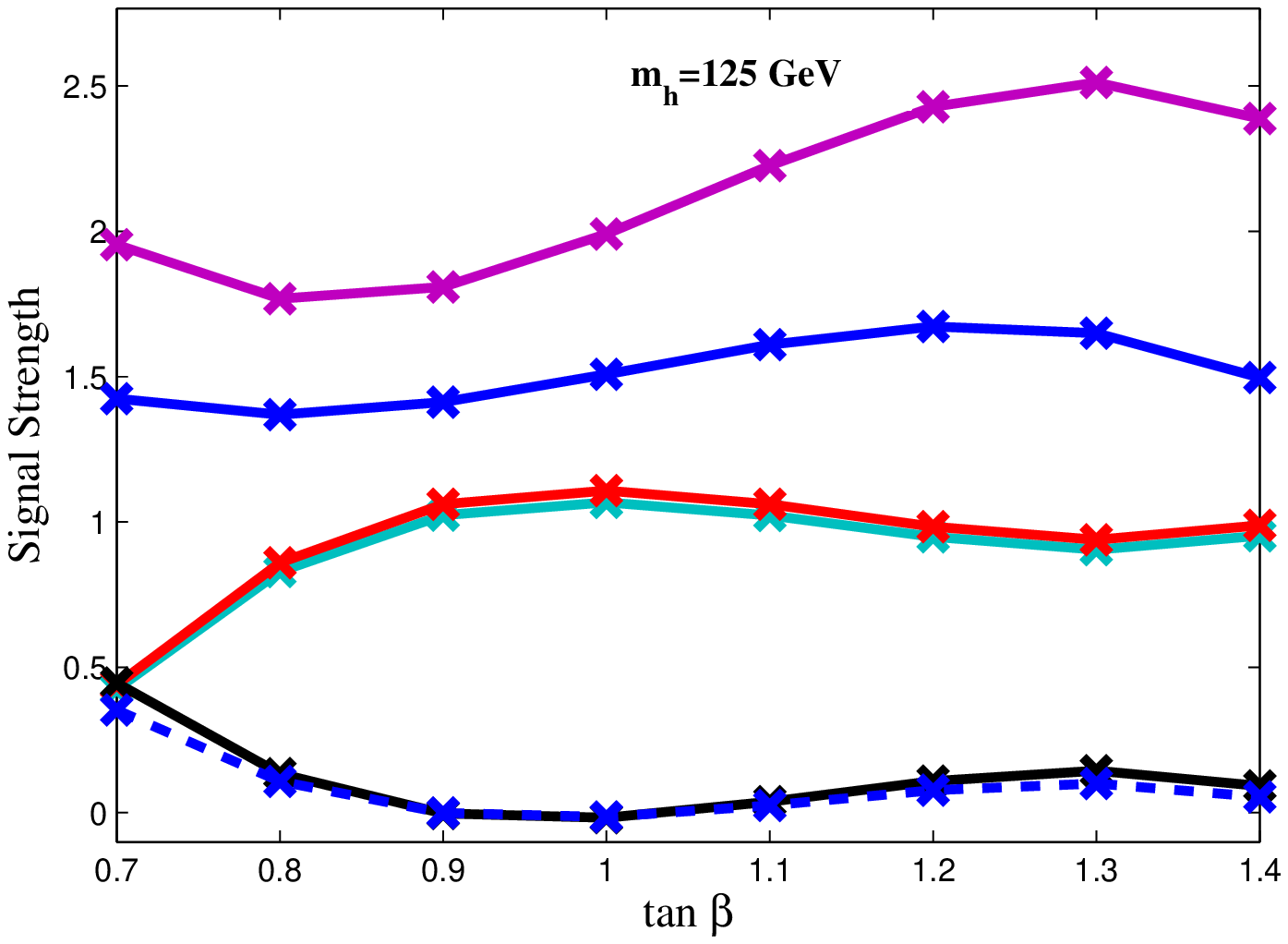,height=6cm,width=8.25cm,angle=0}
\vspace{-0.8cm}
\end{array}$
\end{center}
\caption{\emph{The individual pulls
$\frac{\left(R^{Model}_{XX}-R^{Obs}_{XX}\right)}{\sigma_{XX}}$
(left plot), and the
signal strengths $R^{Model}_{XX}$ (right plot), in the different channels,
that correspond to the best fitted 4G2HDMI curve
with $m_h=125$ GeV and $M_{4G}=400$ GeV,
shown in Fig.~\ref{ChiSqAll2}. Figure taken from \cite{GellerHiggs}.}
\label{indv_pulls}}
\end{figure}

However, more data is required to effectively distinguish between the 4G2HDMI scalars and the SM Higgs.
In particular, in Fig.~\ref{indv_pulls} we show the individual pulls and the signal strengths for the best fitted $h$ signals (i.e., with $m_h=125$ GeV)
in the 4G2HDMI with $M_{4G}=400$ GeV. We can see that appreciable deviations
from the SM are expected in the channels
$gg \to h \to \tau \tau$, $VV \to h \to \gamma \gamma$ and $h V \to bb V$.
In particular,
the most notable effects
are about a $1.5 \sigma$ deviation (from the observed value) in the VBF diphoton channel
$VV\to h \to \gamma\gamma$ and a $2-2.5 \sigma$ deviation in the
$gg \to h \to \tau \tau$ channel. The deviations in these channels
are in fact a prediction of the 4G2HDMI strictly based on the current Higgs data,
which could play a crucial role as data
with higher statistics becomes available. They can be understood as follows:
the channels that dominate the fit (i.e., having a higher statistical
significance due to their smaller errors) are $gg \to h \to \gamma \gamma,ZZ^*,WW^*$.
Thus, since the $gg \to h$ production vertex
is generically enhanced by the $t^\prime$ and $b^\prime$ loops,
the fit then searches for values of the relevant 4G2HDMI parameters which
decrease the $h \to \gamma \gamma,ZZ^*,WW^*$ decays in the appropriate
amount. This in turn leads to an enhanced $gg \to h \to \tau \tau$
(i.e., due to the enhancement in the $gg \to h$ production vertex)
and to a decrease in the $VV \to h \to \gamma \gamma$ and
$p \bar p/pp \to W \to h W \to bb W$, which are
independent of the enhanced $ggh$ vertex
but are sensitive to the decreased $VVh$ one.
It is important to note that some of the characteristics of these
``predictions" can change with more data collected.

Finally, \cite{GellerHiggs} also finds that
for the best fitted 4G2HDMI case,
the heavier CP-even scalar, $H$, is excluded by the current data
(in particular by the $ZZ$ and $WW$ searches) up to $m_H \sim 500$ GeV, whereas
a CP-odd state, $A$, as light as 130 GeV is allowed by current data (for more details see \cite{GellerHiggs}).

\pagebreak

\section{Summary}

We have addressed several fundamental and challenging questions
(that we have outlined in the introduction)
regarding the nature and underlying dynamics of
the physics and phenomenology of 4th generation fermions, if they exist. We have argued
that:

\begin{enumerate}

\item The current stringent bounds on the masses of the 4th generation quarks,
i.e., $m_{q^\prime} \gsim 400$ GeV, are indicative of NP, possibly of a strongly coupled nature,
since such new heavy fermionic degrees of freedom naturally lead to a Landau pole at the nearby TeV-scale,
which may be viewed as the cutoff of 4th generation low-energy theories.

\item The
fact that the 4th generation fermions must be so heavy is, therefore, of no surprise
since their large mass stands out as a strong hint for the widely expected new TeV-scale physics,
where the new heavy fermionic states may be considered to be the agents of EWSB.
\item If indeed the 4th generation fermions are linked to
strong dynamics and/or to compositeness at the nearby TeV-scale, then one is forced to extend
the minimally constructed SM4 framework which is not compatible with this viewpoint and neither with current data.
In particular, in this case
one should expect the sub-TeV particle spectrum to accommodate several new scalar composites of the 4th family fermions.
The challenge in this scenario is to construct a viable theory that can adequately parameterize the physics of TeV-scale
compositeness and that will guide us to the detection of these new states at the LHC.
\end{enumerate}

We have, thus, suggested and reviewed a class of 2HDM's - extended to include a 4th family of fermions -
that can serve as low-energy effective models for the TeV-scale compositeness scenario, and then analyzed/discussed:

\begin{itemize}

\item The constraints on these models from EWPD as well as from low-energy
flavor physics.

\item The expected new phenomenology and the implications for collider searches of the
4th generation heavy fermions as well as of the multi-Higgs states of these models.

\end{itemize}

We have found that it is indeed possible to construct a natural 2HDM framework with heavy 4th generation fermions
with a mass in the range $400 - 600$ GeV, which is consistent with EWPD and which is not excluded by the
recent direct measurements at the current high energy colliders.

In particular, we found that, under the 2HDM frameworks for the 4th generation described in this article,
one can

\begin{itemize}

\item Relax the current mass bounds on the
4th generation quarks.

\item Successfully fit the recently measured 125 GeV Higgs signals,
to the parameters of the 2HDM with roughly similar
quality of fit as the one achieved for the SM with 3 generations.
This result is in sharp contrast to the poor fit obtained with the minimal SM4 setup which
is, therefore, excluded.

\end{itemize}

Finally, we have shown that, if such an extended 4th generation 2HDM setup is realized in nature,
then one should expect to observe further hints for the underlying TeV-scale dynamics
in direct high energy collider signals involving the 4th generation fermions
and the associated new scalars as well as in low energy flavor physics.

\bigskip
\bigskip

{\bf Acknowledgments:} SBS and MG acknowledge research support from the Technion.
The work of AS was supported in part by the U.S. DOE contract
\#DE-AC02-98CH10886(BNL).

\end{document}